\documentclass[a4paper,11pt]{article}
\pdfoutput=1
\usepackage{jheppub}

\usepackage{amsmath}
\usepackage{mathtools}
\usepackage{multirow}
\usepackage{array}
\usepackage{bm}

\usepackage{graphicx}
\usepackage{lscape}
\usepackage{subfig}

\usepackage{bbm}
\usepackage{hyperref}
\usepackage{booktabs}
\usepackage{float}

\newcommand{\zb}{\ensuremath \bar{z}}

\newcommand{\be}{\begin{equation}}
\newcommand{\ee}{\end{equation}}
\newcommand{\bea}{\begin{eqnarray}}
\newcommand{\eea}{\end{eqnarray}}
\newcommand{\bei}{\begin{itemize}}
\newcommand{\eei}{\end{itemize}}
\newcommand{\bean}{\begin{eqnarray*}}
\newcommand{\eean}{\end{eqnarray*}}
\newcommand{\nn}{\nonumber \\}

\def\eps{\epsilon}

\def\top #1{\mathcal{T}_{#1}}

\allowdisplaybreaks[1]

\newcommand\scalemath[2]{\scalebox{#1}{\mbox{\ensuremath{\displaystyle #2}}}}

\newcommand{\X}{X^{0}}
\newcommand{\zz}{z}
\newcommand{\zzb}{\bar{z}}
\newcommand{\xx}{\vec{x}}
\newcommand{\xxi}{\vec{x}_0}
\newcommand{\xxf}{\vec{x}}
\newcommand{\gammaAB}[2]{\ensuremath\gamma}
\newcommand{\gammas}{\ensuremath\gamma_s}
\newcommand{\gammat}{\ensuremath\gamma_t}
\newcommand{\chen}[2]{\ensuremath{\mathcal{C}_{#1}^{\,[#2]}}}

\newcommand{\logxi}[1]{\ensuremath{\log \eta_{#1}(\xxi)}}
\newcommand{\logxf}[1]{\ensuremath{\log \eta_{#1}(\xxf)}}
\newcommand{\logxt}[2]{\ensuremath{\log \eta_{#1}(\xx(#2))}}

\newcommand{\dA}{\ensuremath d\AA}
\newcommand{\pathord}{\ensuremath \mathcal{P}}

\newcommand{\dAk}[1]{\ensuremath \underbrace{\dA \ldots \dA}_{\text{#1 times}}}
\newcommand{\ie}{i.e.\ }
\newcommand{\eg}{e.g.\ }
\newcommand{\Den}{\ensuremath D}
\newcommand{\dd}{\ensuremath \mathrm{d}}
\DeclareMathOperator{\dlog}{\mathit{d}log}

\newcommand{\minus}{\ensuremath \scalebox{0.5}[1.0]{\( - \)}}
\newcommand{\pminus}{\hphantom{\minus}}
\newcommand{\FF}{\ensuremath \text{F}}
\newcommand{\FFvec}{\ensuremath \mathbf{F}}
\newcommand{\GG}{\ensuremath \text{I}}
\newcommand{\GGvec}{\ensuremath \mathbf{I}}

\newcommand{\KK}{\ensuremath \mathbb{K}}
\newcommand{\MM}{\ensuremath \mathbb{M}}
\renewcommand{\AA}{\ensuremath \mathbb{A}}

\newcommand{\unipd}{Dipartimento di Fisica ed Astronomia, Universit\`a di Padova, Via Marzolo 8, 35131 Padova, Italy}
\newcommand{\pdinfn}{INFN, Sezione di Padova, Via Marzolo 8, 35131 Padova, Italy}
\newcommand{\desy}{DESY, Notkestra\ss e 85, D-22607 Hamburg, Germany}
\newcommand{\argonne}{High Energy Physics Division, Argonne National Laboratory, Argonne, IL 60439, USA}

\allowdisplaybreaks

\author[a]{Stefano Di Vita,}
\author[b,c]{Pierpaolo Mastrolia,}
\author[b,c]{Amedeo Primo,}
\author[d]{Ulrich Schubert}

\affiliation[a]{\desy}
\affiliation[b]{\unipd}
\affiliation[c]{\pdinfn}
\affiliation[d]{\argonne}

\preprint{\texttt{DESY 17-032}}

\emailAdd{stefano.divita@desy.de}
\emailAdd{pierpaolo.mastrolia@pd.infn.it}
\emailAdd{amedeo.primo@pd.infn.it}
\emailAdd{schubertmielnik@anl.gov}

\title{Two-loop master integrals for the leading QCD corrections to
  the Higgs coupling to a $\bm{W}$ pair and to the triple gauge
  couplings $\bm{ZWW}$ and $\bm{\gamma^*WW}$}


\abstract{ We compute the two-loop master integrals required for the
  leading QCD corrections to the interaction vertex of a massive
  neutral boson $\X$, \eg $H,Z$ or $\gamma^{*}$, with a pair of $W$
  bosons, mediated by a $SU(2)_L$ quark doublet composed of one
  massive and one massless flavor.  All the external legs are allowed
  to have arbitrary invariant masses.
  The Magnus exponential is employed to identify a set of master
  integrals that, around $d=4$ space-time dimensions, obey a canonical
  system of differential equations.  The canonical master integrals
  are given as a Taylor series in $\epsilon = (4-d)/2$, up to order
  four, with coefficients written as combination of Goncharov
  polylogarithms, respectively up to weight four.
  In the context of the Standard Model, our results are relevant for
  the mixed EW-QCD corrections to the Higgs decay to a $W$ pair, as
  well as to the production channels obtained by crossing, and to the
  triple gauge boson vertices $ZWW$ and $\gamma^*WW$.}

\begin{document}

\maketitle
\flushbottom

\section{Introduction}
The calculation of Feynman integrals with massive particles both as
external legs and as internal lines is not an easy task. Indeed, the
combination of external kinematic invariants and internal masses may
give rise to physical and non-physical singularities which require the
use of special functions, with non-trivial arguments, embedding
them. Identifying the rise of such functions within the parametric
representation of Feynman integrals is very challenging.  Therefore,
rather than by direct integration, multivariate Feynman integrals may
be more simply determined by solving differential equations
(DEs)~\cite{Kotikov:1990kg,Remiddi:1997ny,Gehrmann:1999as}.

In general, Feynman integrals in dimensional regularization obey
relations that can be used, on the one side, to identify a basis of
independent integrals, dubbed {\it master integrals} (MIs), and, on
the other side, to write special equations satisfied by the MIs
themselves. MIs are found to obey systems of linear, first-order
partial DEs in the kinematic variables.  Solving these equations,
provided that their values or behavior at special points is known, becomes
a method to completely determine MIs, hence to compute Feynman
integrals alternatively to their direct integration
(as reviewed in~\cite{Argeri:2007up,Henn:2014qga}).

The entries of the matrix associated to the system of DEs depend, in
general, on the kinematic invariants and on the space-time dimension
$d$. Although it is a mathematically interesting problem, finding the
expression of the MIs for arbitrary values of $d$ is not always
possible and, according to the physical context, it may be sufficient
to know the MIs around a critical dimension $d_c$, with
$d = d_c + \eps$ and $\eps \to 0$.  In a perturbative approach, the
solution of the system around $d=d_c$ may admit a representation in
terms of iterated integrals (as reviewed
in~\cite{Brown:2009lectures,Duhr:2014woa}), where the matrix
associated to the system constitutes the integration kernel.
Therefore, the structure of such a matrix has a direct impact on the
form of the solutions, namely on the functions required to classify
them: simplifying the matrix means simplifying the solutions.

The idea of finding MIs that obey \textit{canonical} systems of DEs,
i.e. systems with an associated matrix where the dependence on the
space-time dimensions is decoupled from the
kinematics~\cite{Henn:2013pwa,Henn:2013woa}, has led to a substantial
improvement of the system-solving
strategy~\cite{Gehrmann:2014bfa,Magnus, Lee:2014ioa,
  Papadopoulos:2014lla,Remiddi:2016gno,Meyer:2016slj,Primo:2016ebd,Prausa:2017ltv,Gituliar:2017vzm,Adams:2017tga},
and to the availability of many novel results.  In the case of Feynman
integrals that depend on several scales, we have shown that the Magnus
exponential~\cite{Magnus} is an efficient tool to derive MIs obeying
canonical systems starting from a basis of MIs that obey systems of DEs whose matrix has a
\textit{linear} dependence on the space-time
dimension~\cite{Argeri:2014qva,DiVita:2014pza,Bonciani:2016ypc}.

\begin{figure}
  \centering
  \subfloat{%
    \includegraphics[width=0.23\textwidth]{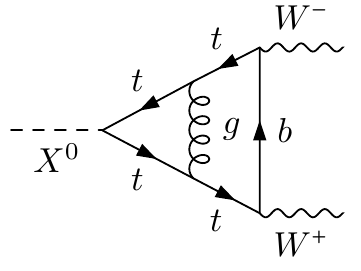}
  }
  \subfloat{%
    \includegraphics[width=0.23\textwidth]{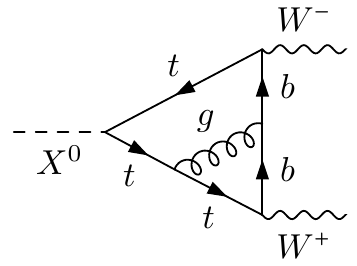}
  }
  \subfloat{%
    \includegraphics[width=0.23\textwidth]{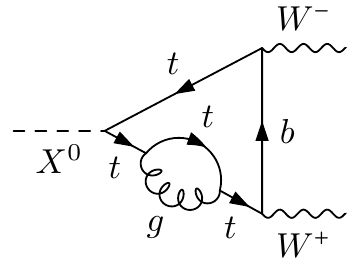}
  }
  \subfloat
  {%
    \includegraphics[width=0.23\textwidth]{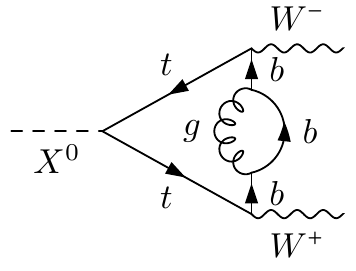}
  }
  \caption{Representative two-loop Feynman diagrams contributing to
    the $\X\, W^{+}W^{-}$ interaction, where
    $\X=H,Z,\gamma^{*}$. Similar diagrams where $t$ and $b$ quarks are
    exchanged are also taken into account. The diagrams have been
    generated using \texttt{FeynArts}~\cite{Hahn:2000kx}.}
  \label{fig:diags}
\end{figure}

In the Standard Model, the coupling between two $W$ bosons and one
neutral boson $\X=H,Z,\gamma^*$ is present in the tree-level
Lagrangian\footnote{Several motivated extensions of the Standard
  Model feature an extended Higgs sector with Yukawa couplings to the
  $SU(2)$ fermion doublets. In particular, one or more neutral
  pseudoscalar bosons might be part of the spectrum, together with
  other neutral scalar bosons. While we do not refer explicitly to
  this possibility, our results would also be applicable to the case
  $X^0 = S^0,\,A^0$, where we schematically indicate with $S^0$ the
  scalars and with $A^0$ the pseudoscalars.}. At one-loop, the
$\X\, W^{+}W^{-}$ interaction receives electro-weak (EW) corrections,
either \emph{via} bosonic- or \emph{via} fermionic-loop. Strong (QCD)
corrections must proceed through a closed quark-loop so that they can
first occur at the two-loop level.
In this article, we present the calculation of the two-loop
three-point integrals required for the determination of the leading
QCD corrections to the interaction vertex between a neutral boson $\X$
with arbitrary mass and a pair of $W$ bosons of arbitrary squared
four-momenta ($\X\, W^{+}W^{-}$), mediated by a fermion loop of a
$SU(2)_L$ quark doublet, with one massive and one massless flavors.
In what follows, we refer to the massive flavor as to the \textit{top}
($m_t = m$), and to the massless one as to the \textit{bottom}
($m_b = 0$).  Representative Feynman graphs for the considered
integrals are shown in figure~\ref{fig:diags}.
The MIs for the case in which only massless quarks propagate in the
loops has been previously studied
in~\cite{Usyukina:1994iw,Birthwright:2004kk,Chavez:2012kn}.

Our results represent the full set of MIs needed to compute the
${\cal O}(\alpha \alpha_s)$ corrections to the Higgs decay into a pair
of $W$ bosons, and to the triple gauge boson processes $Z^*WW$ and
$\gamma^*WW$, with leptonic final states, at $e^+ e^-$ colliders.
  As for the latter process with semi-leptonic or hadronic final
  states, our MIs would only be a subset of
  the needed MIs.
They are also a subset of the MIs needed for the computation of the
two-loop mixed EW-QCD corrections to the Higgs production cross
section either in the $WW$-fusion channel or in association with a $W$
boson, and to $WW$ production in higher multiplicity processes.
The same MIs would also be needed for the computation of a class of
diagrams entering the NNLO EW corrections. Except for the
first-generation quarks (that are approximately degenerate), the
fermionic one-loop diagrams always involve an $SU(2)_L$ doublet with a
(nearly or exactly) massless flavor.  Is is then clear that the
corrections due to photon exchange between the fermionic lines share
the same topologies as the ones of the leading QCD corrections.

We distinguish two sets of integrals, according to the flavor that
couples to the $\X$ boson, \ie either the massive or the massless one,
for which integration-by-parts reduction returns 24 and 23 MIs,
respectively.  The calculation of the MIs proceeds according to the
following strategy. We identify a set of MIs that obey systems of DEs
whose matrix has a \textit{linear} dependence on $d = 4 - 2\eps$, and,
by means of the Magnus exponential, we derive a canonical set of
master integrals.  The matrices associated to the canonical systems
admit a logarithmic-differential form ($\dlog$) with rational
arguments, therefore, the canonical MIs can be cast in Taylor series
around $d=4$ with coefficients written as combinations of Goncharov
polylogarithms
(GPLs)~\cite{Goncharov:polylog,Remiddi:1999ew,Gehrmann:2001pz,Vollinga:2004sn}.
Boundary conditions are imposed by requiring the regularity of the
solutions at special kinematics points, and by using simpler integrals
as independent input.  The analytic expressions of the MIs have been
numerically evaluated with the help of
\texttt{GiNaC}~\cite{Bauer:2000cp} and successfully tested against the
values provided by the public computer code
\texttt{SecDec}~\cite{Borowka:2015mxa}. The package
\texttt{Reduze}~\cite{vonManteuffel:2012np} has been used throughout
the calculations.

The paper is organized as follows. In section~\ref{notations} we fix
our notation and conventions. In section~\ref{sec:diffeq} we discuss
the general features of the systems of DEs satisfied by the master
integrals and its general solution in terms of iterated integrals. In
section~\ref{sec:WWMIs} we describe the computation of the two-loop
MIs for $\X\, W^{+}W^{-}$ in Euclidean kinematics and in
section~\ref{sec:continuation} we discuss the analytic continuation of
our result. Conclusions are given in section~\ref{sec:conclusions}. In
appendix~\ref{sec:chen} we recall the main properties of iterated
integrals. In appendix~\ref{sec:gMIcoeff} we list the coefficients of
the linear combinations of MIs that satisfy a canonical system of DEs
and finally, in appendix~\ref{sec:dlog-form}, we give the expressions
of the $\dlog$-form of the matrices associated to such systems. The
analytic expressions of the canonical MIs up to ${\cal O}(\eps^5)$ are
attached to the \texttt{arXiv} version of the manuscript as ancillary
files.

\section{Notation and conventions \label{notations}}
In this paper we will consider the two-loop three-point functions of a
$\X$ boson with momentum $q$, and two W bosons with momenta $p_1$ and
$p_2$,
\begin{align}
  \X (q) \to W^+(p_1) + W^-(p_2)
\end{align}
where
\begin{align}
  s = q^2=(p_1+p_2)^2  \qquad {\rm and} \qquad  p_1^2\neq p_2^2 \neq 0 \ .
\end{align}
The calculation involves the evaluation of Feynman integrals in
$d=4-2\eps$ dimensions of the type
\begin{gather}
  \int \widetilde{\dd^d k_1}\widetilde{\dd^d k_2}\,
  \frac{1}{\Den_{a_1}^{n_1} \ldots \Den_{a_p}^{n_p}}.
\end{gather}
In our conventions, the integration measure is defined as
\begin{equation}
  \widetilde{\dd^dk_i} = \frac{\dd^d k_i}{(2\pi)^d} \left(\frac{i \, S_\eps}{16 \pi^2} \right)^{-1} \left( \frac{m^2}{\mu^2} \right)^{\eps}, 
\label{eq:intmeasure1}
\end{equation}
where $m^2$ is the mass of the \textit{top} quark circulating in the loops, $\mu$ the 't Hooft scale of dimensional regularization and
\begin{equation}
  S_\eps = (4\pi)^\eps \, \Gamma(1+\eps) \, .
  \label{eq:intmeasure2}
\end{equation}
 	

\section{System of differential equations for master integrals}
\label{sec:diffeq}

In this section we briefly discuss the general features of the systems
of DEs obeyed by the MIs and the properties of the corresponding
solutions. The details of the calculations of the MIs for
$\X\, W^{+}W^{-}$ are described in section~\ref{sec:WWMIs}.

\begin{figure}
	\centering
 	\subfloat[\label{fig:FigTop1WWbb}]{%
 		\includegraphics[width=0.23\textwidth]{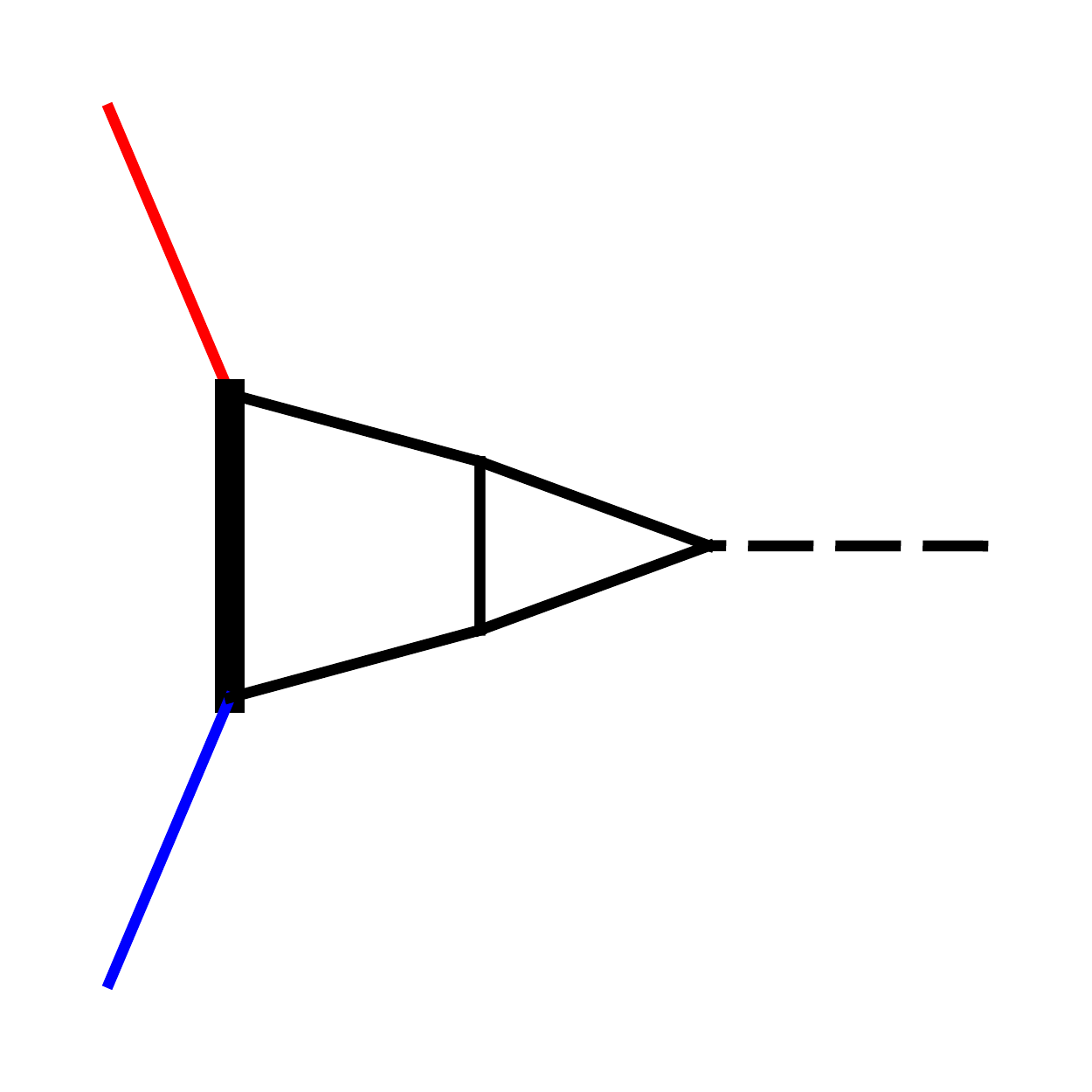}
 	}
 	\subfloat[\label{fig:FigTop2WWbb}]{%
 		\includegraphics[width=0.23\textwidth]{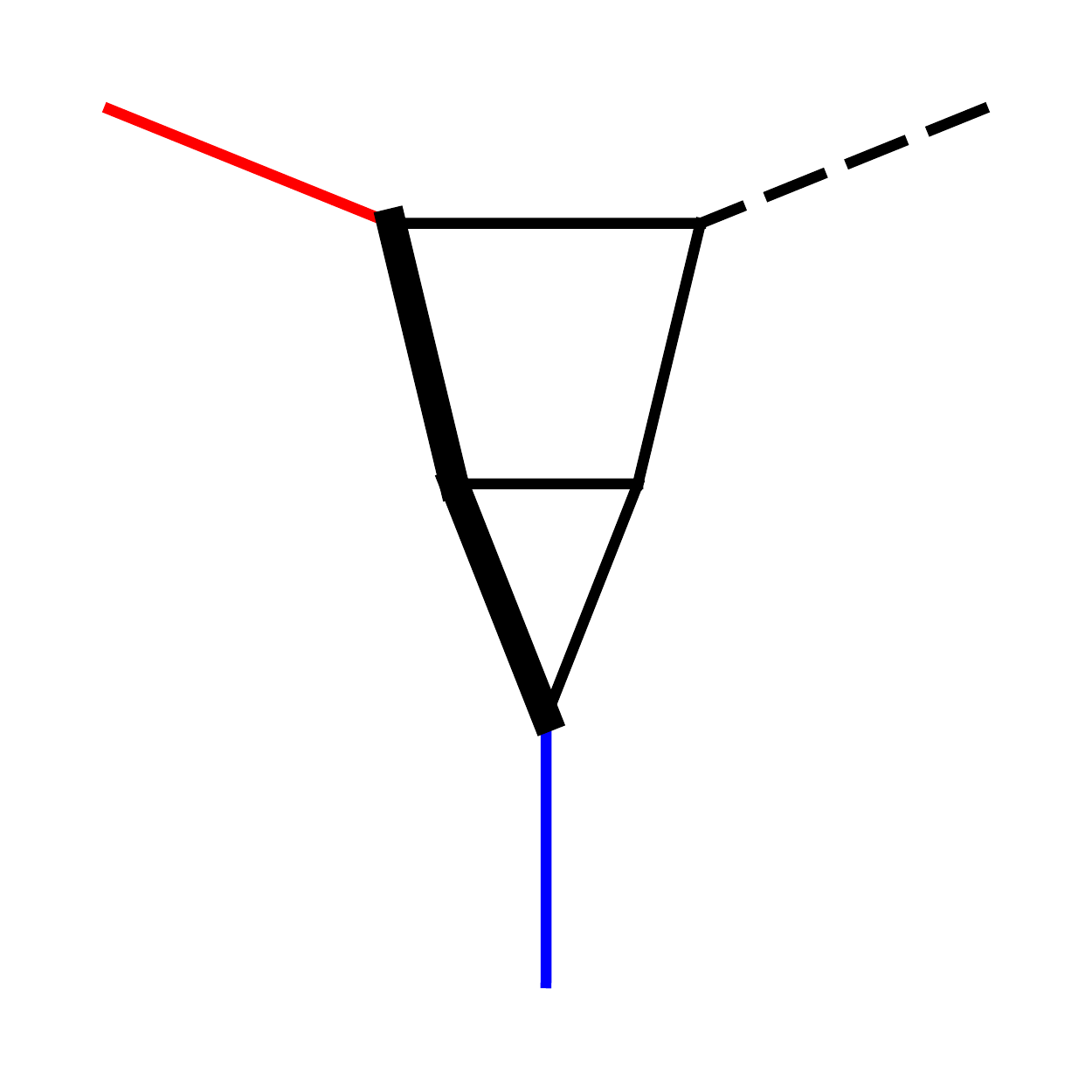}
 	}
 	\centering
 	\subfloat[\label{fig:FigTop1WW}]{%
 		\includegraphics[width=0.23\textwidth]{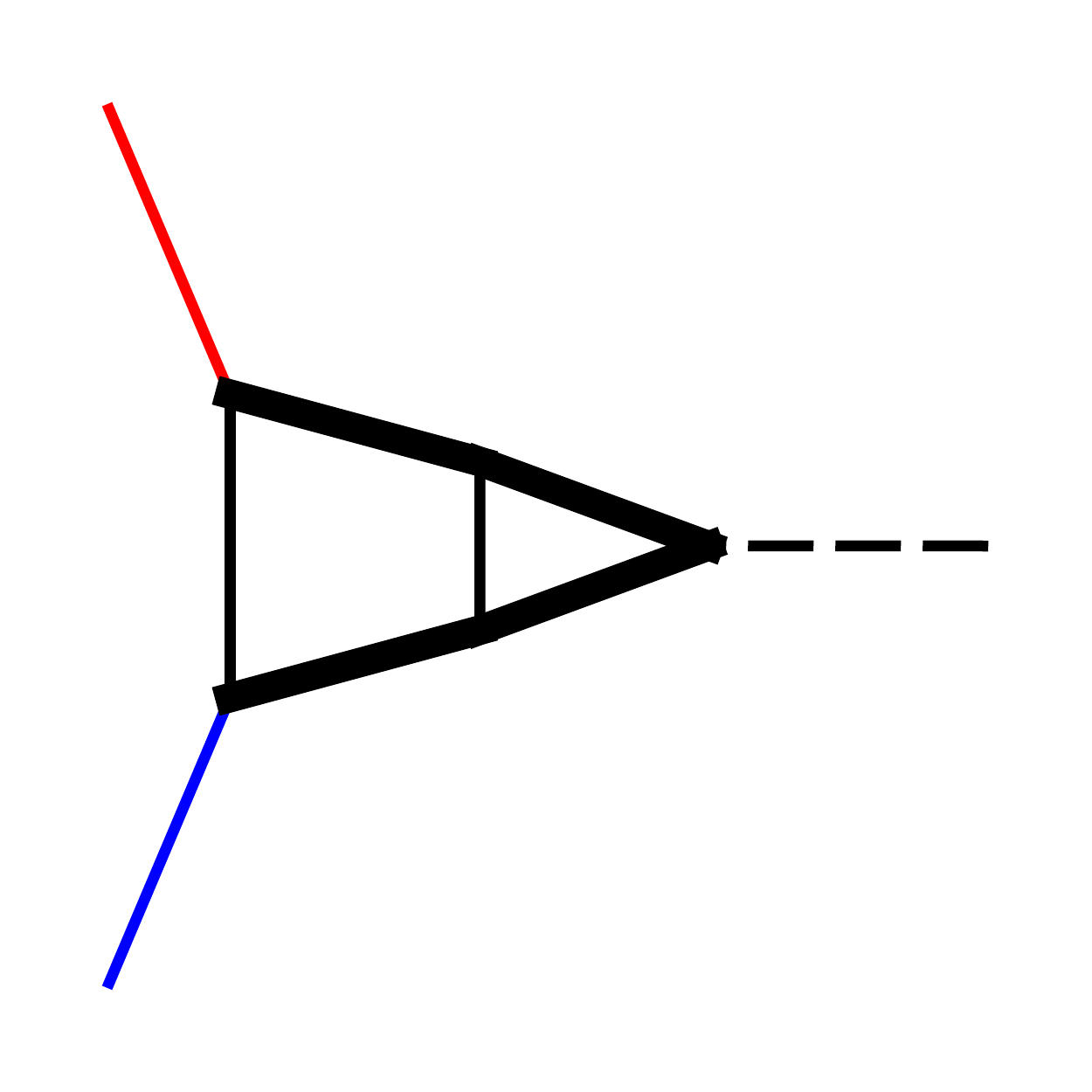}
 	}
 	\subfloat[\label{fig:FigTop2WW}]{%
 		\includegraphics[width=0.23\textwidth]{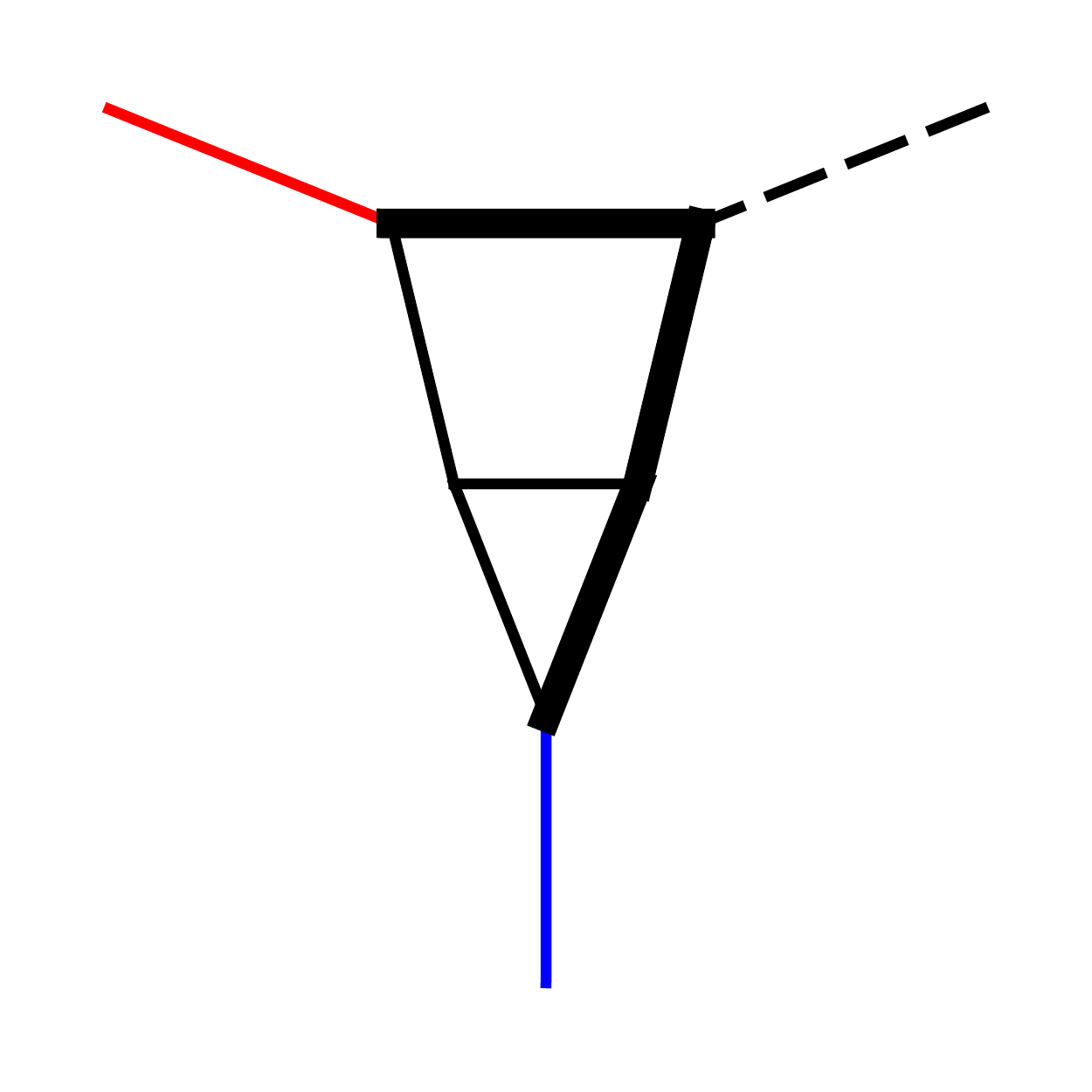}
 	}

  	\caption{Two-loop topologies for $\X\, W^{+}W^{-}$
          interactions. Thin lines represent massless propagators and
          thick lines stand for massive ones. The dashed external line
          corresponds to the off-shell leg with squared momentum equal
          to $s$ whereas the red and blue lines represent the two
          external vector bosons with off-shell momenta $p_1^2$ and
          $p_2^2$ respectively.}
 	\label{fig:TopoWW}
 \end{figure}

The two-loop Feynman diagrams contributing to $\X\, W^{+}W^{-}$ can be
reduced to four parent topologies, which are depicted in
figure~\ref{fig:TopoWW}. The integrals belonging to these topologies
depend on the three external invariants
\begin{equation}
p_1^2\,,\quad p_2^2\,, \quad s,
\end{equation}
as well as on the top mass $m^2$.  These four dimensionful parameters
can be combined into three independent dimensionless variables,
$\xxf=(u,\zz,\zzb)$ for topologies
\protect\subref{fig:FigTop1WWbb}-\protect\subref{fig:FigTop2WWbb} and
$\xxf=(v,\zz,\zzb)$ for topologies
\protect\subref{fig:FigTop1WW}-\protect\subref{fig:FigTop2WW}, whose
explicit definition will be later specified.  The MIs satisfy a linear
system of partial DEs in these variables, which, if we organize the
MIs into a vector $\FFvec$, can be combined into a matrix equation for
the total differential of $\FFvec$,
\begin{align}
  \label{eq:noncanonicalDE}
d \FFvec = \KK \, \FFvec \ .
\end{align}
In general, the matrix-valued differential form
$\KK = \KK_a dx^a$ ($a=1,2,3$) depends \emph{both} on the
kinematic variables and on the spacetime dimension $d=4-2\eps$.  Since
the left-hand side of the system in eq.~\eqref{eq:noncanonicalDE} is a
total differential by construction, it is easy to show that $\KK$
satisfies the (matrix) integrability condition
\begin{equation}
  \partial_a \KK_b - \partial_b \KK_a - [\KK_a,\KK_b] = 0\,, \quad a,b=1,2,3\,.
\end{equation}

 Starting from a basis of MIs associated to a matrix $\KK$ with a
 linear dependence on $\eps$, one can use the \emph{Magnus
   exponential}~\cite{Magnus,Argeri:2014qva} and apply the procedure
 outlined in section\ 2 of \cite{DiVita:2014pza} in order to perform a
 basis transformation and obtain a \emph{canonical} set of
 MIs~\cite{Henn:2013pwa} $\GGvec$ enjoying two remarkable features:
 first, the canonical basis $\GGvec$ obeys a system of DEs where the
 dependence on $\eps$ is factorized from the kinematics and, in
 addition, the kinematic matrices can be organized into a
 logarithmic differential form, referred to as canonical
 $\dlog$-form. Thus, the canonical basis $\GGvec$ satisfies a system
 of equations of the form
\begin{align}
 \label{eq:canonicalDE}
d \GGvec = \eps \, \dA \, \GGvec \ ,
\end{align}
where
\begin{align}
\dA = \sum_{i=1}^n \MM_i \, \dlog \eta_i\,,
\label{dlog-form}
\end{align}
is the $\dlog$ matrix written in terms of the differentials
$\dlog \eta_i$, whose arguments $\eta_i = \eta_i(\xxf)$ solely enclose
the kinematic dependence and form the so called \emph{alphabet} of the
problem. The coefficient matrices $\MM_i$ have rational-number
entries. Due to the $\eps$-factorization, the integrability condition
for eq.~\eqref{eq:canonicalDE} splits into
\begin{align}
  \label{eq:canonicalintegrability}
  \partial_a \partial_b \AA - \partial_a \partial_b \AA = 0\,, \qquad [\partial_a \AA,\partial_b \AA]=0\,, \quad a,b=1,2,3\,.
\end{align}

\subsection{General solution}
At any point $\xxf$, the general solution of the canonical system of
DEs~\eqref{eq:canonicalDE} can be expressed in terms of \emph{Chen's
  iterated integrals}~\cite{Chen:1977oja} as the \emph{path-ordered}
exponential
\begin{align}
  \label{eq:canonicalsolution}
  \GGvec(\eps,\xxf) = {} & \pathord \exp\left\{\epsilon \int_\gamma \dA\right\} \GGvec(\eps,\xxi)\,,
\end{align}
where $\GGvec(\eps, \xxi)$ is a constant vector depending on $\eps$
only and $\gamma$ is a piecewise-smooth path connecting $\vec{x}_0$ to
$\vec{x}$,
\begin{align}
\begin{cases}
\gammaAB{\xxi}{\xxf}:[0,1]\ni t \mapsto  \gamma(t) = (\gamma^1(t),\gamma^2(t),\gamma^3(t))\\
\gamma(0)=\xxi\\
\gamma(1)=\xxf.
\end{cases}
\label{eq:gamma}
\end{align}
In the limit $\xxf\to\xxi$ the integration path $\gamma$ shrinks to a
point and $\GGvec(\eps,\xxf)\to\GGvec(\eps,\xxi)$. In this
perspective, the integration constants $\GGvec(\eps,\xxi)$ which,
together with $\dA$, completely specify the solution, can be thought
as the initial values of the MIs, which then evolve to arbitrary
points $\vec{x}$ under the action of the path-ordered exponential. It
can be proven that, whenever $\gamma$ does not cross any singularity
or branch cuts of $\dA$ (but at its endpoints), the path-ordered
exponential is independent of the explicit choice of the path.  By
choosing a proper normalization, we can assume all canonical MIs to be
finite in the $\eps\to 0$ limit, in such a way that $\GGvec(\xx)$
admits a Taylor expansion in $\epsilon$,
\begin{align}
  \GGvec(\eps,\xx) =  \GGvec^{(0)}(\xx) + \epsilon\, \GGvec^{(1)}(\xx) + \epsilon^2 \GGvec^{(2)}(\xx) + \ldots\,
\end{align}
and, according to eq.~\eqref{eq:canonicalsolution}, the $n$-th order
coefficient is given by
\begin{align}
\GGvec^{(n)}(\vec{x}) = \sum_{i=0}^n \Delta^{(n-i)} (\vec{x}, \vec{x}_0)  \GGvec^{(i)}(\vec{x}_0),
\end{align}
where we introduced the weight-$k$ operator
\begin{align}
\label{eq:chendAk}
\Delta^{(k)} (\vec{x}, \vec{x}_0) =\int_\gamma \dAk{k},\qquad \Delta^{(0)} (\vec{x}, \vec{x}_0) = 1\,,
\end{align}
which iterates $k$ ordered integrations of the matrix-valued 1-form
$\dA$ along the path $\gamma$. According to eq.~\eqref{dlog-form},
each entry of $\Delta^{(k)}$ is a linear combination of Chen's
iterated integrals of the type
\begin{align}
  \label{eq:chendlog}
  \chen{i_k,\ldots,i_1}{\gamma} = \int_\gamma \dlog \eta_{i_k} \ldots \dlog \eta_{i_1}= {} & \int_{0\leq t_1 \leq \ldots \leq t_k \leq 1} g^\gamma_{i_k}(t_k) \ldots g^\gamma_{i_1}(t_1) \,dt_1 \ldots \,dt_k\,,
\end{align}
being
\begin{align}
  \label{eq:dlogweight}
  g^\gamma_i(t) = {} & \frac{d}{dt} \log \eta_i(\gamma(t))\,.
\end{align}
It should be remarked that, as explicitly indicated in
\eqref{eq:chendlog}, individual Chen's iterated integrals are, in
general, \emph{functionals} of the path and that only the full
combinations appearing as entries of $\Delta^{(k)}$ are independent of
the particular choice of $\gamma$. The most relevant properties of
Chen's iterated integrals are summarized in appendix~\ref{sec:chen}.


\section{Two-loop master integrals for $\X\, W^{+}W^{-}$}
\label{sec:WWMIs}
In this section we present the solution of the system of DEs for the
MIs associated to the four topologies
\protect\subref{fig:FigTop1WWbb}-\protect\subref{fig:FigTop2WW}. Since
topologies
\protect\subref{fig:FigTop1WWbb}-\protect\subref{fig:FigTop2WWbb} and
\protect\subref{fig:FigTop1WW}-\protect\subref{fig:FigTop2WW} belong
to two distinct integral families, we discuss their evaluation
separately.
\subsection{Topologies (a)-(b)}
\label{sec:WWMIsb}
The two topologies \protect\subref{fig:FigTop1WWbb} and
\protect\subref{fig:FigTop2WWbb} belong to the 7-denominator integral
family identified by the set of denominators
\begin{gather}
\Den_1 = k_1^2,\quad
\Den_2 = k_2^2,\quad
\Den_3 = (k_1-p_2)^2-m^2,\quad
\Den_4 = (k_2-p_2)^2-m^2, \nonumber\\
\Den_5 = (k_1-p_1-p_2)^2,\quad
\Den_6 = (k_2-p_1-p_2)^2,\quad
\Den_7 = (k_1-k_2)^2,
\end{gather} 
where $k_1$ and $k_2$ are the two loop momenta.  \,The integrals
belonging to this family can be reduced to a set of 29 MIs which are
conveniently expressed in terms of the dimensionless variables $u$,
$\zz$ and $\zzb$ defined by
\begin{align}
  -\frac{s}{m^2} = u\,,\quad \frac{p_1^2}{s} = \,\zz\,\zzb\,, \quad \frac{p_2^2}{s}=(1-\zz)(1-\zzb)\,.
  \label{eq:paramab}
\end{align}
The same parametrization for $p_1^2$ and $p_2^2$ was used also for the
massless triangles considered in~\cite{Chavez:2012kn}.  The following
set of MIs obeys a system of DEs which is linear in $\eps$ :
\begin{align*}
\FF_{1}&=\eps^2 \, \top{1}\,,  &
\FF_{2}&=\eps^2 \, \top{2}\,,  &
\FF_{3}&=\eps^2 \, \top{3}\,,  \\
\FF_{4}&=\eps^2 \, \top{4}\,,  &
\FF_{5}&=\eps^2 \, \top{5}\,,  &
\FF_{6}&=\eps^2 \, \top{6}\,,  \\
\FF_{7}&=\eps^2 \, \top{7}\,,  &
\FF_{8}&=\eps^2 \, \top{8}\,,  &
\FF_{9}&=\eps^2 \, \top{9}\,,  \\
\FF_{10}&=\eps^3 \, \top{10}\,,  &
\FF_{11}&=\eps^2 \, \top{11}\,,  &
\FF_{12}&=\eps^2 \, \top{12}\,,  \\
\FF_{13}&=\eps^2 \, \top{13}\,,  &
\FF_{14}&=\eps^2 \, \top{14}\,,  &
\FF_{15}&=\eps^2 \, \top{15}\,,  \\
\FF_{16}&=\eps^3\, \top{16}\,,  &
\FF_{17}&=\eps^2\, \top{17}\,,  &
\FF_{18}&=\eps^3 \, \top{18}\,,  \\
\FF_{19}&=\eps^3 \, \top{19}\,,  &
\FF_{20}&=\eps^2 \, \top{20}\,,  &
\FF_{21}&=\eps^3 \, \top{21}\,,  \\
\FF_{22}&=\eps^2 \, \top{22}\,,  &
\FF_{23}&=\eps^3 \, \top{23}\,, &
\FF_{24}&=\eps^3 \, \top{24}\,,  \\
\FF_{25}&=\eps^4\, \top{25}\,,  &
\FF_{26}&=\eps^4 \, \top{26}\,, &
\FF_{27}&=\eps^3 \, \top{27}\,,  \\
\FF_{28}&=\eps^3\, \top{28}\,,  &
\FF_{29}&=\eps^2 \, \top{29}\,,\stepcounter{equation}\tag{\theequation}
\label{def:1M1LBasisMIsb}
\end{align*}
\begin{figure}[h]
  \centering
  \captionsetup[subfigure]{labelformat=empty}
  \subfloat[$\mathcal{T}_1$]{%
    \includegraphics[width=0.14\textwidth]{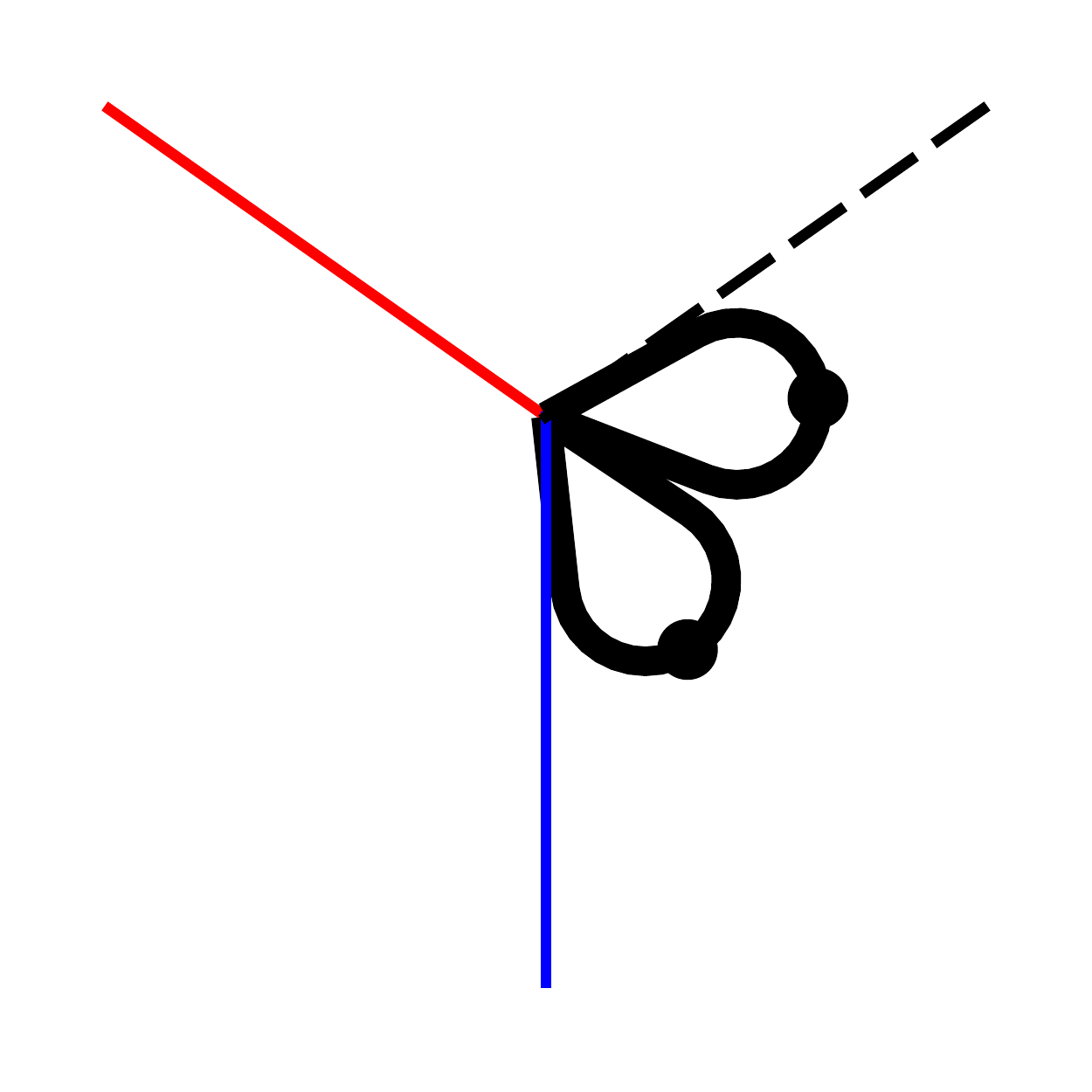}
  }
  \subfloat[$\mathcal{T}_2$]{%
    \includegraphics[width=0.14\textwidth]{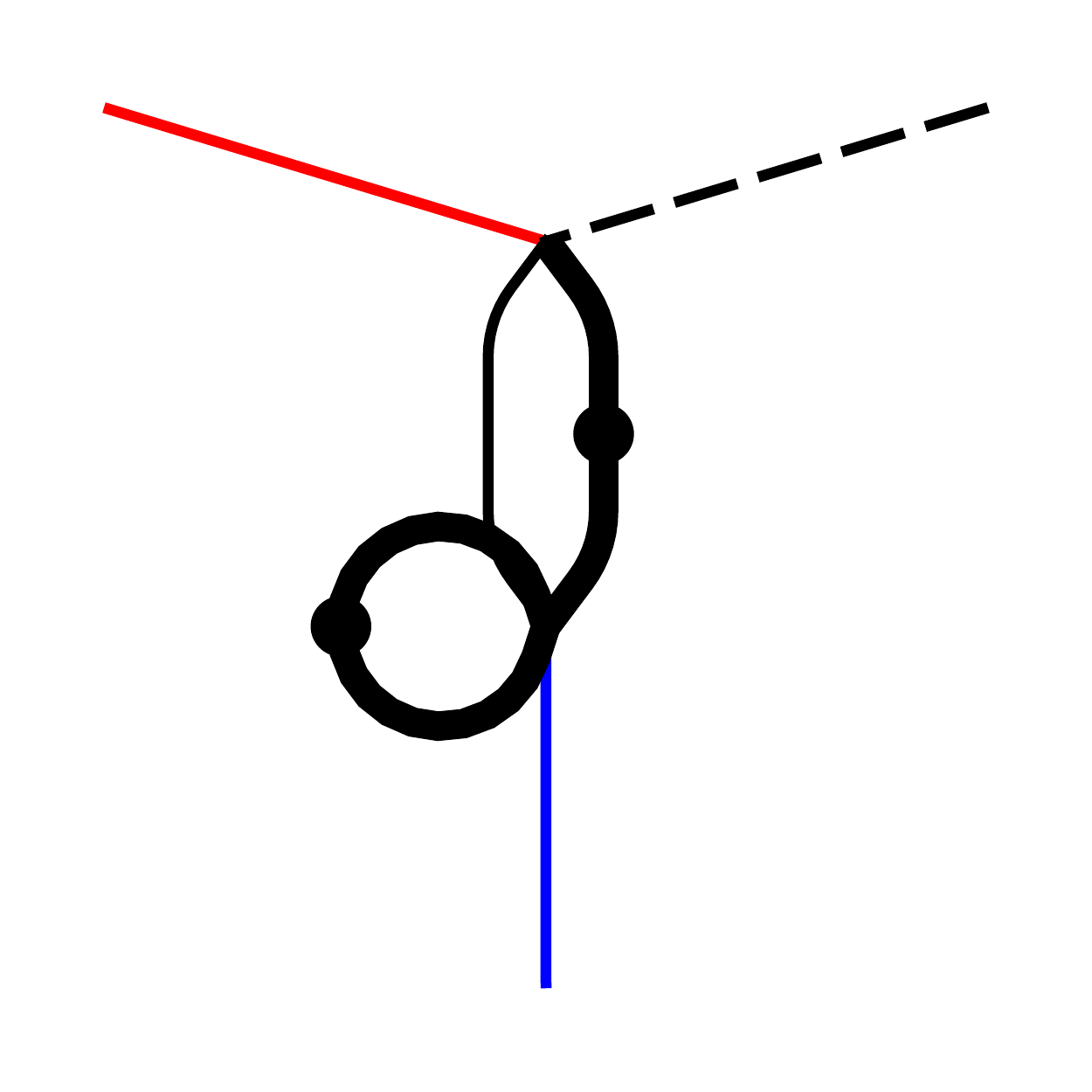}
  }
  \subfloat[$\mathcal{T}_3$]{%
    \includegraphics[width=0.14\textwidth]{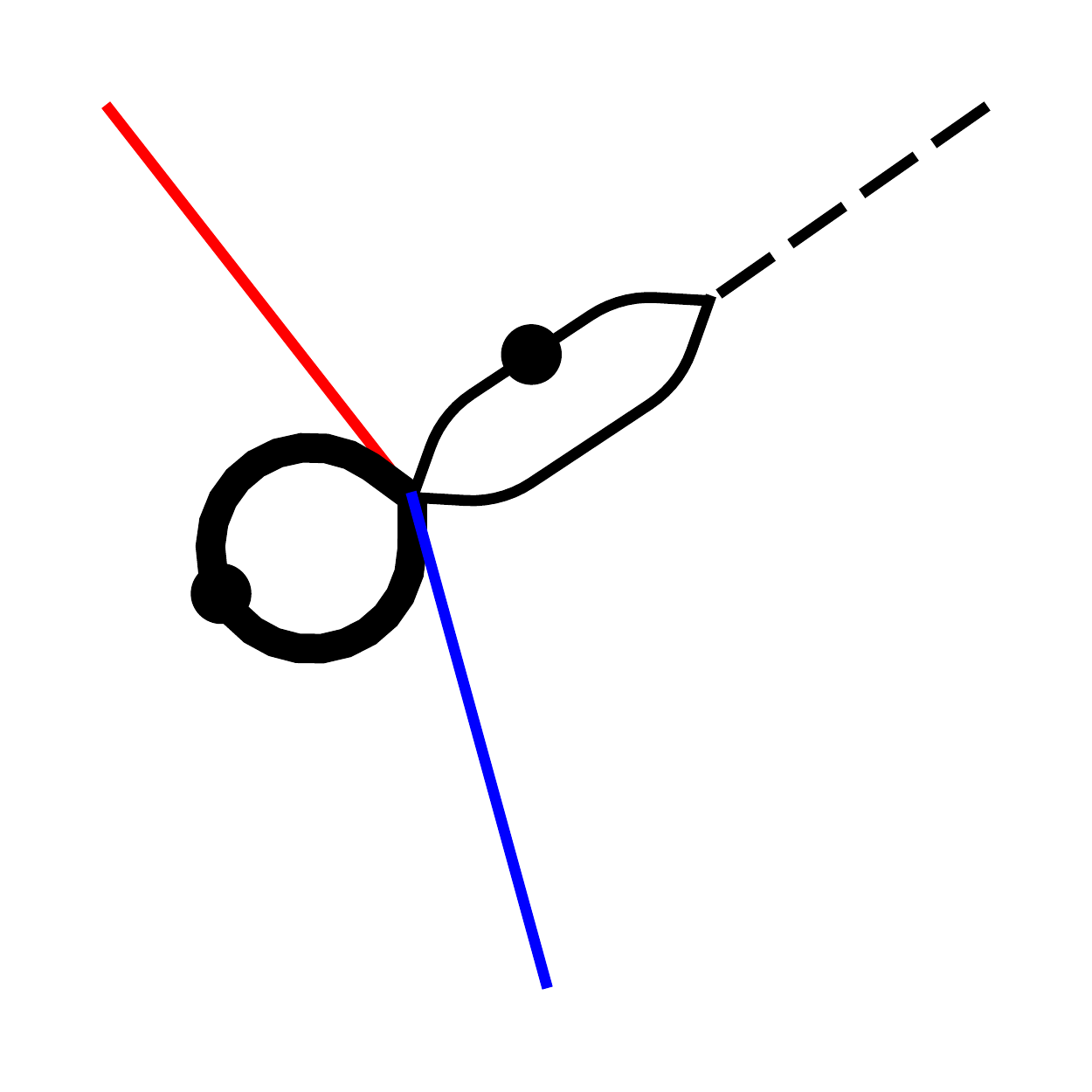}
  }
  \subfloat[$\mathcal{T}_4$]{%
    \includegraphics[width=0.14\textwidth]{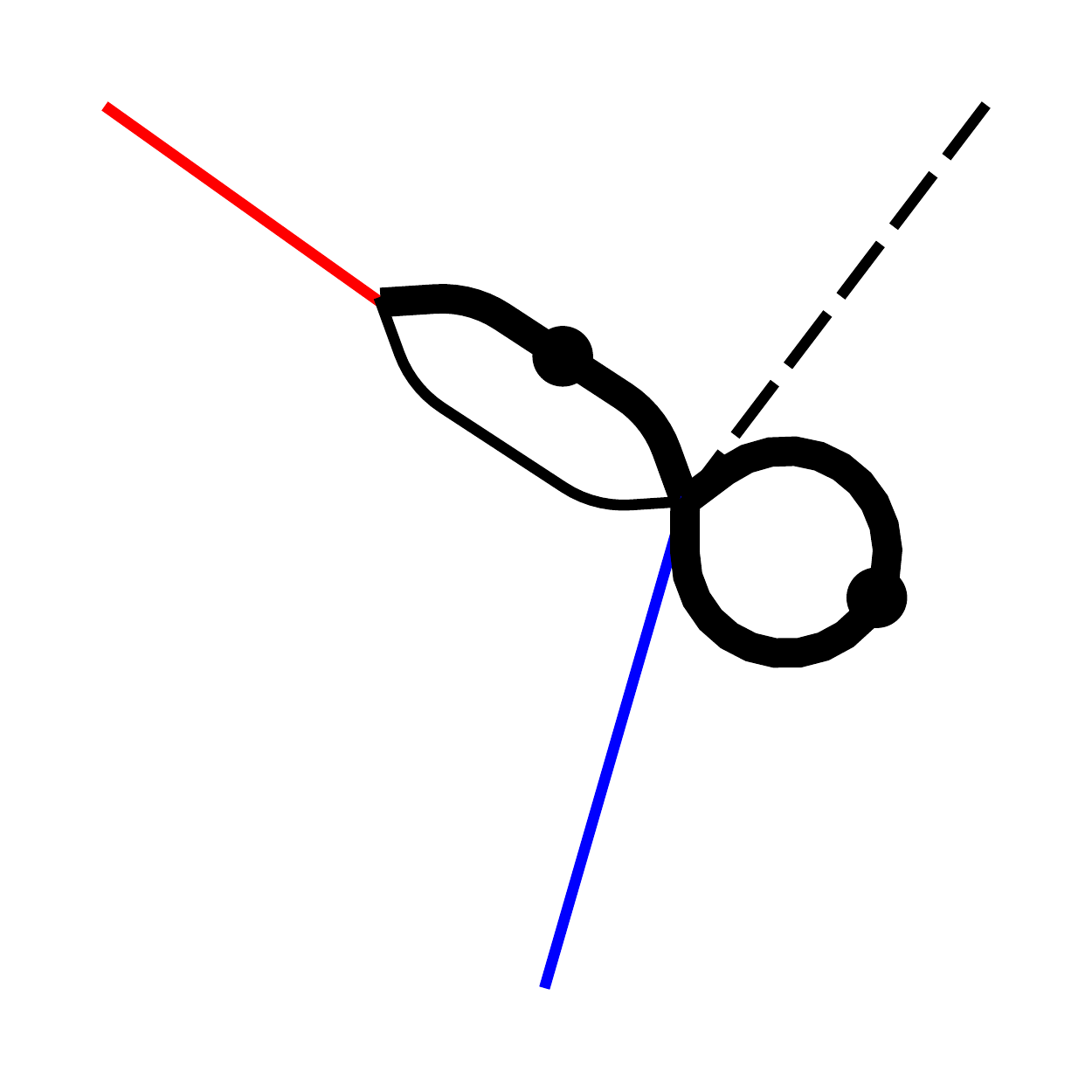}
  }
  \subfloat[$\mathcal{T}_5$]{%
    \includegraphics[width=0.14\textwidth]{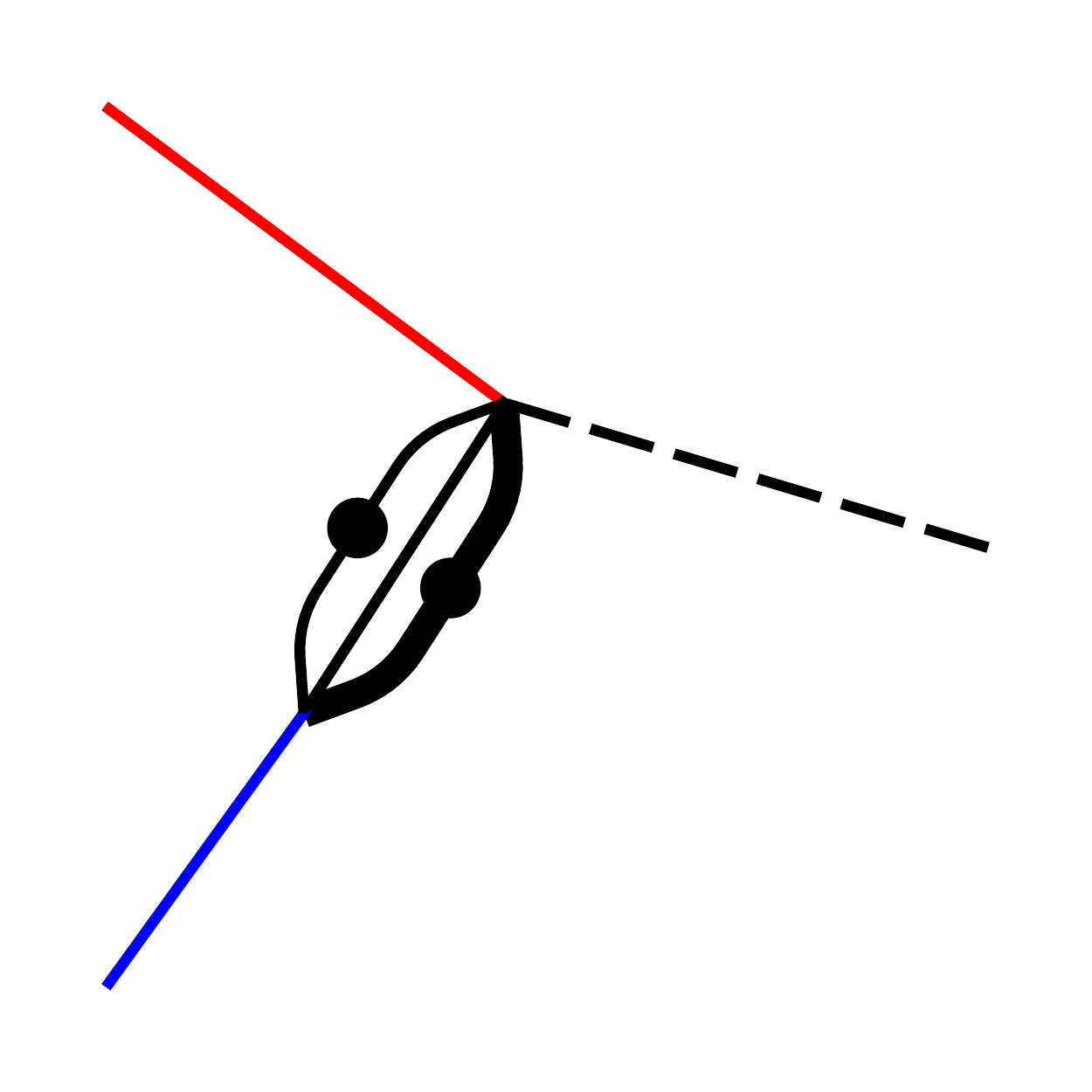}
  }
  \subfloat[$\mathcal{T}_6$]{%
    \includegraphics[width=0.14\textwidth]{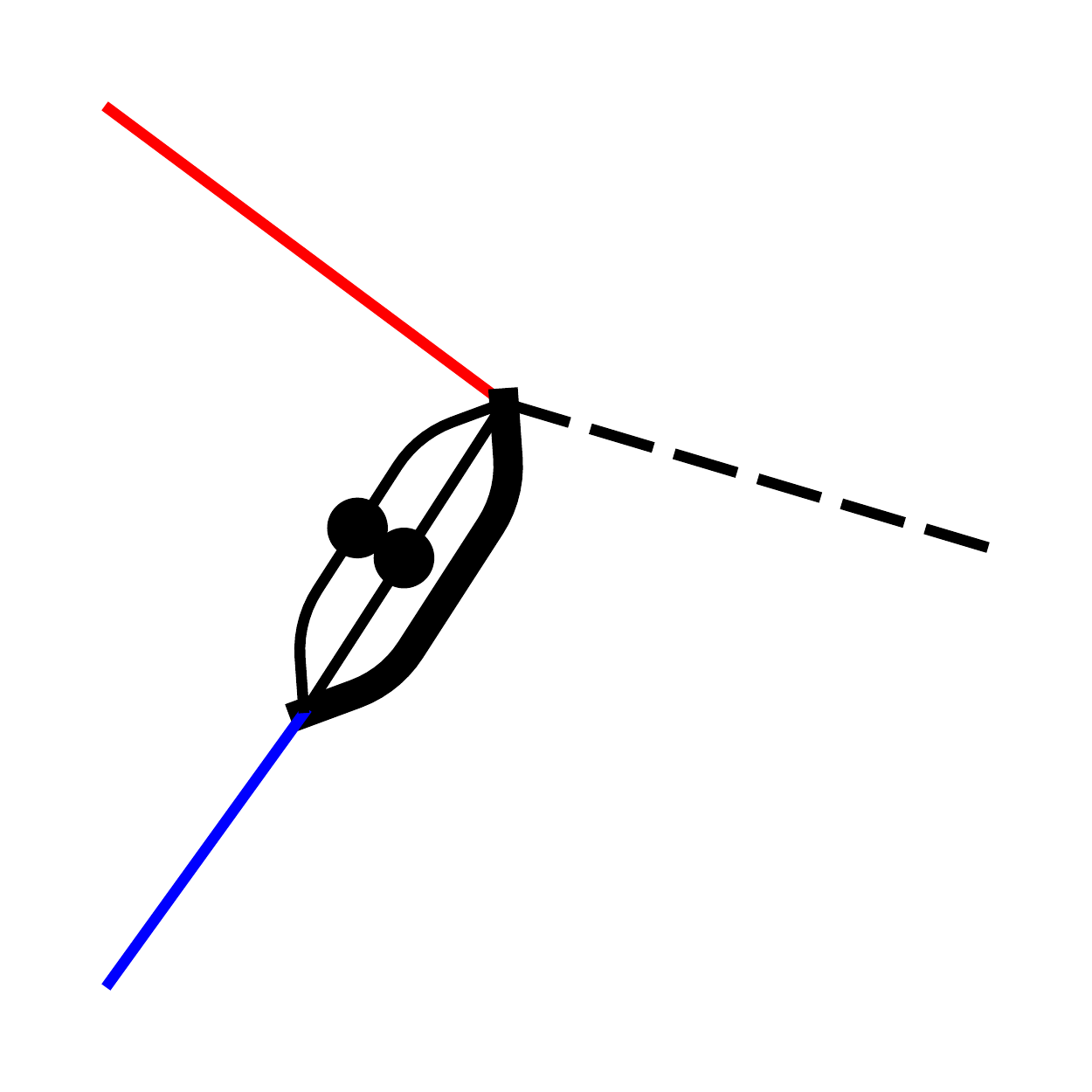}
  }
  \\
  \subfloat[$\mathcal{T}_7$]{%
    \includegraphics[width=0.14\textwidth]{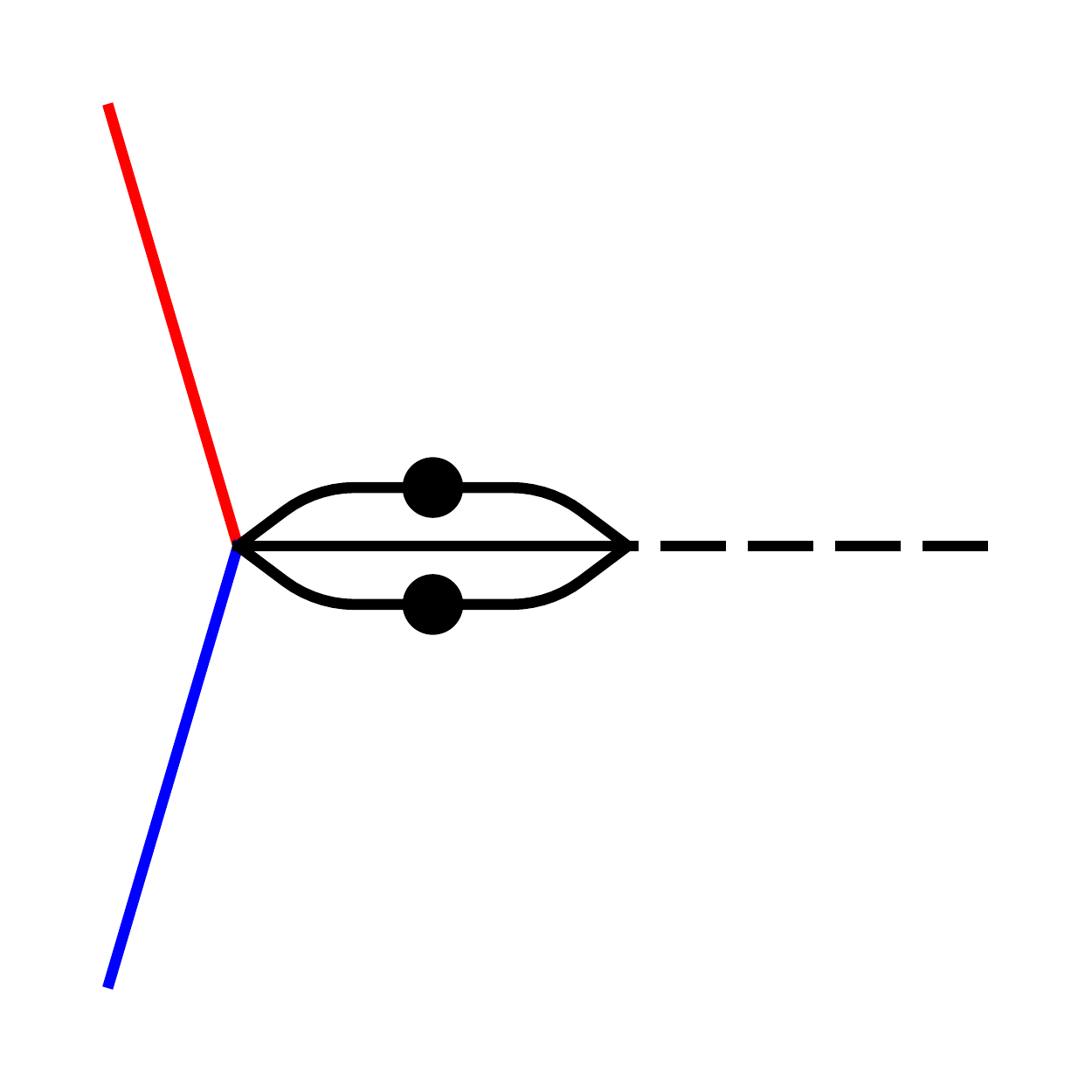}
  }
  \subfloat[$\mathcal{T}_8$]{%
    \includegraphics[width=0.14\textwidth]{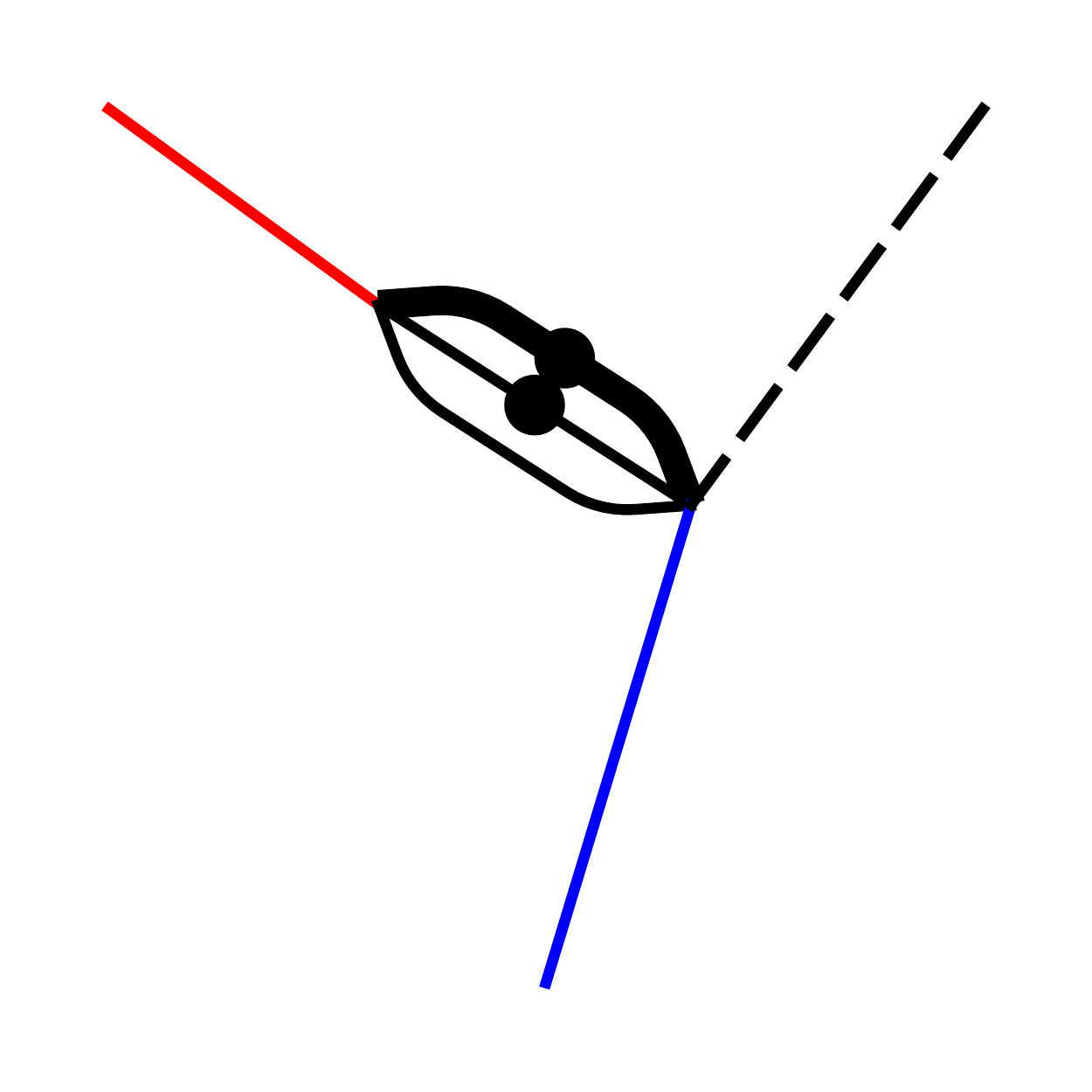}
  }
  \subfloat[$\mathcal{T}_9$]{%
    \includegraphics[width=0.14\textwidth]{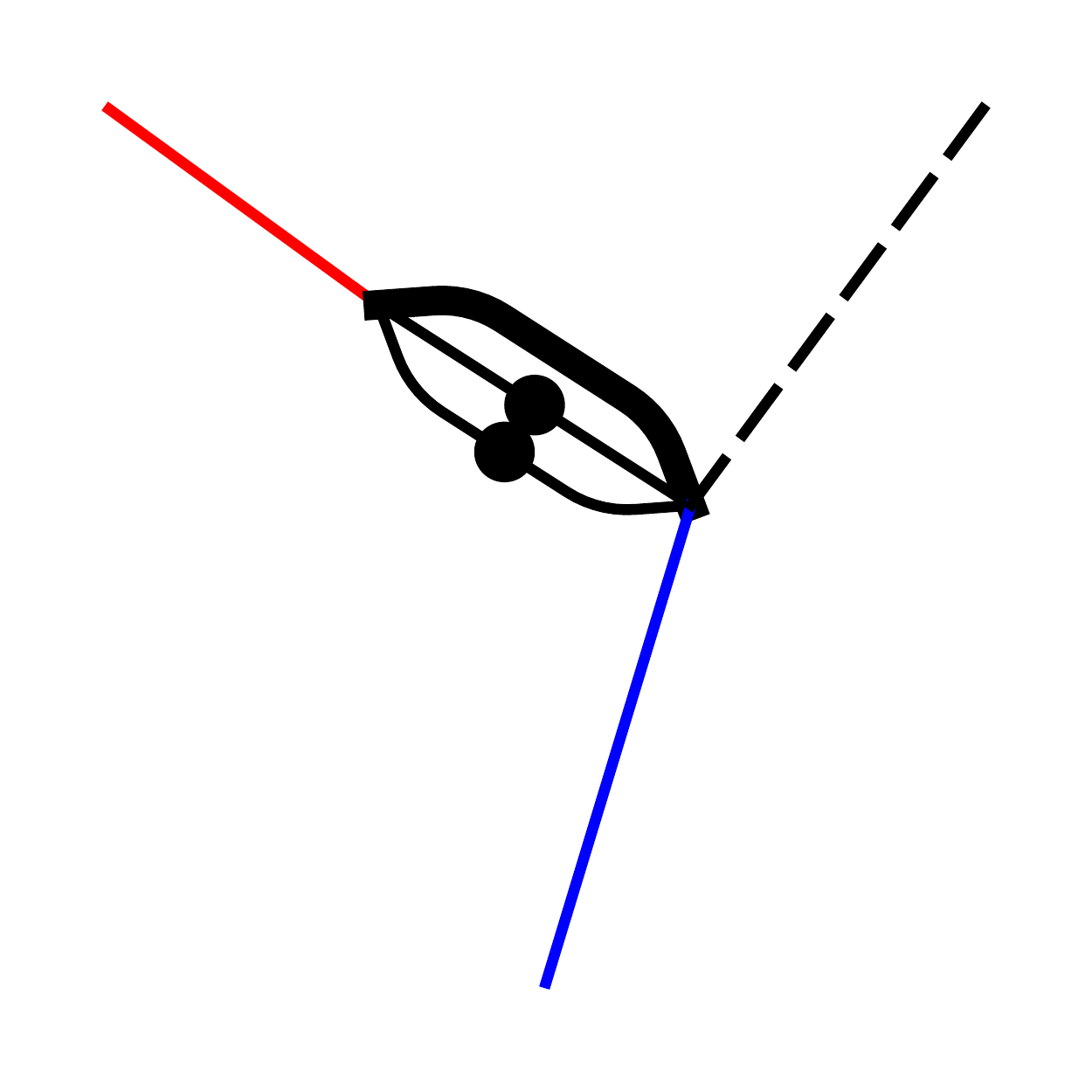}
  }
  \subfloat[$\mathcal{T}_{10}$]{%
    \includegraphics[width=0.14\textwidth]{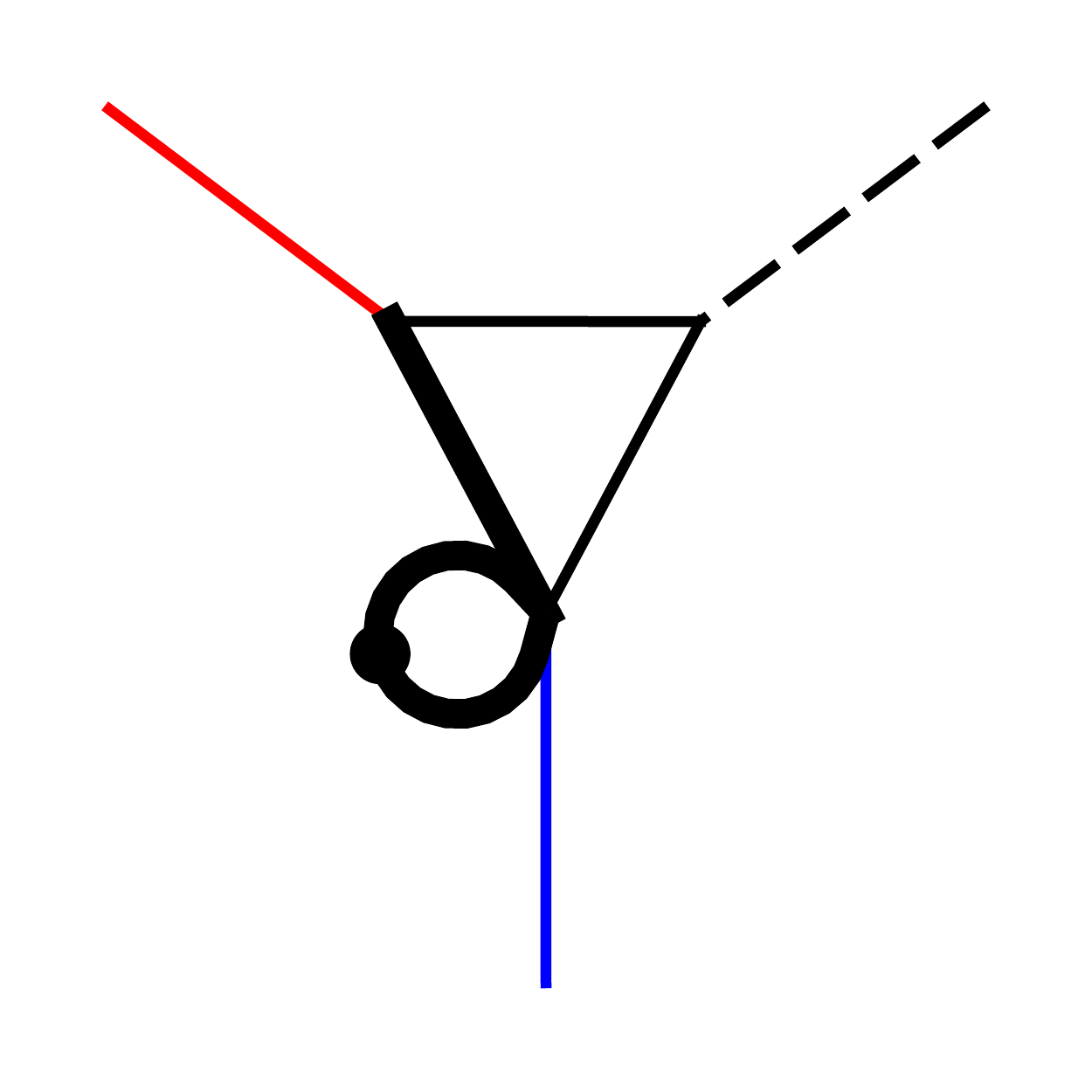}
  }
  \subfloat[$\mathcal{T}_{11}$]{%
    \includegraphics[width=0.14\textwidth]{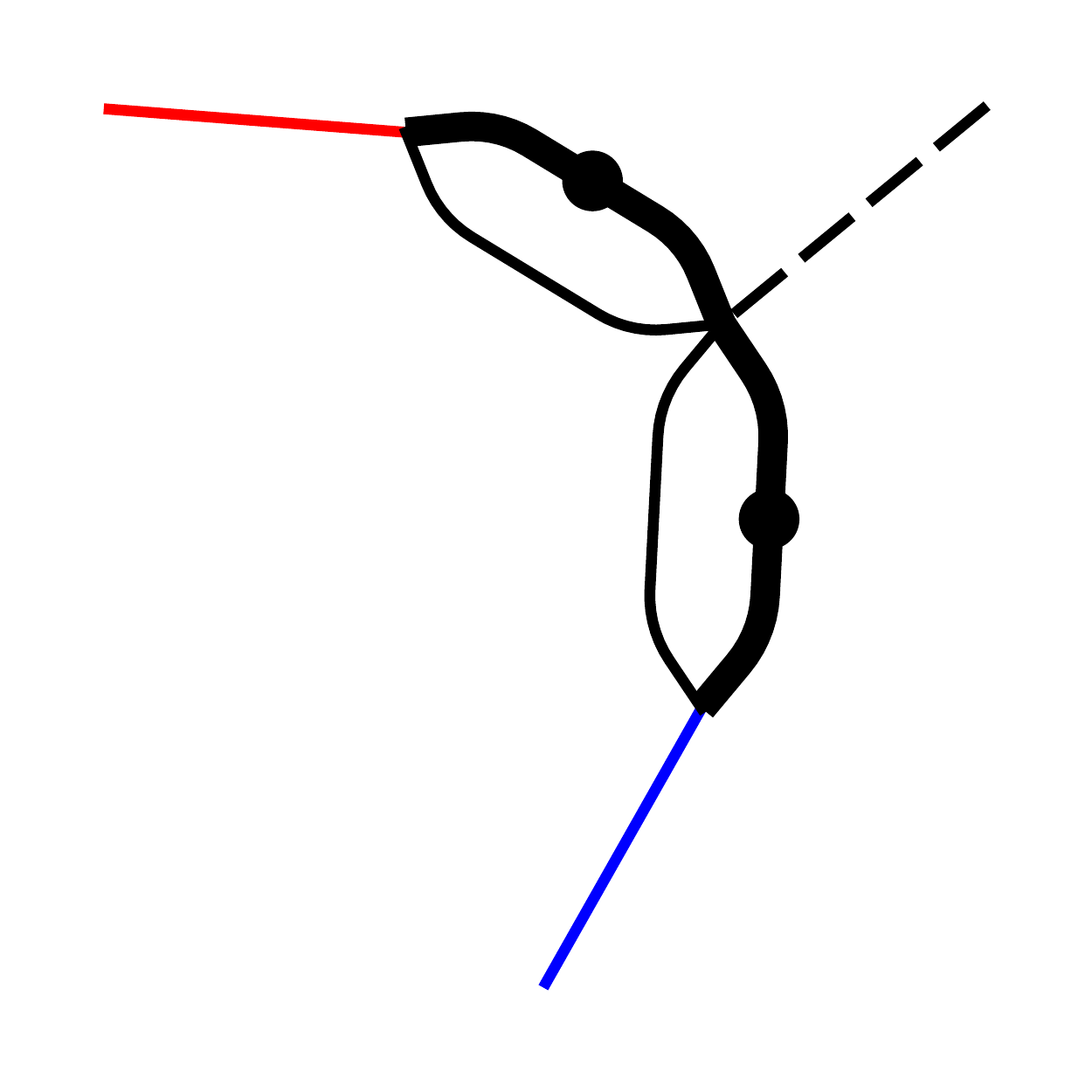}
  }
  \subfloat[$\mathcal{T}_{12}$]{%
    \includegraphics[width=0.14\textwidth]{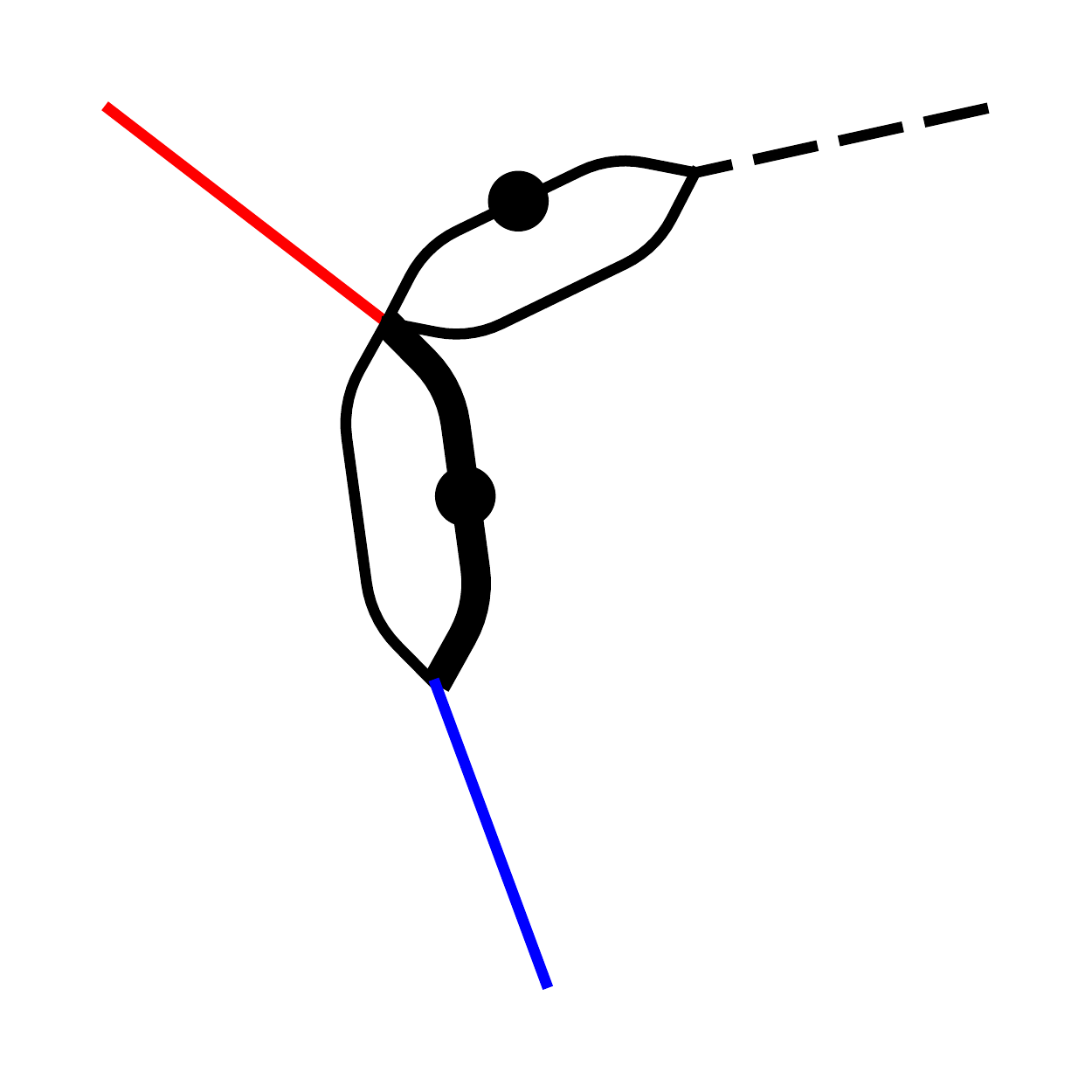}
  }
  \\
  \subfloat[$\mathcal{T}_{13}$]{%
    \includegraphics[width=0.14\textwidth]{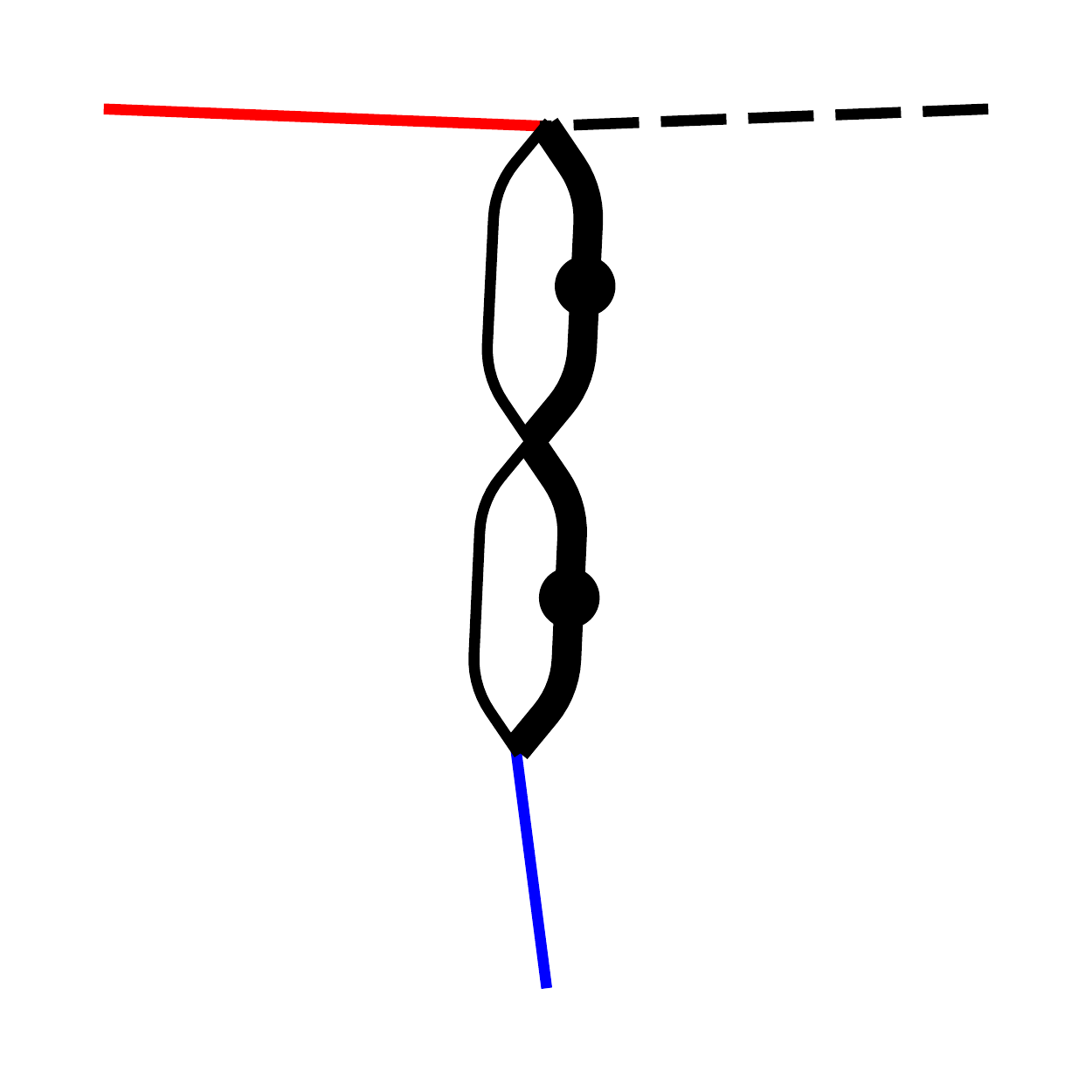}
  }
  \subfloat[$\mathcal{T}_{14}$]{%
    \includegraphics[width=0.14\textwidth]{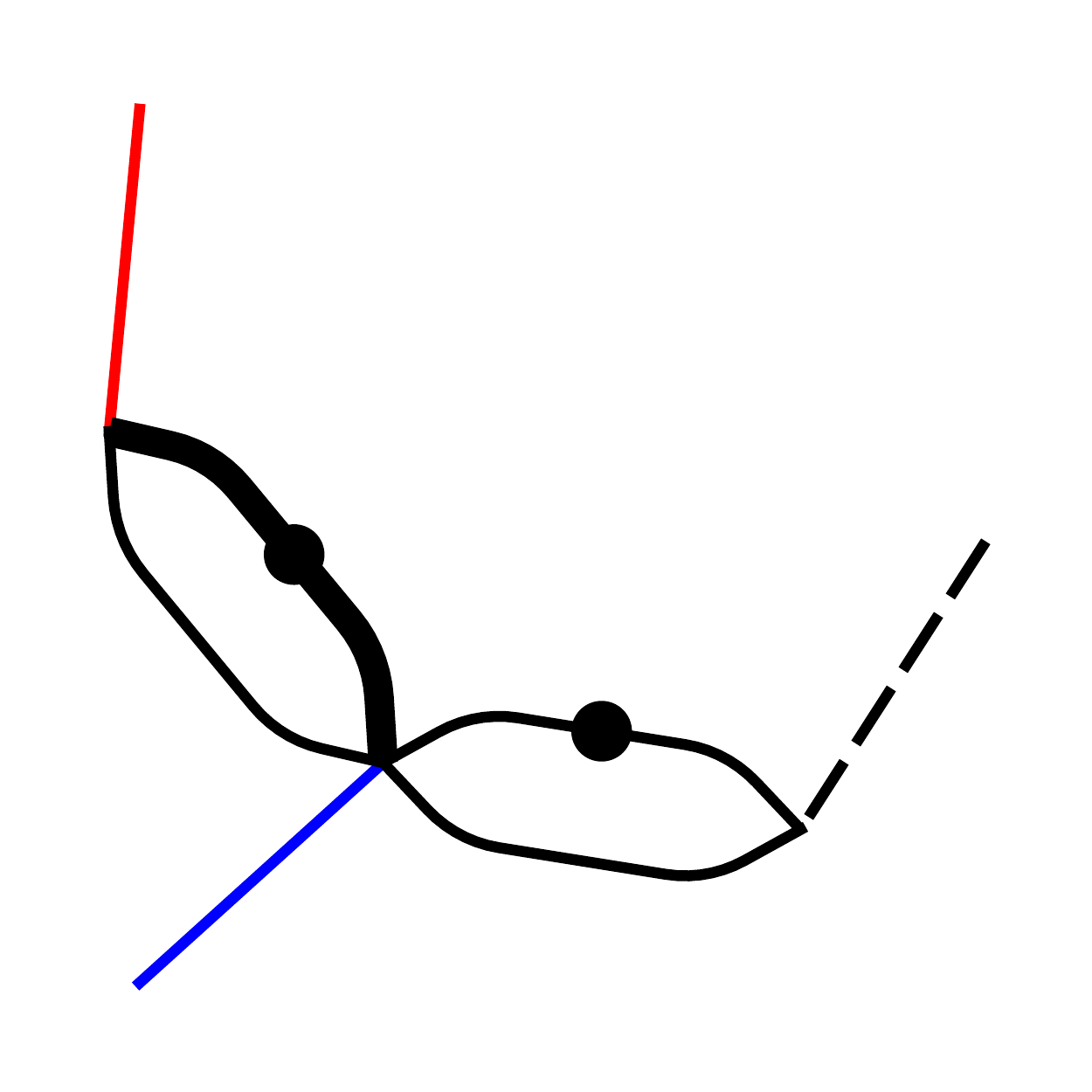}
  }
  \subfloat[$\mathcal{T}_{15}$]{%
    \includegraphics[width=0.14\textwidth]{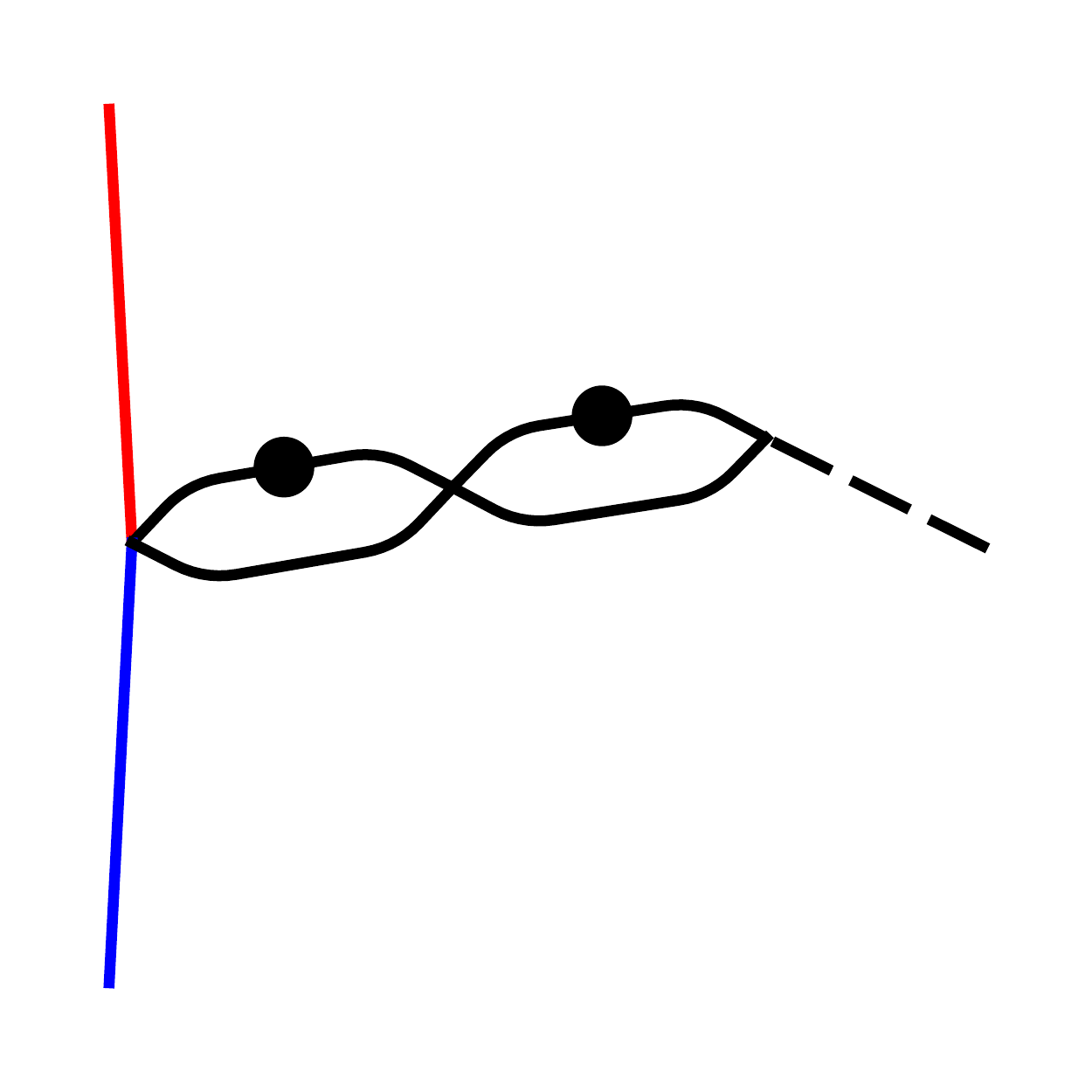}
  }
  \subfloat[$\mathcal{T}_{16}$]{%
    \includegraphics[width=0.14\textwidth]{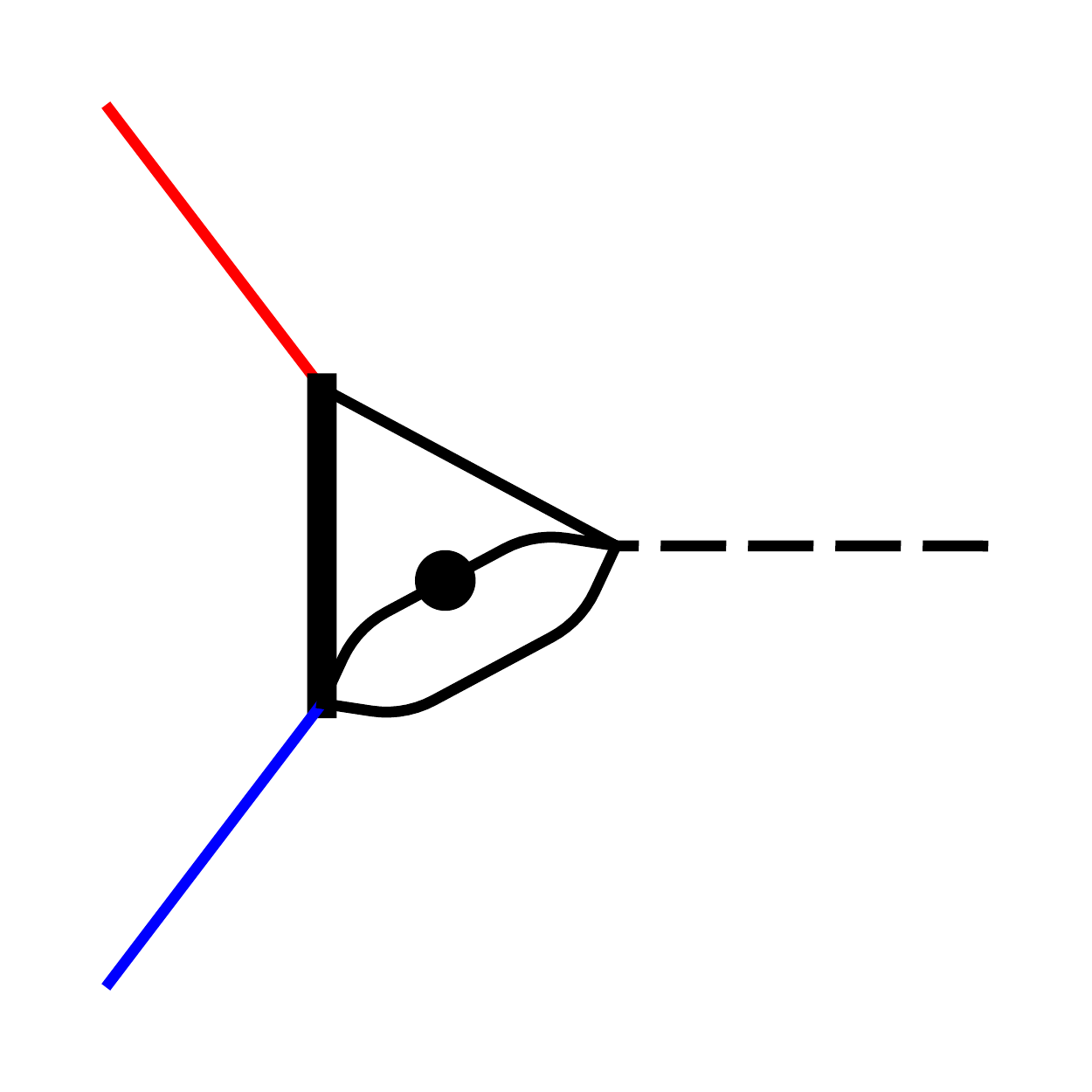}
  }
  \subfloat[$\mathcal{T}_{17}$]{%
    \includegraphics[width=0.14\textwidth]{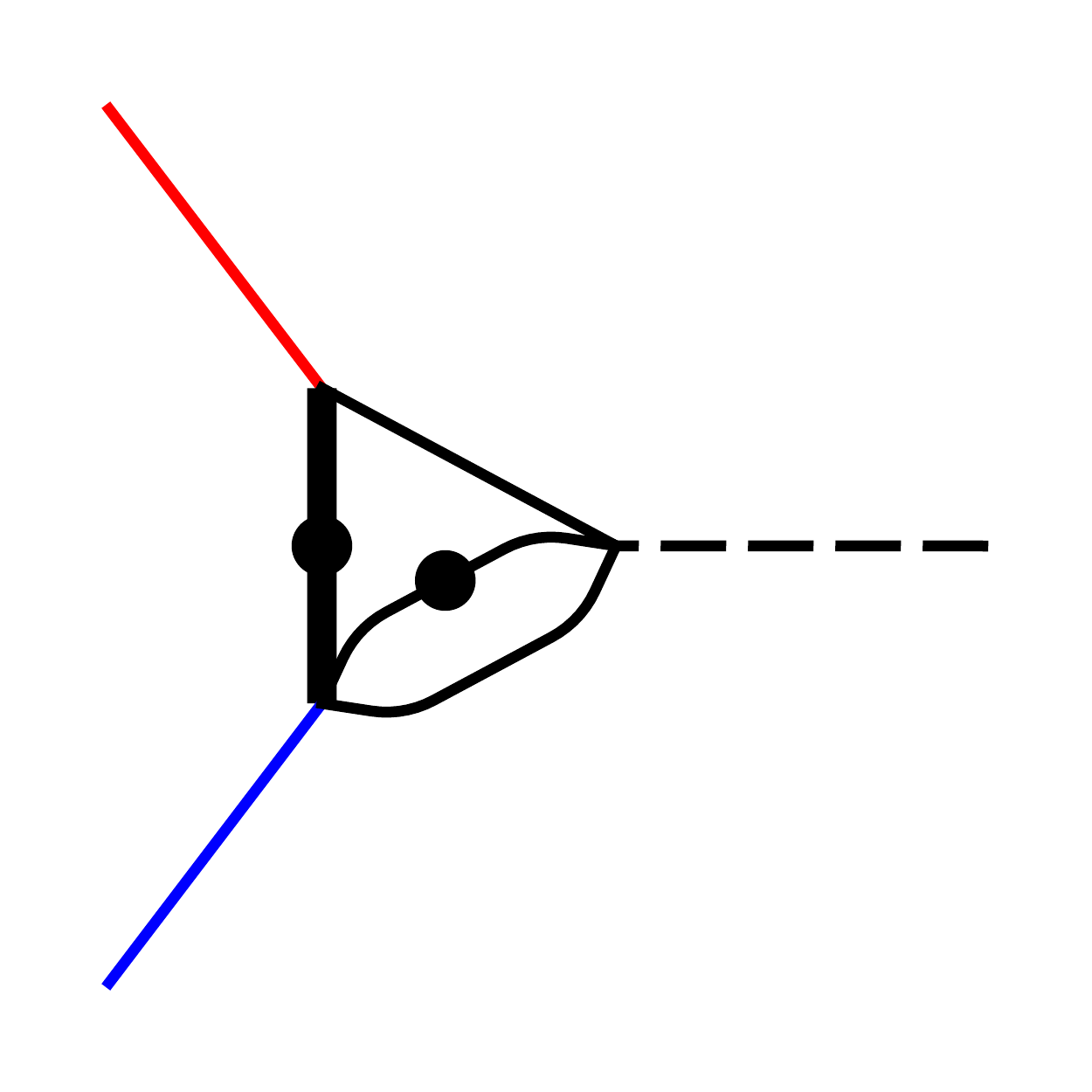}
  }
  \subfloat[$\mathcal{T}_{18}$]{%
    \includegraphics[width=0.14\textwidth]{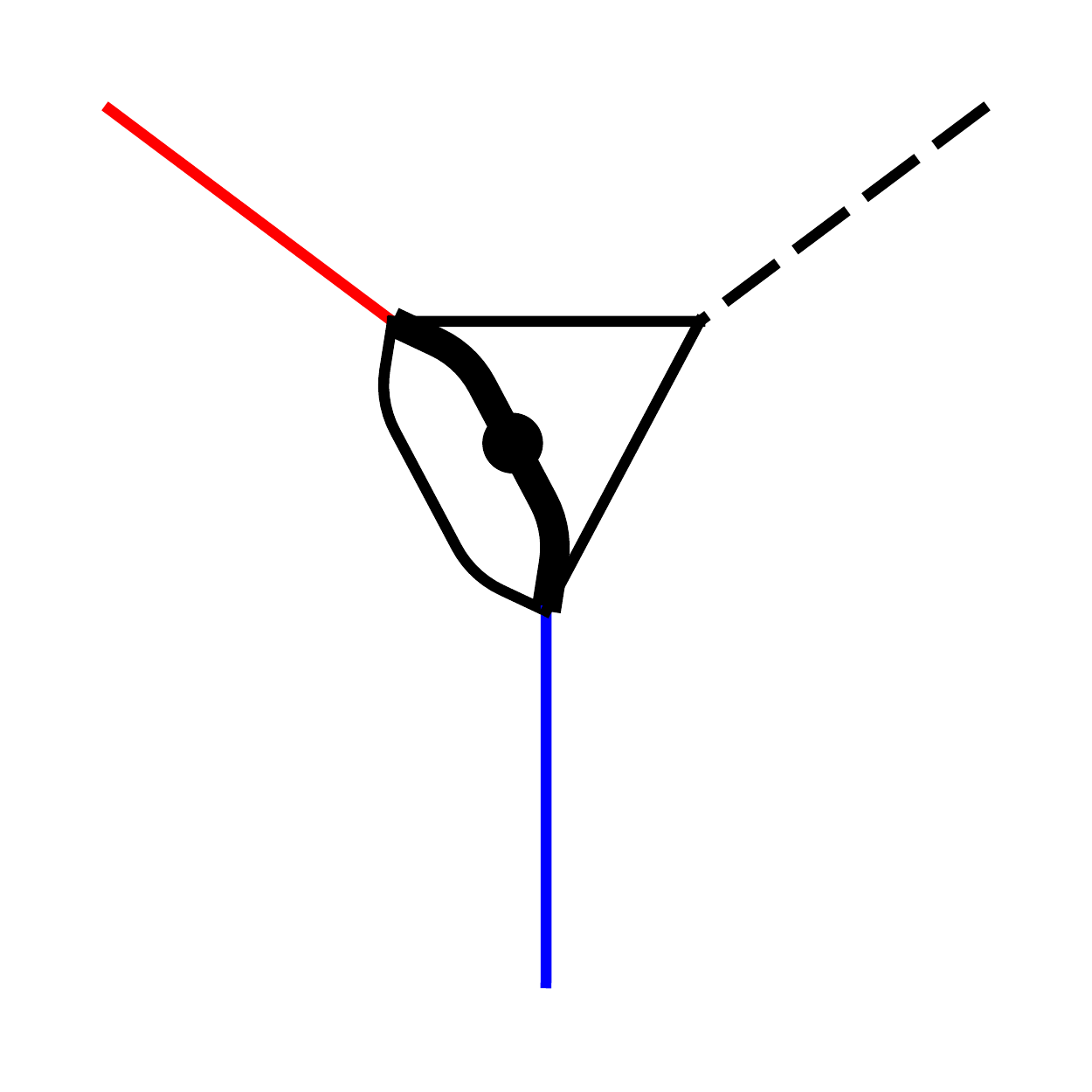}
  }
  \\
  \subfloat[$\mathcal{T}_{19}$]{%
    \includegraphics[width=0.14\textwidth]{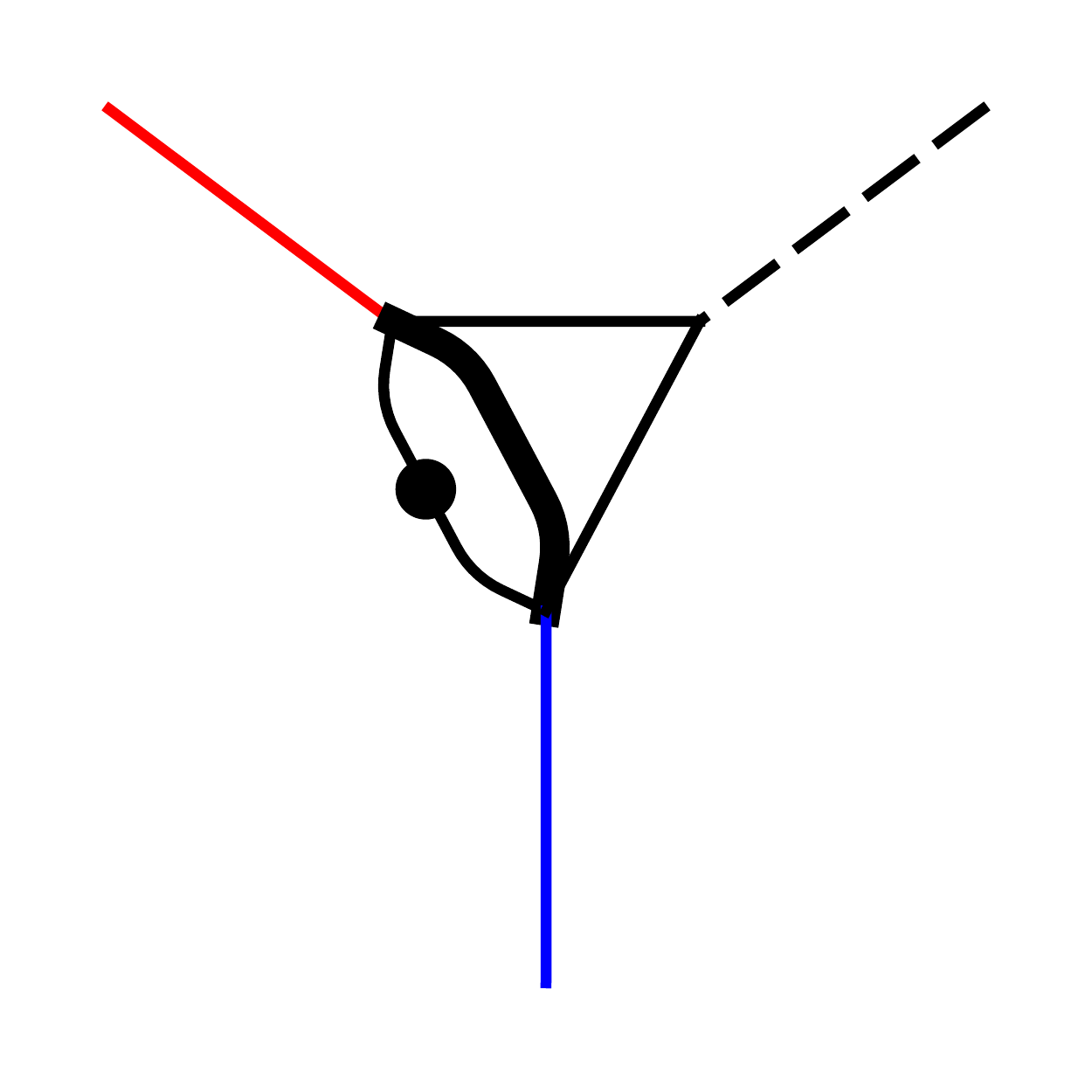}
  }
  \subfloat[$\mathcal{T}_{20}$]{%
    \includegraphics[width=0.14\textwidth]{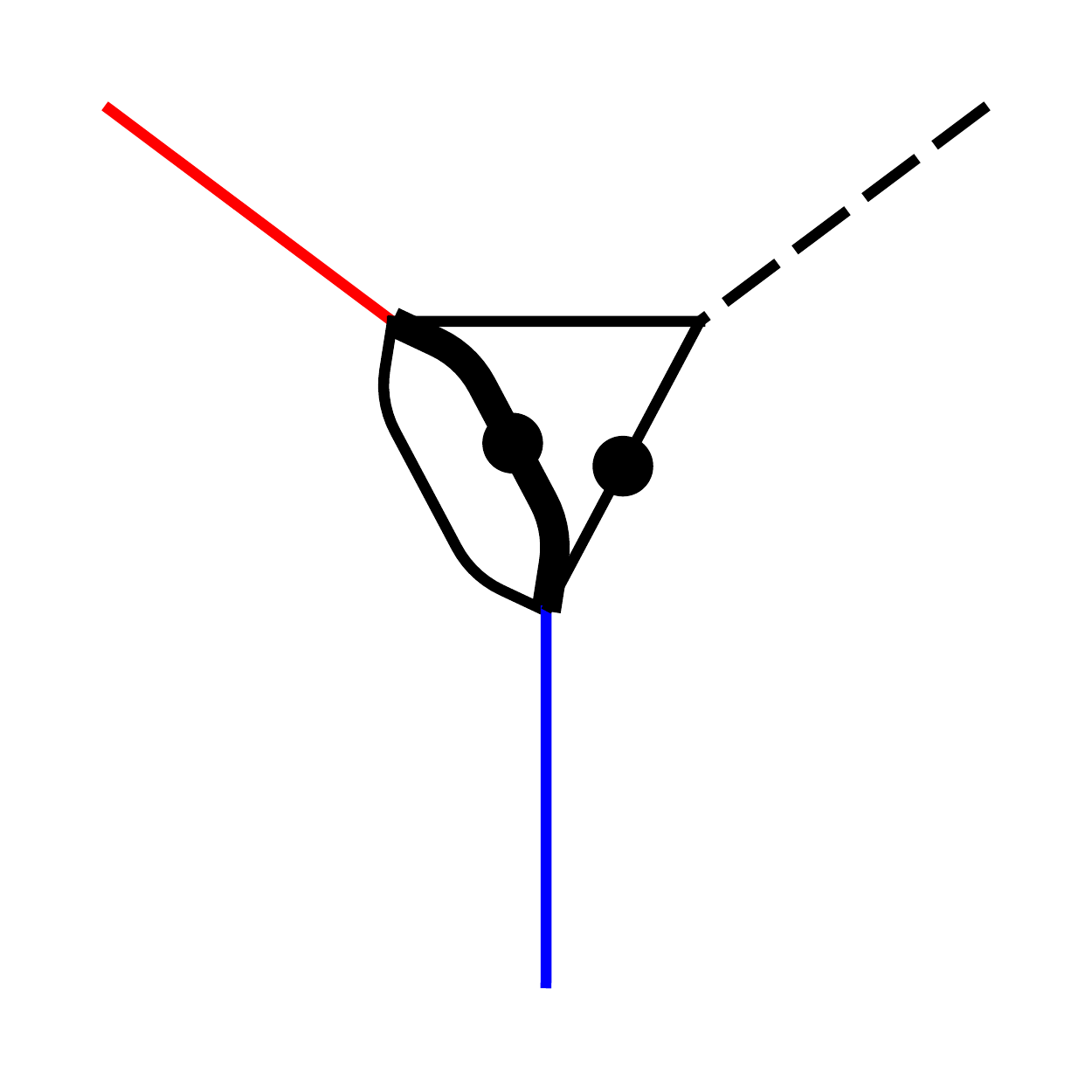}
  }
  \subfloat[$\mathcal{T}_{21}$]{%
    \includegraphics[width=0.14\textwidth]{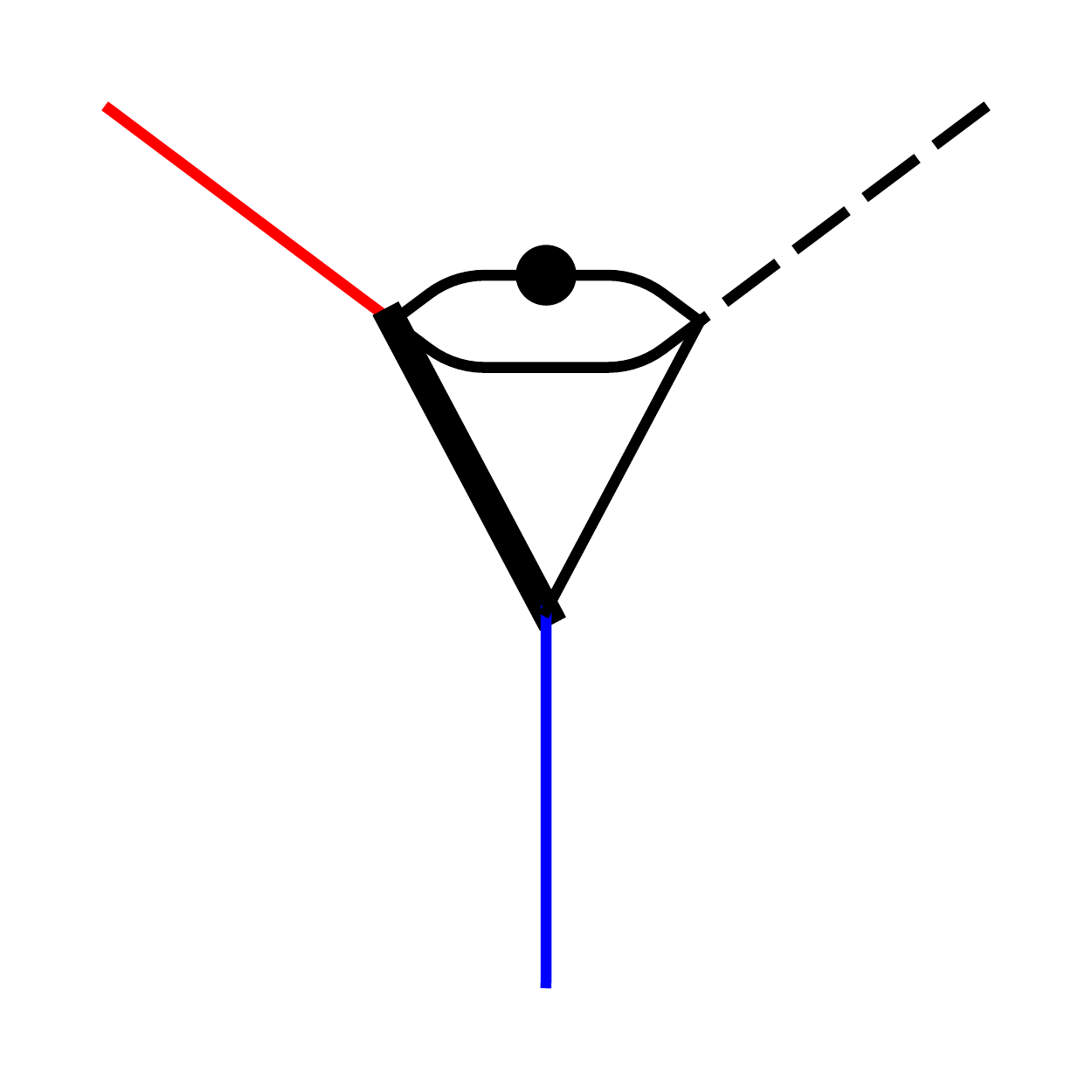}
  }
  \subfloat[$\mathcal{T}_{22}$]{%
    \includegraphics[width=0.14\textwidth]{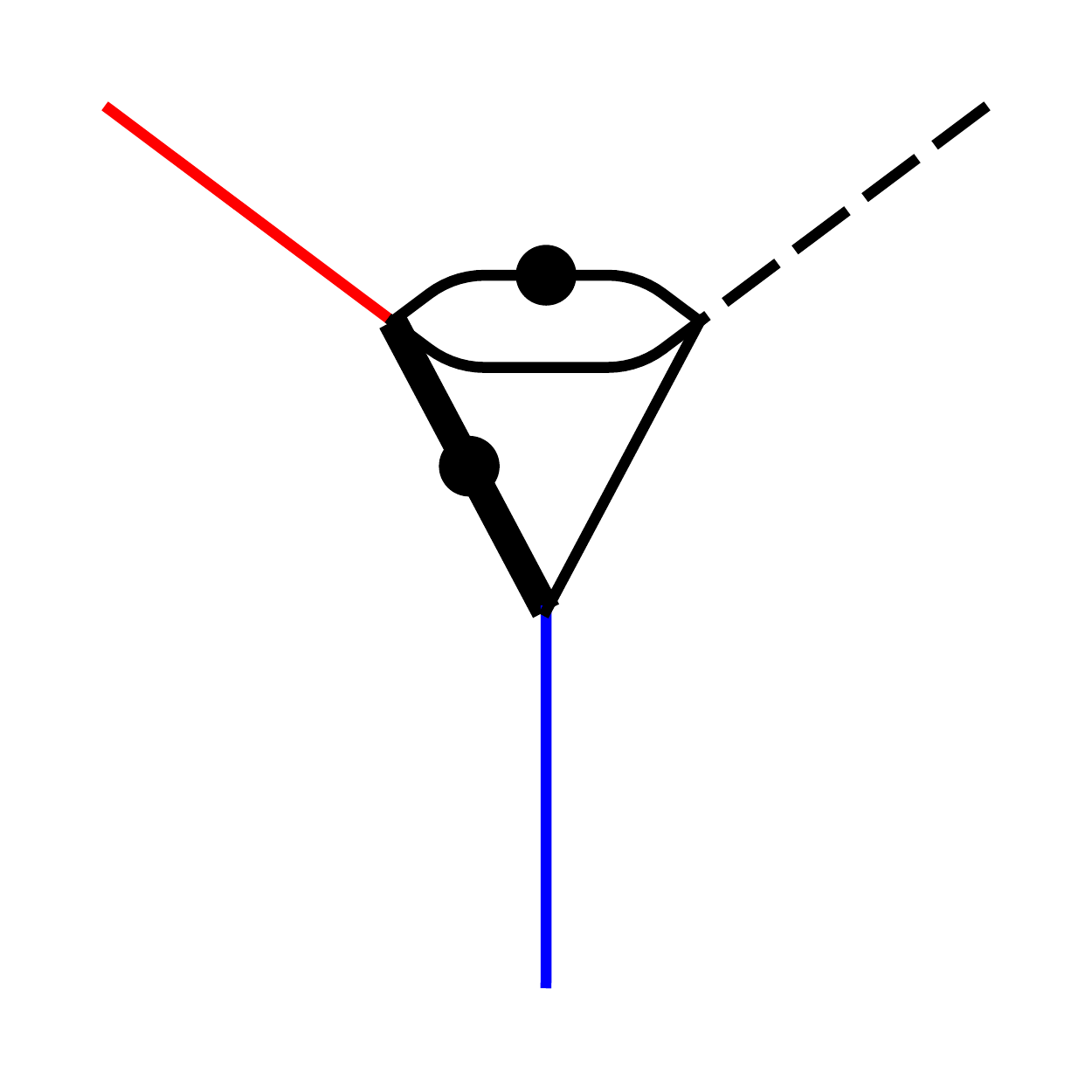}
  }
  \subfloat[$\mathcal{T}_{23}$]{%
    \includegraphics[width=0.14\textwidth]{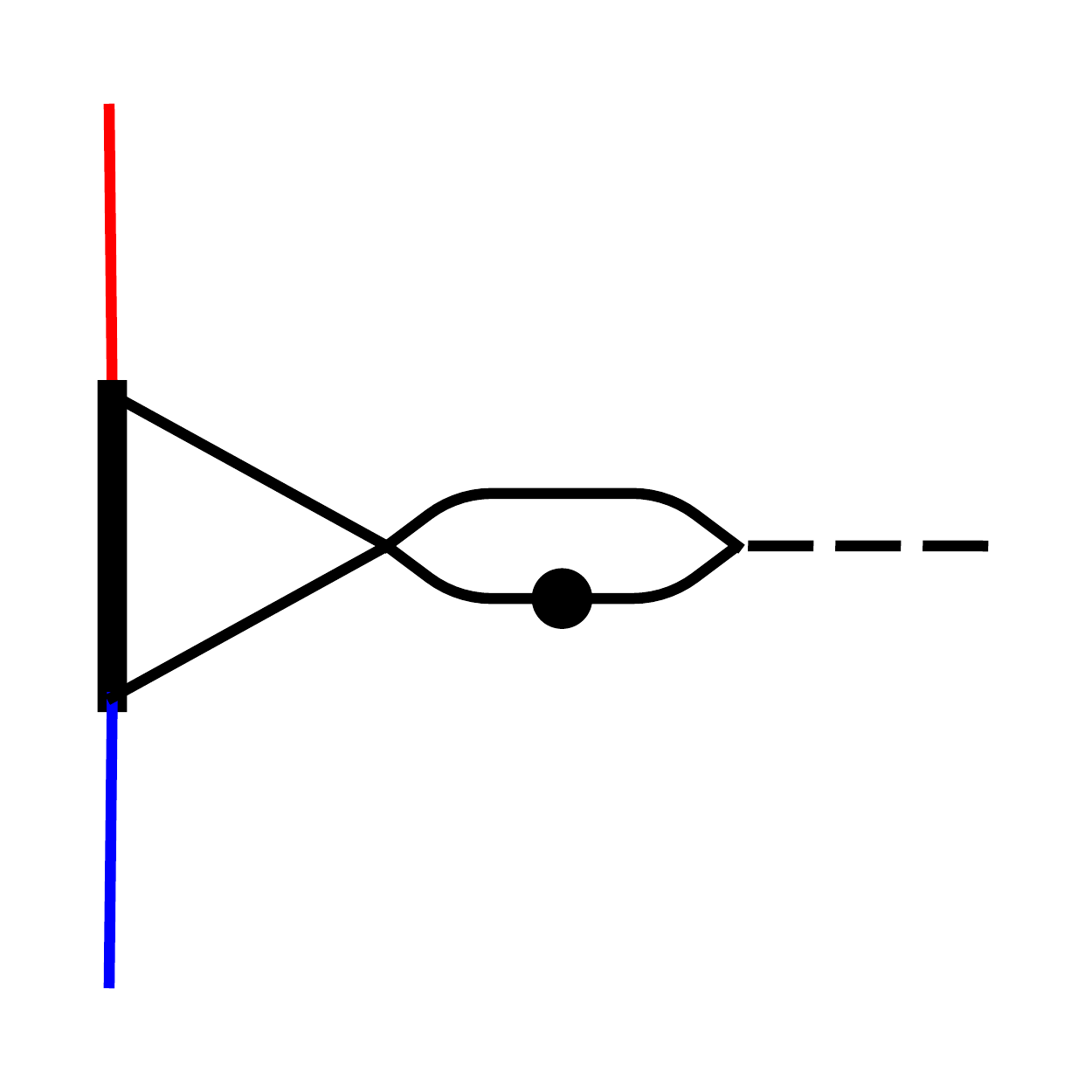}
  }
  \subfloat[$\mathcal{T}_{24}$]{%
    \includegraphics[width=0.14\textwidth]{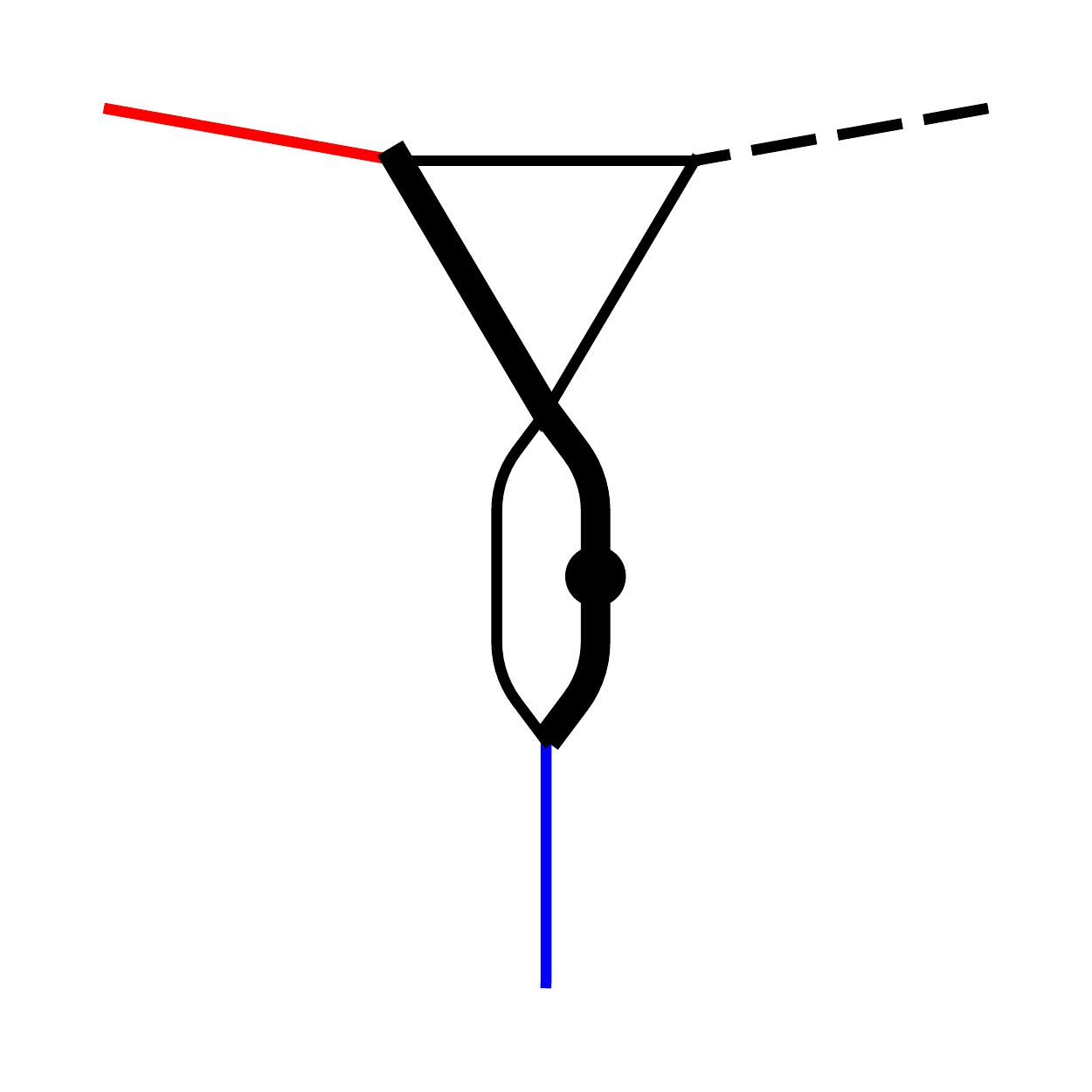}
  }
  \\
  \subfloat[$\mathcal{T}_{25}$]{%
    \includegraphics[width=0.14\textwidth]{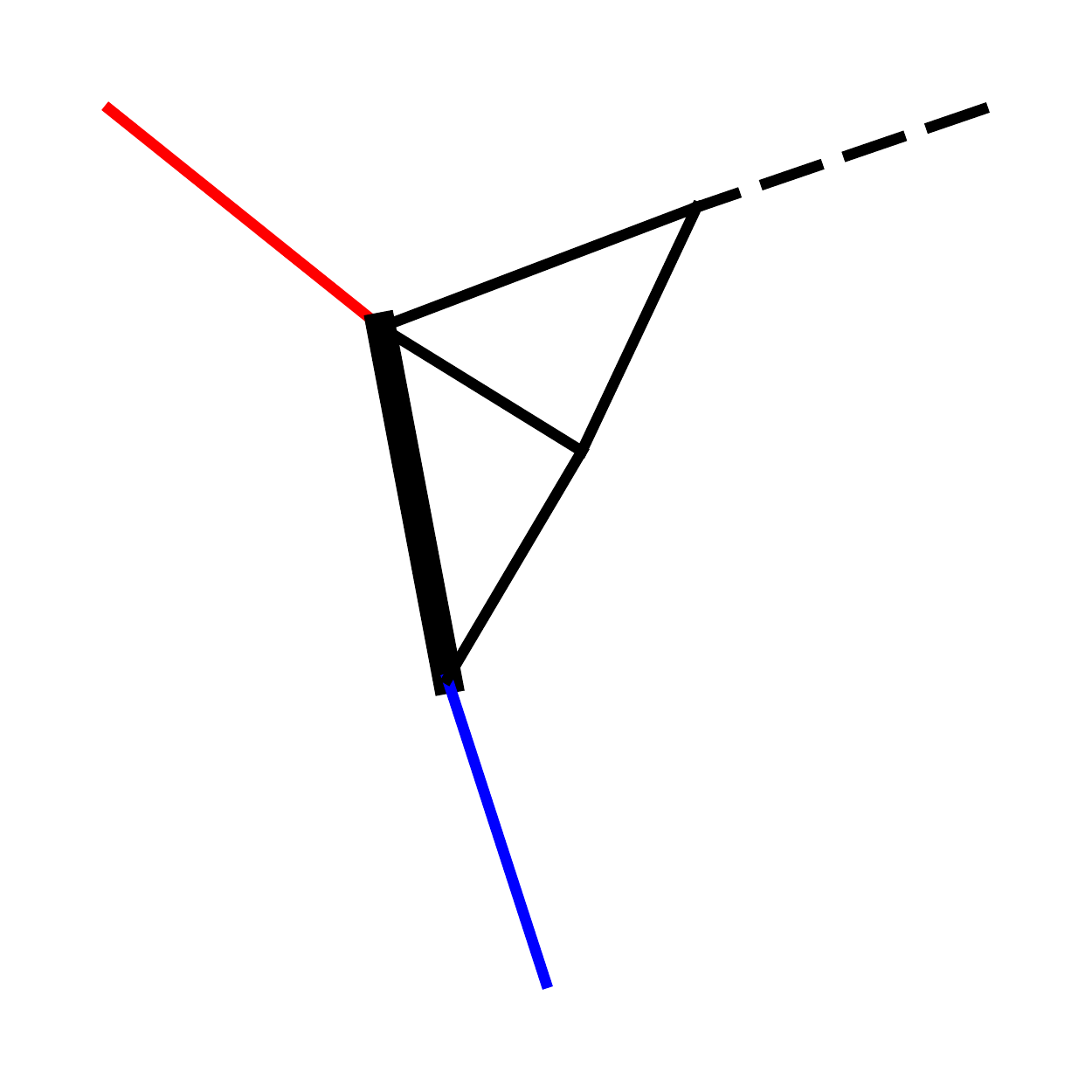}
  }
  \subfloat[$\mathcal{T}_{26}$]{%
    \includegraphics[width=0.14\textwidth]{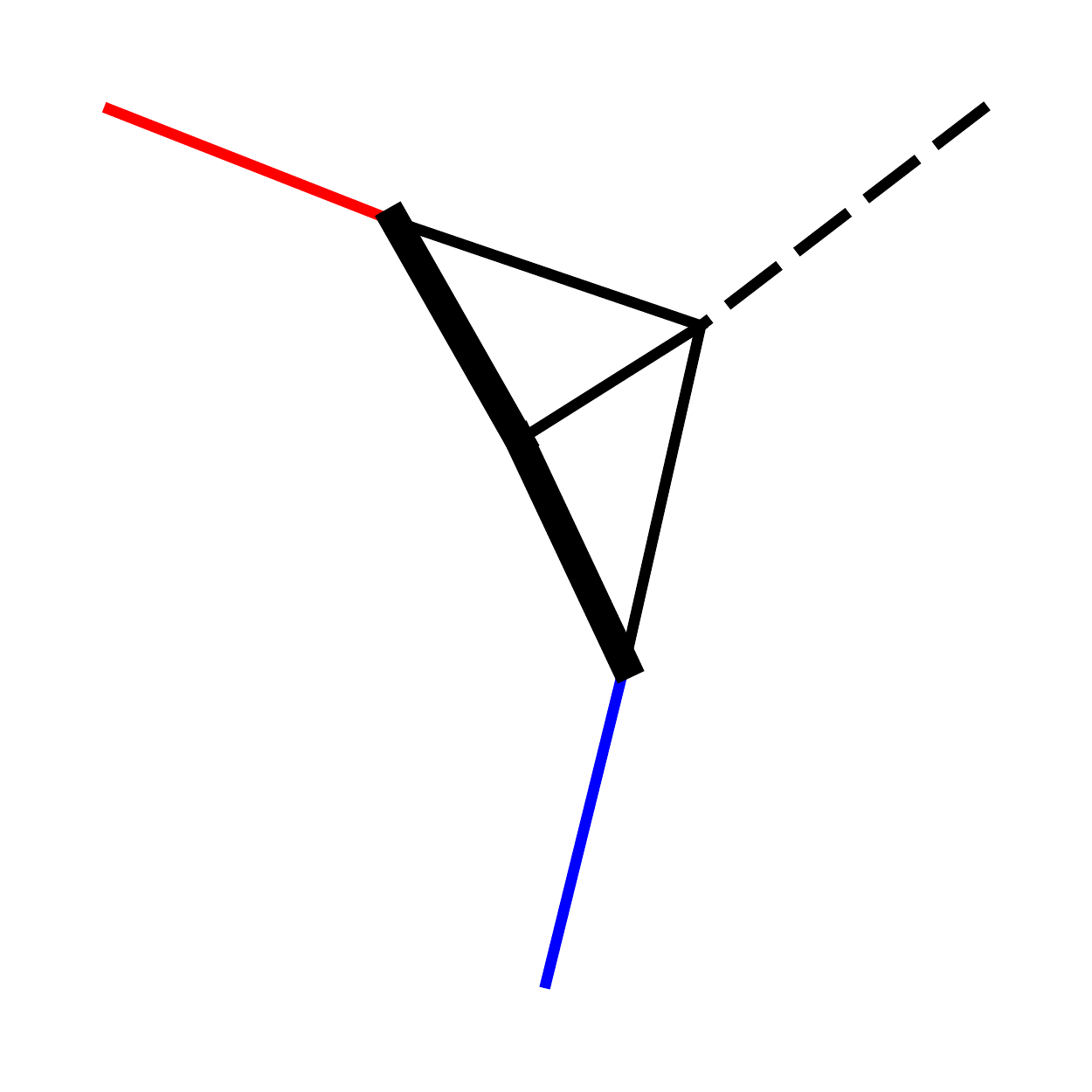}
  }
  \subfloat[$\mathcal{T}_{27}$]{%
    \includegraphics[width=0.14\textwidth]{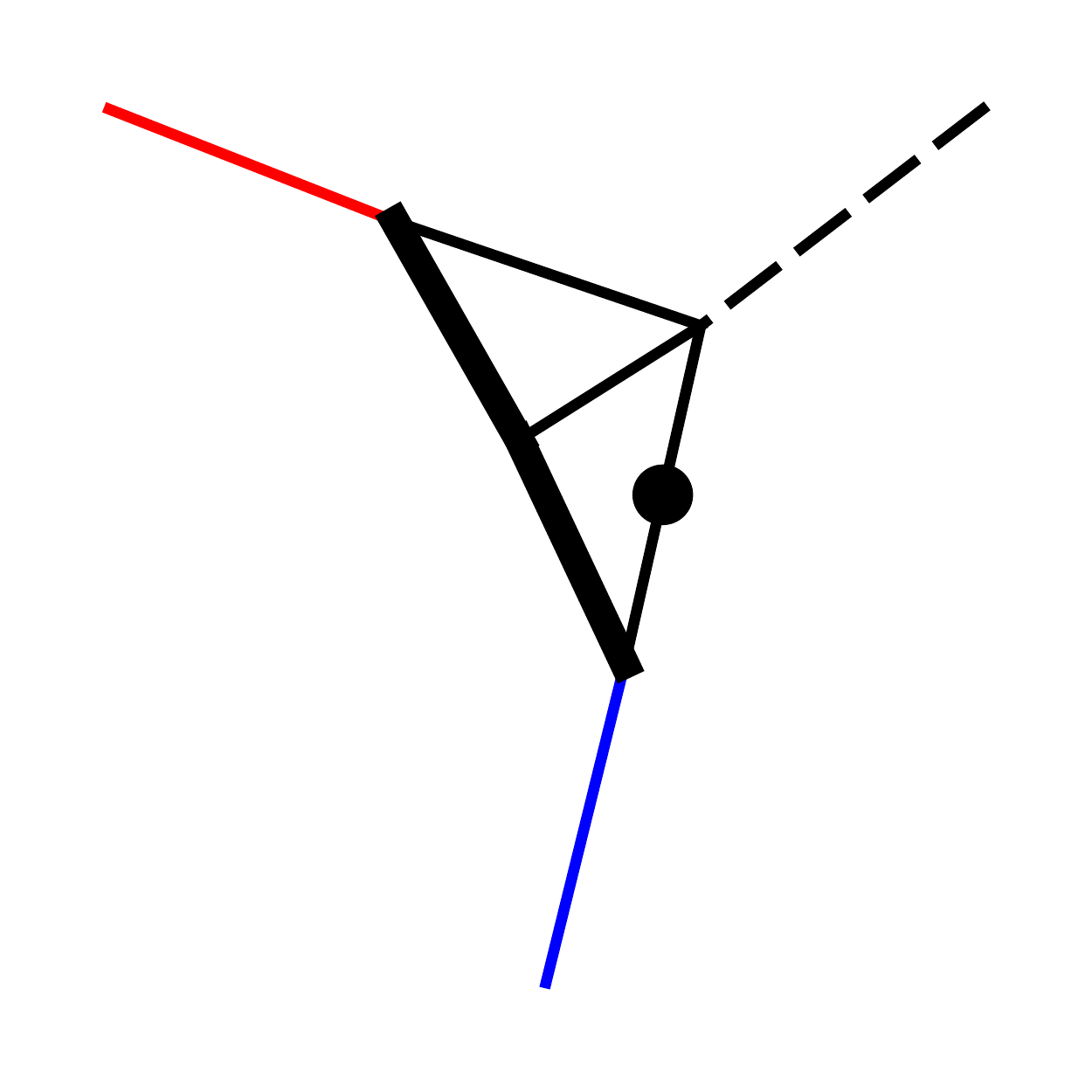}
  }
  \subfloat[$\mathcal{T}_{28}$]{%
    \includegraphics[width=0.14\textwidth]{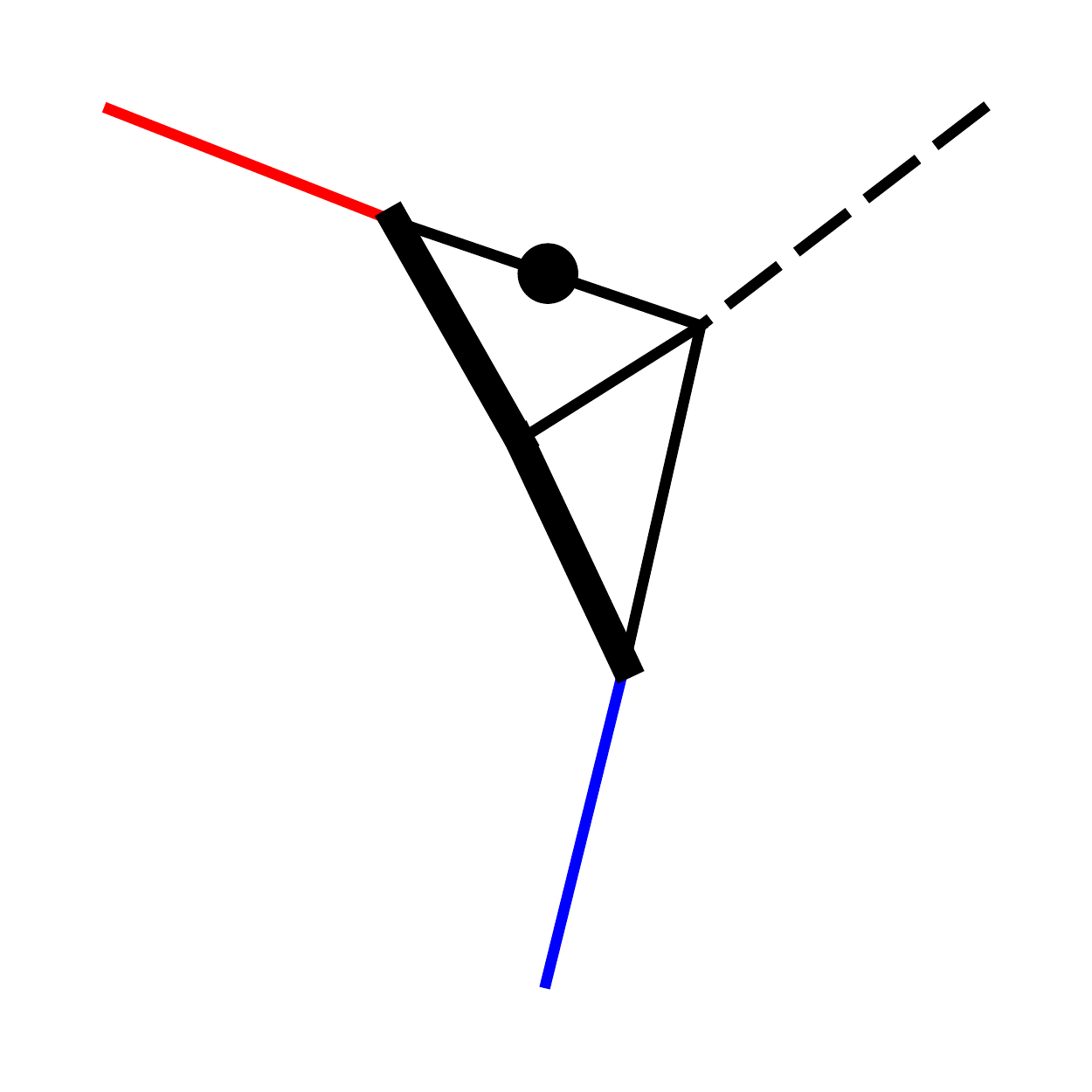}
  }
  \subfloat[$\mathcal{T}_{29}$]{%
    \includegraphics[width=0.14\textwidth]{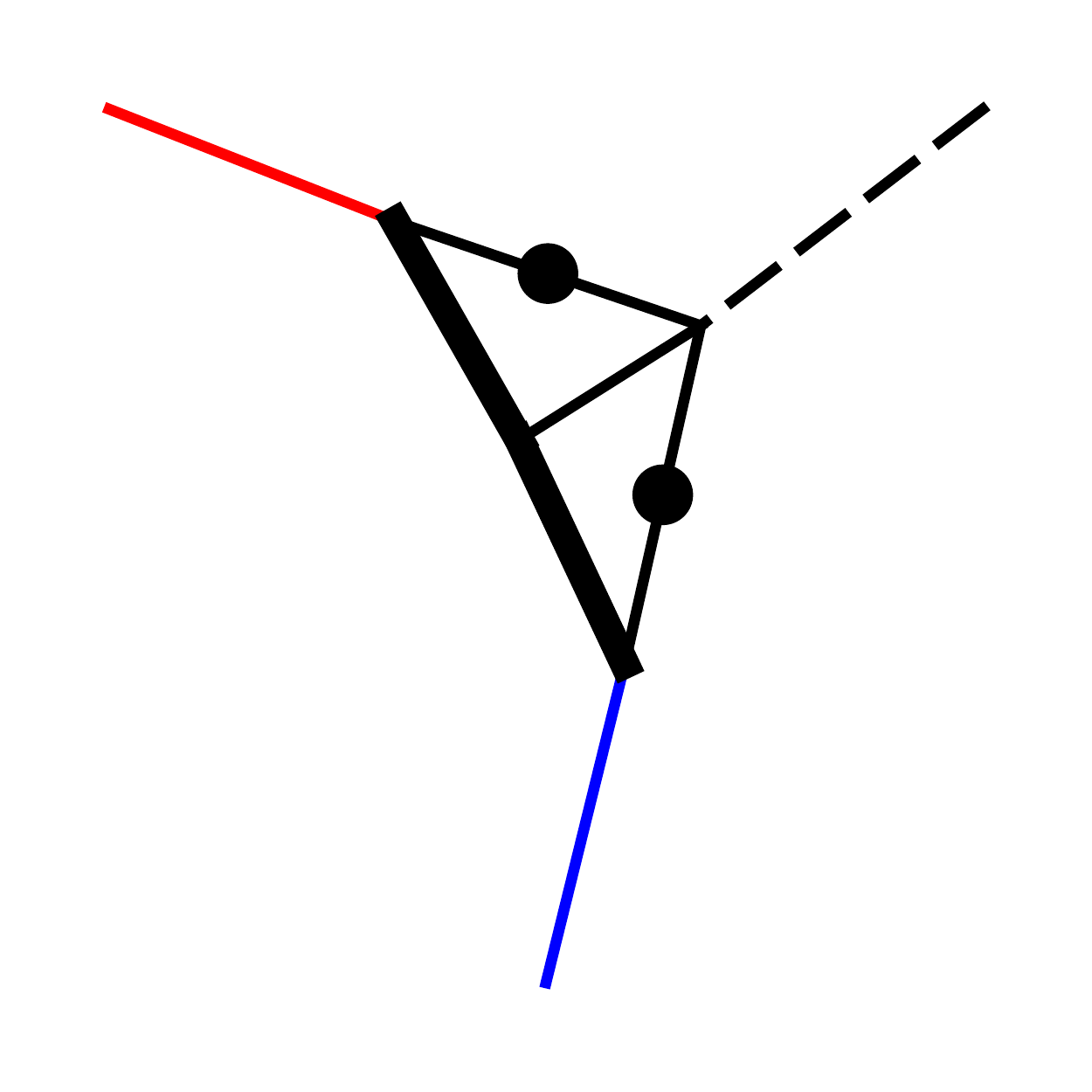}
  }
  \caption{Two-loop MIs $\mathcal{T}_{1,\ldots,29}$ for topologies
    \protect\subref{fig:FigTop1WWbb}-\protect\subref{fig:FigTop2WWbb}. Graphical
    conventions are the same as in figure~\ref{fig:TopoWW}. Dots
    indicate squared propagators.}
  \label{fig:WWMIsbbbb}
\end{figure}	 

where the $\mathcal{T}_i$ are depicted in
figure~\ref{fig:WWMIsbbbb}. We observe that some of integrals
$\mathcal{T}_i$ are trivially related by $p_1^2 \leftrightarrow p_2^2$
symmetry,
\begin{align}
  \top{4} \leftrightarrow  \top{2}\,,\quad  \top{8} \leftrightarrow  \top{5}\,, \quad \top{9} \leftrightarrow  \top{6}\,, \quad  \top{14} \leftrightarrow  \top{12}\,, \quad  \top{21} \leftrightarrow  \top{16}\,, \quad  \top{22} \leftrightarrow  \top{17}\,,
  \label{eq:Tabsymm}
\end{align}
so that the actual number of independent integrals is reduced to
23. However, in order to determine the solution of the DEs with the
method described in section~\ref{sec:diffeq}, i.e. by
simultaneously integrating the whole system of equations, one has to
consider the full set of integrals given in
eq.~\eqref{def:1M1LBasisMIsb}.

The Magnus exponential allow us to obtain a set of canonical MIs
obeying a system of equations of the form \eqref{eq:canonicalDE}

\begin{align*}
  \begin{alignedat}{2}
  \GG_{1}&=   \FF_1\,, & \qquad
  \GG_{2}&= -p_2^2  \,  \FF_2\,,\nn
  \GG_{3}&= -s\, \FF_3\,,  & \qquad
  \GG_{4}&= -p_1^2 \FF_4 \,,\nn
  \GG_{5}&= -p_2^2 \, \FF_5\,, &\qquad
  \GG_{6}&= 2m^2\,\FF_5+(m^2-p_2^2) \,  \FF_6\,,  \nn
  \GG_{7}&=-s\,\FF_7\,, & \qquad
  \GG_{8}&= -p_1^2\FF_8\,,\nn
  \GG_{9}&=2m^2\,\FF_{8}+(m^2-p_1^2)\,\FF_{9}\,,& \qquad
  \GG_{10}&= -\sqrt{\lambda}\,\FF_{10}\,,\nn
  \GG_{11}&= p_1^2\,p_2^2\, \FF_{11}\,,  & \qquad  
  \GG_{12}&= p_1^2\,s\, \FF_{12}\,,\nn
  \GG_{13}&= p_2^4\, \FF_{13}\,,  & \qquad  
  \GG_{14}&= p_1^2\,s \, \FF_{14}\,,\nn
  \GG_{15}&=s^2 \, \FF_{15}\,,  & \qquad  
  \GG_{16}&=-\sqrt{\lambda} \, \FF_{16}\,, \nn 
  \end{alignedat}
  \\
  \begin{alignedat}{2}
  \GG_{17}&= c_{16,\,17}\, \FF_{16}+ c_{17,\,17}\, \FF_{17}\,,  & \qquad 
  \GG_{18}&= -\sqrt{\lambda} \, \FF_{18}\,, \nn
  \GG_{19}&= -\sqrt{\lambda} \, \FF_{19}\,, & \qquad 
  \GG_{20}&=c_{18,\,20}\, \FF_{18}+c_{19,\,20}\, \FF_{19}+c_{20,\,20}\, \FF_{20}\,, \nn
    \GG_{21}&= -\sqrt{\lambda} \, \FF_{21}\,,  & \qquad 
   \GG_{22}&= c_{21,\,22}\, \FF_{21}+c_{22,\,22}\, \FF_{22}\,, \nn
  \GG_{23}&= s \sqrt{\lambda} \, \FF_{23}\,, &\qquad
   \GG_{24}&= p_2^2\sqrt{\lambda} \, \FF_{24}\,, \nn
   \GG_{25}&= -\sqrt{\lambda}\, \FF_{25}\,,& \qquad
   \GG_{26}&= -\sqrt{\lambda}\, \FF_{26}\,,\nn
   \GG_{27}&= (p_2^2-m^2)\sqrt{\lambda} \, \FF_{27}\,,  & \qquad 
   \GG_{28}&= (p_1^2-m^2)\sqrt{\lambda}\, \FF_{28}\,,\nn
   \end{alignedat}
   \nn
   \begin{alignedat}{1}
  \GG_{29}&= c_{1,\,29}\, \FF_{1}+c_{2,\,29}\, \FF_{2}+c_{4,\,29}\, \FF_{4}+c_{11,\,29}\, \FF_{11}+c_{27,\,29}\, \FF_{27}+c_{28,\,29}\, \FF_{28}+c_{29,\,29}\, \FF_{29}\,,
   \label{def:WWCanonicalMIsbb}
   \end{alignedat}\stepcounter{equation}\tag{\theequation}
 \end{align*}
 where $\lambda$ is the K{\"a}ll\'{e}n function related to the
 external kinematics,
 \begin{align}
   \lambda \equiv \lambda(s,p_1^2,p_2^2)=(s - p_1^2 -p_2^2)^2-4\,p_1^2\,p_2^2.
   \label{eq:lambda}
 \end{align}
 Explicit expressions for the coefficients $c_{i,\,j}$ are given in
 appendix~\ref{sec:gMIcoeffb}. The alphabet of the corresponding
 $\dlog$-form contains the following 10 letters:
 \begin{align}
\eta_1 & =u\,,&
\eta_2 & =\zz\,, \nn 
\eta_3 & =1-\zz\,, &
\eta_4 & =\zzb\,, \nn
\eta_5 & =1-\zzb\,,&
\eta_6 & =\zz-\zzb\,,\nn
\eta_7 & =1+u\,\zz\,\zzb\,,&
\eta_8 & =1-u\,\zz(1-\zzb),\nn
\eta_9 & =1-u\,\zzb(1-\zz)\,,&
\eta_{10} & =1+u\, (1-\zz)(1-\zzb)\,.
\label{alphabet:MWWb}
\end{align}
The coefficient matrices $\MM_i$ are collected in the
appendix~\ref{dlogWW2Lbb}.  It can be easily checked that all letters
are real and positive in the region 
\begin{equation}
  \label{eq:positivityabuzzb}
  0<\zz<1\,, \quad 0<\zzb<\zz\,, \quad 0<u<\frac{1}{\zz(1-\zzb)}\,.
\end{equation}
If one fixes $m^2>0$, this corresponds to a patch of the Euclidean
region, $s\,,p_1^2\,, p_2^2 < 0$, defined by the following
constraints
\begin{gather}
  \label{eq:positivityabsp1p2}
  \sqrt{-p_1^2} \sqrt{-p_2^2} > m^2 \,, \nn
  -\frac{\left(p_1^2 - m^2\right) \left(p_2^2 - m^2\right)}{m^2} < s < p_1^2+p_2^2-2 \sqrt{-p_1^2} \sqrt{-p_2^2}\,.
\end{gather}
Since the alphabet is rational, the solution can be directly expressed
in terms of GPLs with argument depending on the kinematics variables
$u$, $\zz$ and $\zzb$. The prescriptions for the analytic continuation
to the other patches of the Euclidean region ($s,p_1^2,p_2^2<0$) and
to the physical regions are given in section~\ref{sec:continuation}.

Imposing the regularity of our solutions at the unphysical thresholds,
$\zz,\zzb=0$ (corresponding to $p_1^2=0$) and $\zz,\zzb=1$
(corresponding to $p_2^2=0$) entails relations between the boundary
constants. These relations allow us to derive all boundary constants
from five simpler integrals $\GG_{1,3,6,7,15}$, which are obtained in the
following way:

\begin{itemize}
\item $\GG_1$ is a constant to be determined by direct integration
  and, due to the normalization of the integration measure
  \eqref{eq:intmeasure1}, it is simply set to
  \begin{align}
    \GG_1(\eps,\vec{x})=1.
    \label{bound_i1_top}
  \end{align}	
\item $\GG_{3}$ can be obtained by direct integration
  \begin{equation}
    \GG_{3}(\eps,\vec{x})=  \frac{\Gamma(1-\eps)^2}{\Gamma(1-2\eps)} u^{-\eps} \, , 
  \end{equation}
\item Besides being regular in the massless kinematic limit $\zz\to1$
  ($p_2^2 \rightarrow 0$), $\GG_6$ is reduced, through IBPs, to a
  two-loop vacuum diagram,
  \begin{align}
    \GG_{6}(\eps,\zz=1)= -\frac{2\eps^2(1-\eps)(1-2\eps)}{m^2} \raisebox{-19pt}{\includegraphics[scale=0.11]{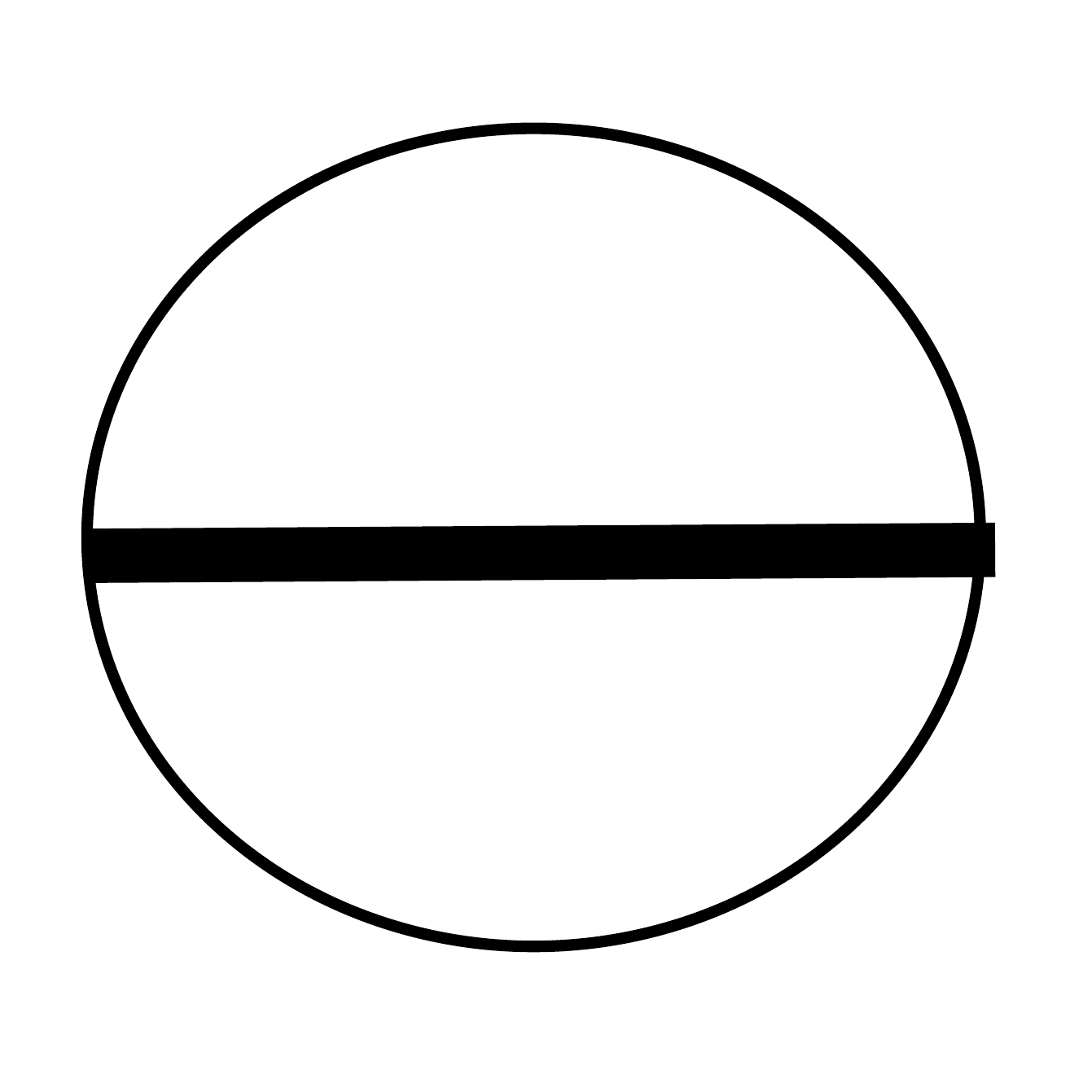}}.
    \label{eq:IBPMI5}
  \end{align}
  Therefore, by using as an input the analytic expression of the
  two-loop vacuum graph,
  \begin{align}
    \raisebox{-19pt}{\includegraphics[scale=0.11]{figures/FigVac.pdf}}=-\frac{m^2 \Gamma(-\eps)\Gamma(-1+2\eps)}{(1-\eps) \Gamma(1+\eps)},
  \end{align}
  we can fix the boundary constants by matching the $z\to1$
  limit of the expression of $\GG_6$ obtained from the solution
  of the DE against the $\eps$-expansion of
  eq.~\eqref{eq:IBPMI5},
  \begin{align}
    \GG_{6}(\eps,\zz=1)=-1-\frac{1}{3}\pi^2\eps^2+2\zeta_3\eps^3-\frac{1}{10}\pi^4\eps^4+\mathcal{O}(\eps^5).
    \label{bound_i5_top}
  \end{align}
\item $\GG_{7}$ and $\GG_{15}$ can be directly integrated 
  \begin{align}
    \GG_{7}(\eps,\vec{x})= {} & - \frac{\Gamma(1-\eps)^3\Gamma(1+2\eps)}{\Gamma(1-3\eps)\Gamma(1+\eps)^2 }  \, u^{-2\eps} , \\
    \GG_{15}(\eps,\vec{x})= {} &  \frac{\Gamma(1-\eps)^4}{\Gamma(1-2\eps)^2} u^{-2\eps} \, .
  \end{align}
\end{itemize}
The results have been numerically checked, in both the Euclidean and
the physical regions, with the help of the public computer codes
\texttt{GiNaC} and \texttt{SecDec 3.0}, and their analytic expressions
are given in electronic form in the ancillary files attached to the
\texttt{arXiv} version of the manuscript.

 \subsection{Topologies (c)-(d)}
\label{sec:WWMIst}
The topologies \protect\subref{fig:FigTop1WW} and
\protect\subref{fig:FigTop2WW} belong to the 7-denominator family
defined by the set of denominators
\begin{gather}
\Den_1 = k_1^2-m^2,\quad
\Den_2 = k_2^2-m^2,\quad
\Den_3 = (k_1-p_2)^2,\quad
\Den_4 = (k_2-p_2)^2, \nonumber\\
\Den_5 = (k_1-p_1-p_2)^2-m^2,\quad
\Den_6 = (k_2-p_1-p_2)^2-m^2,\quad
\Den_7 = (k_1-k_2)^2,
\end{gather}
where $k_1$ and $k_2$ are the two loop momenta.  The integrals
belonging to this integral family can be reduced to a set of 31 MIs
which are conveniently expressed in terms of the variables $v$, $\zz$
and $\zzb$, defined by
\begin{align}
-\frac{s}{m^2} = & \frac{(1-v)^2}{v},\qquad \frac{p_1^2}{s} = \zz\zzb,\qquad \frac{p_2^2}{s} = (1-\zz)(1-\zzb).
\label{eq:paramcd}
\end{align}
\begin{figure}
  \centering
  \captionsetup[subfigure]{labelformat=empty}
  \subfloat[$\mathcal{T}_1$]{%
    \includegraphics[width=0.14\textwidth]{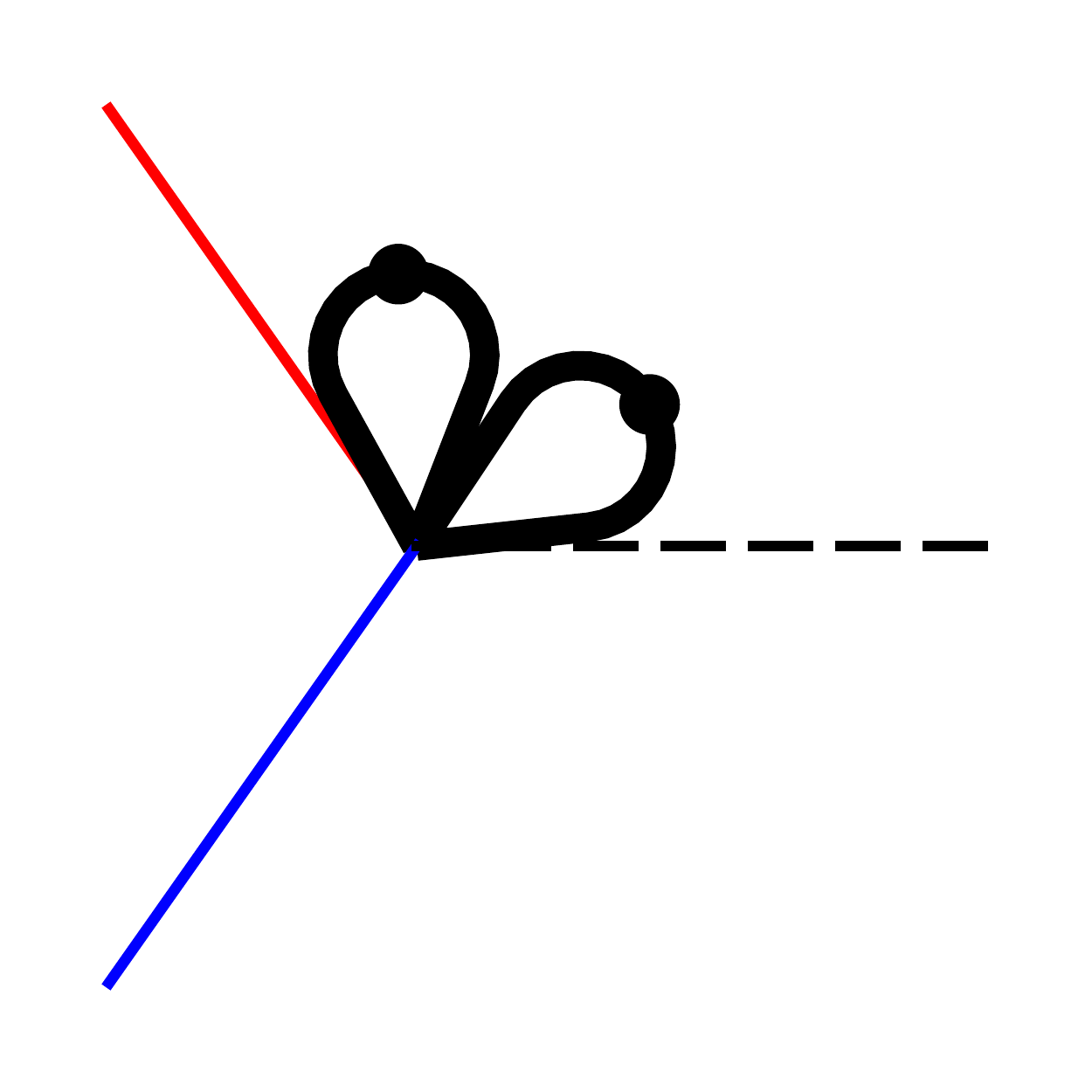}
  }
  \subfloat[$\mathcal{T}_2$]{%
    \includegraphics[width=0.14\textwidth]{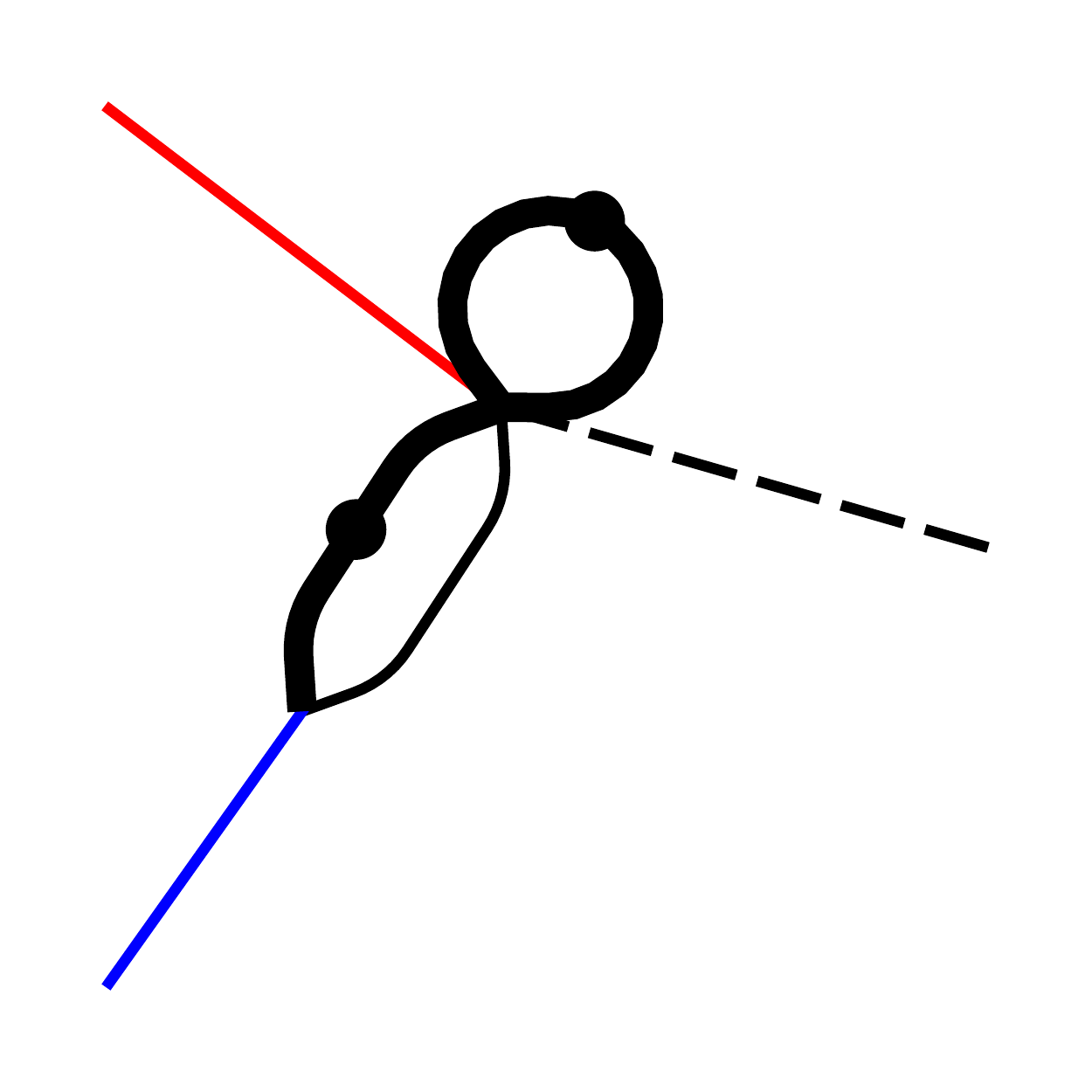}
  }
  \subfloat[$\mathcal{T}_3$]{%
    \includegraphics[width=0.14\textwidth]{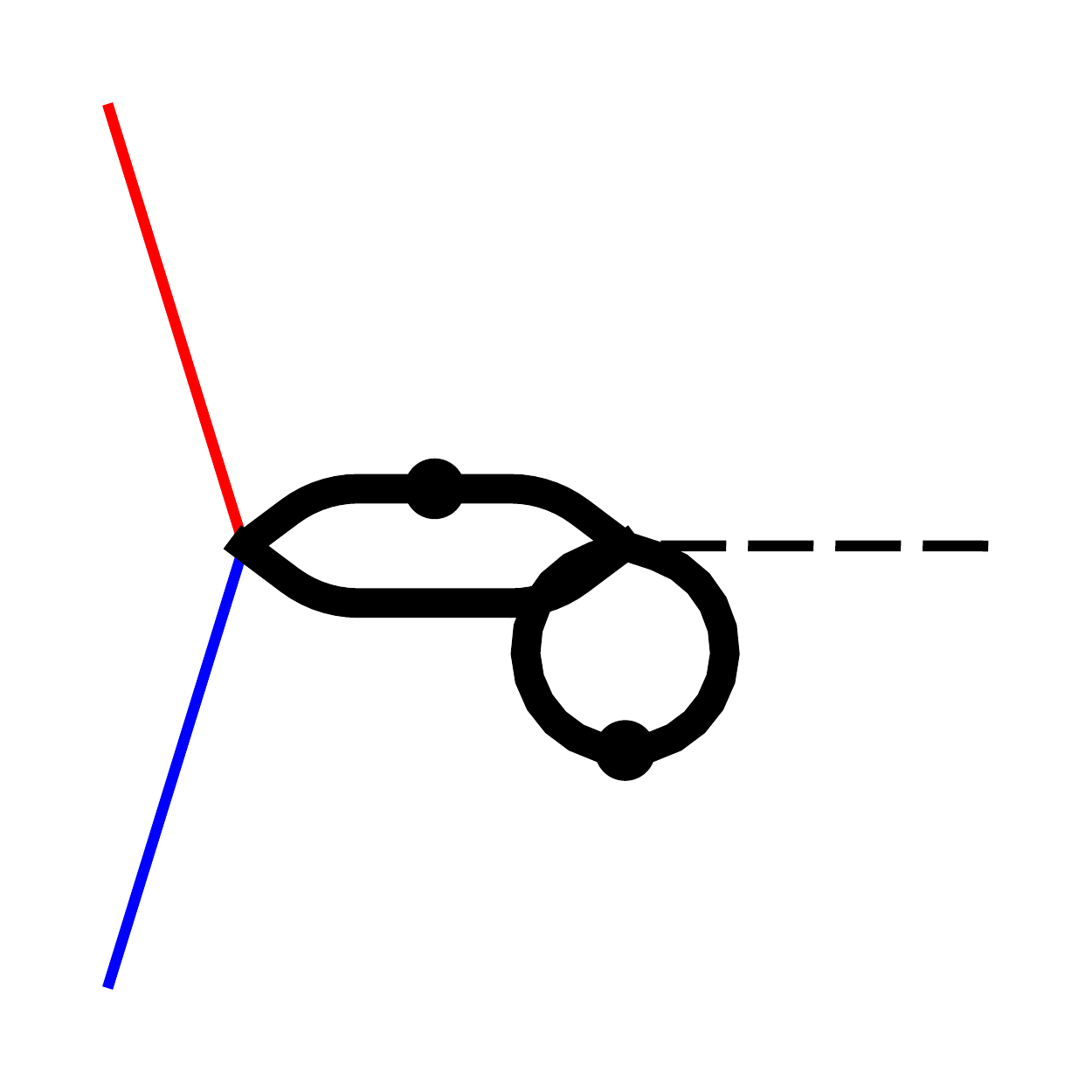}
  }
  \subfloat[$\mathcal{T}_4$]{%
    \includegraphics[width=0.14\textwidth]{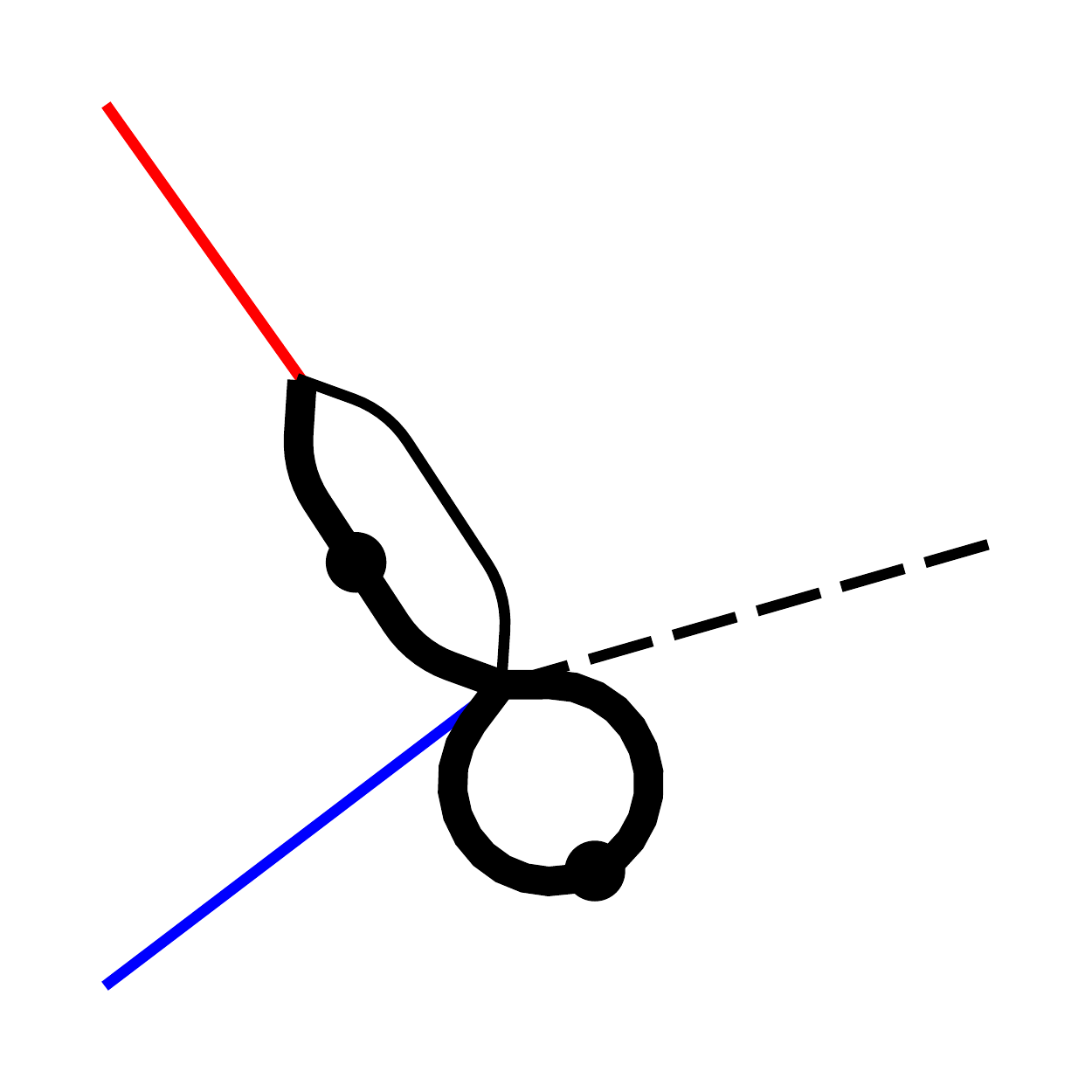}
  }
  \subfloat[$\mathcal{T}_5$]{%
    \includegraphics[width=0.14\textwidth]{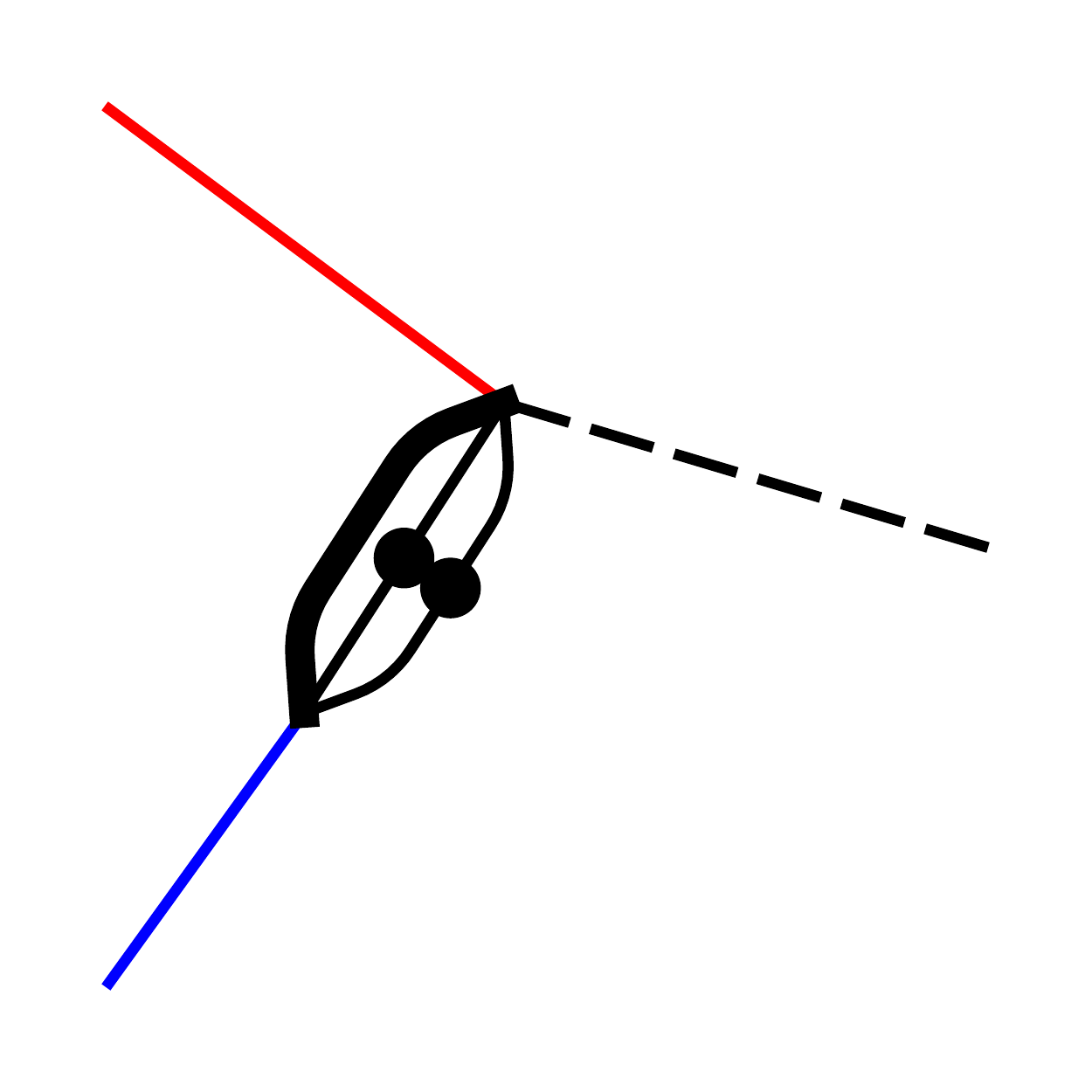}
  }
  \subfloat[$\mathcal{T}_6$]{%
    \includegraphics[width=0.14\textwidth]{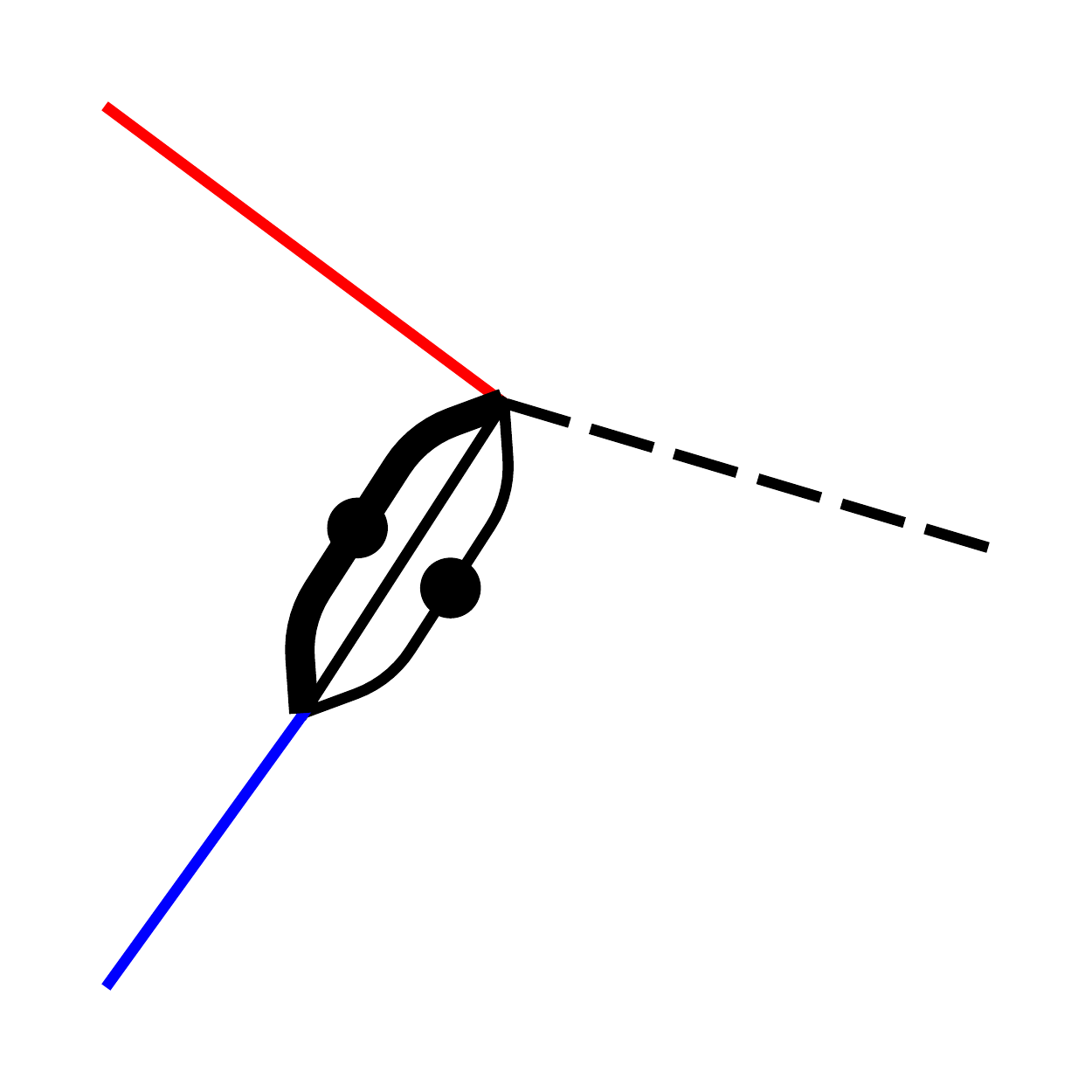}
  }
  \\
  \subfloat[$\mathcal{T}_7$]{%
    \includegraphics[width=0.14\textwidth]{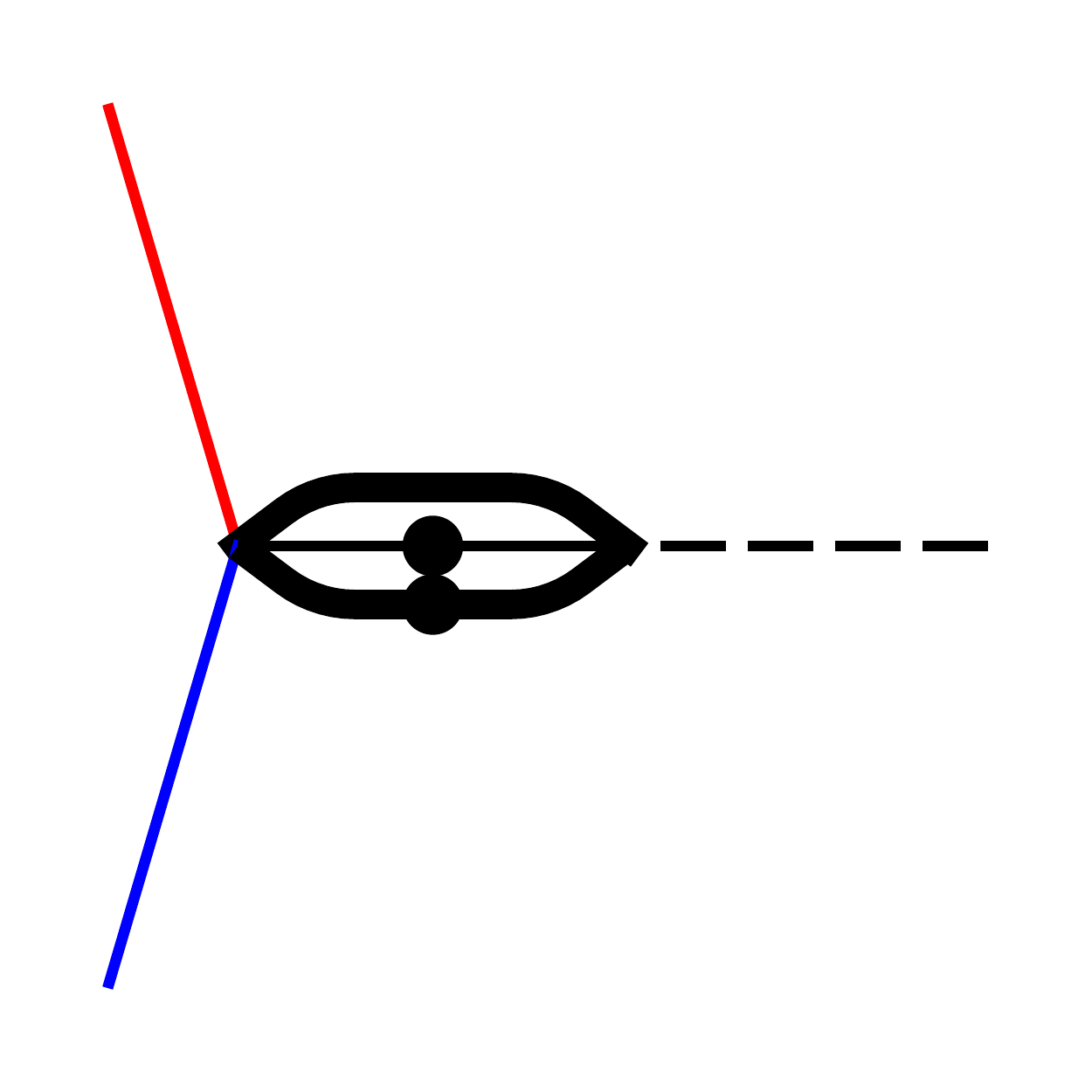}
  }
  \subfloat[$\mathcal{T}_8$]{%
    \includegraphics[width=0.14\textwidth]{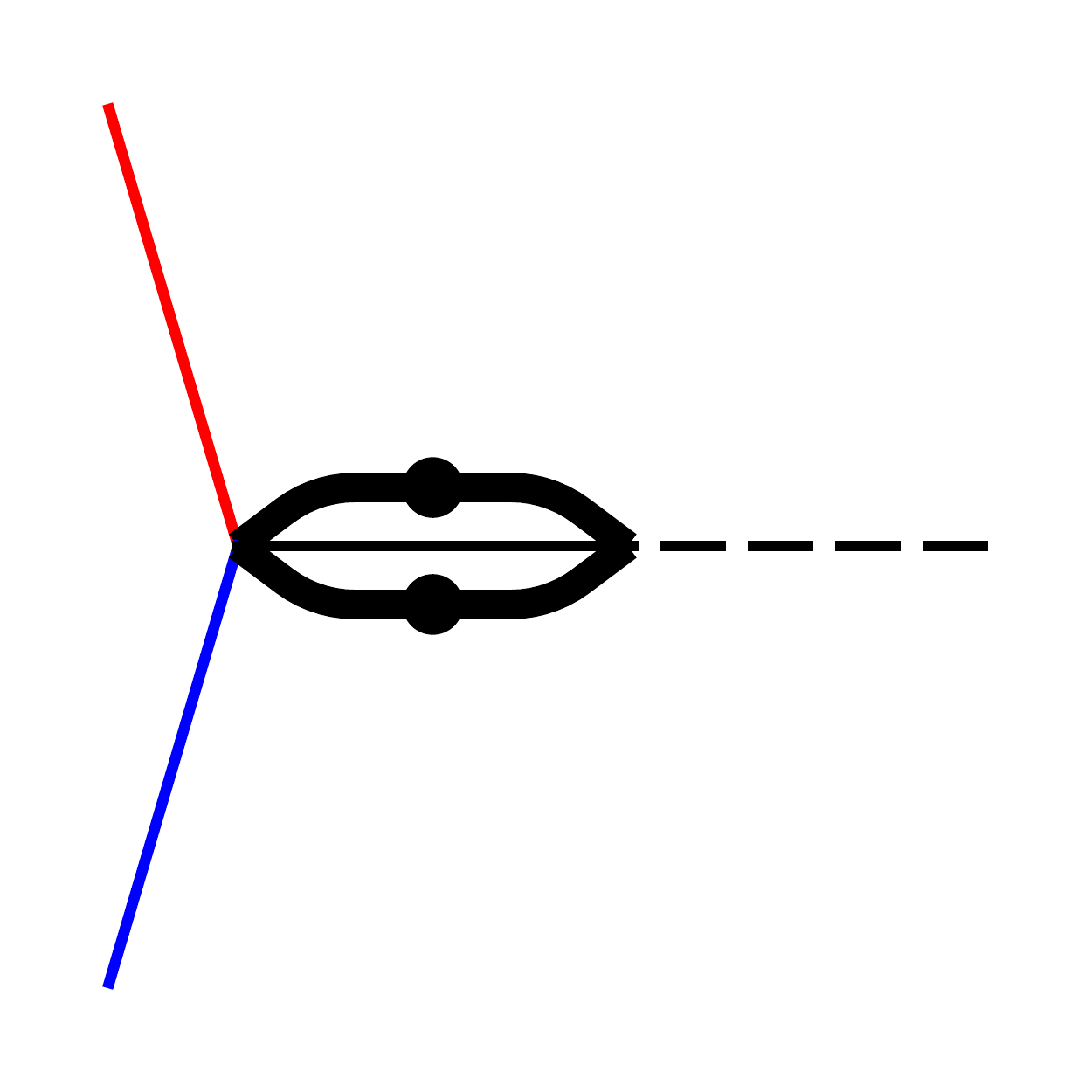}
  }
  \subfloat[$\mathcal{T}_9$]{%
    \includegraphics[width=0.14\textwidth]{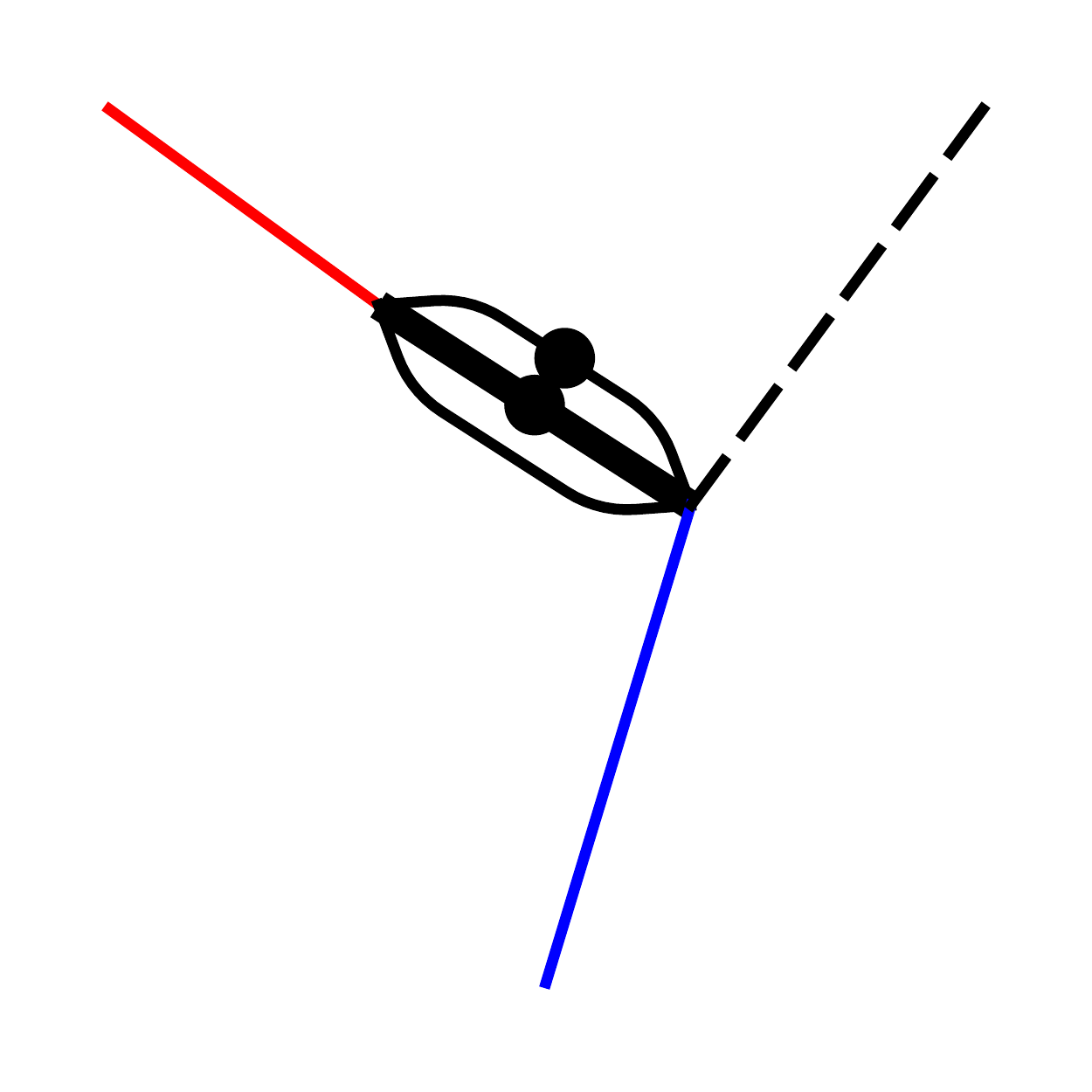}
  }
  \subfloat[$\mathcal{T}_{10}$]{%
    \includegraphics[width=0.14\textwidth]{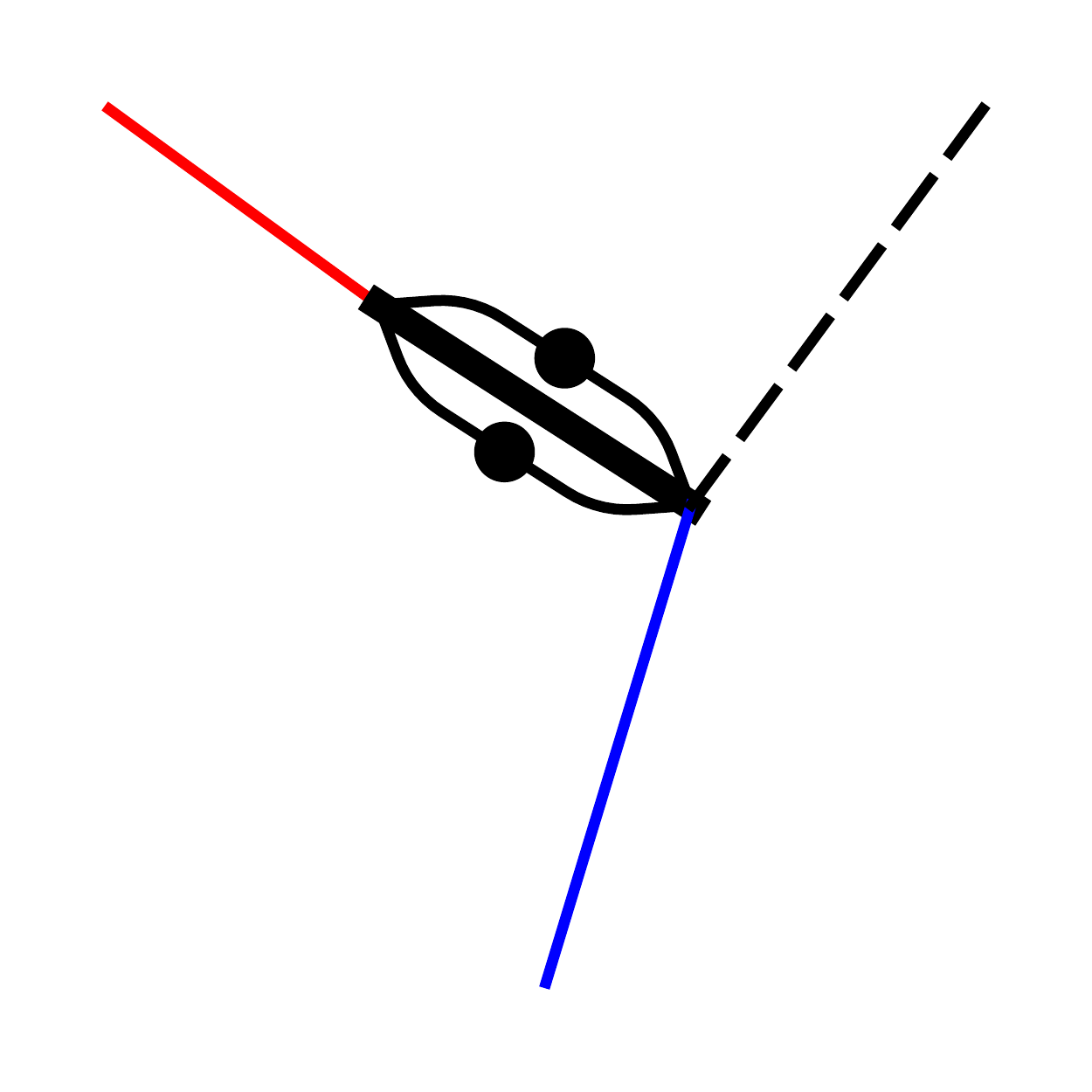}
  }
  \subfloat[$\mathcal{T}_{11}$]{%
    \includegraphics[width=0.14\textwidth]{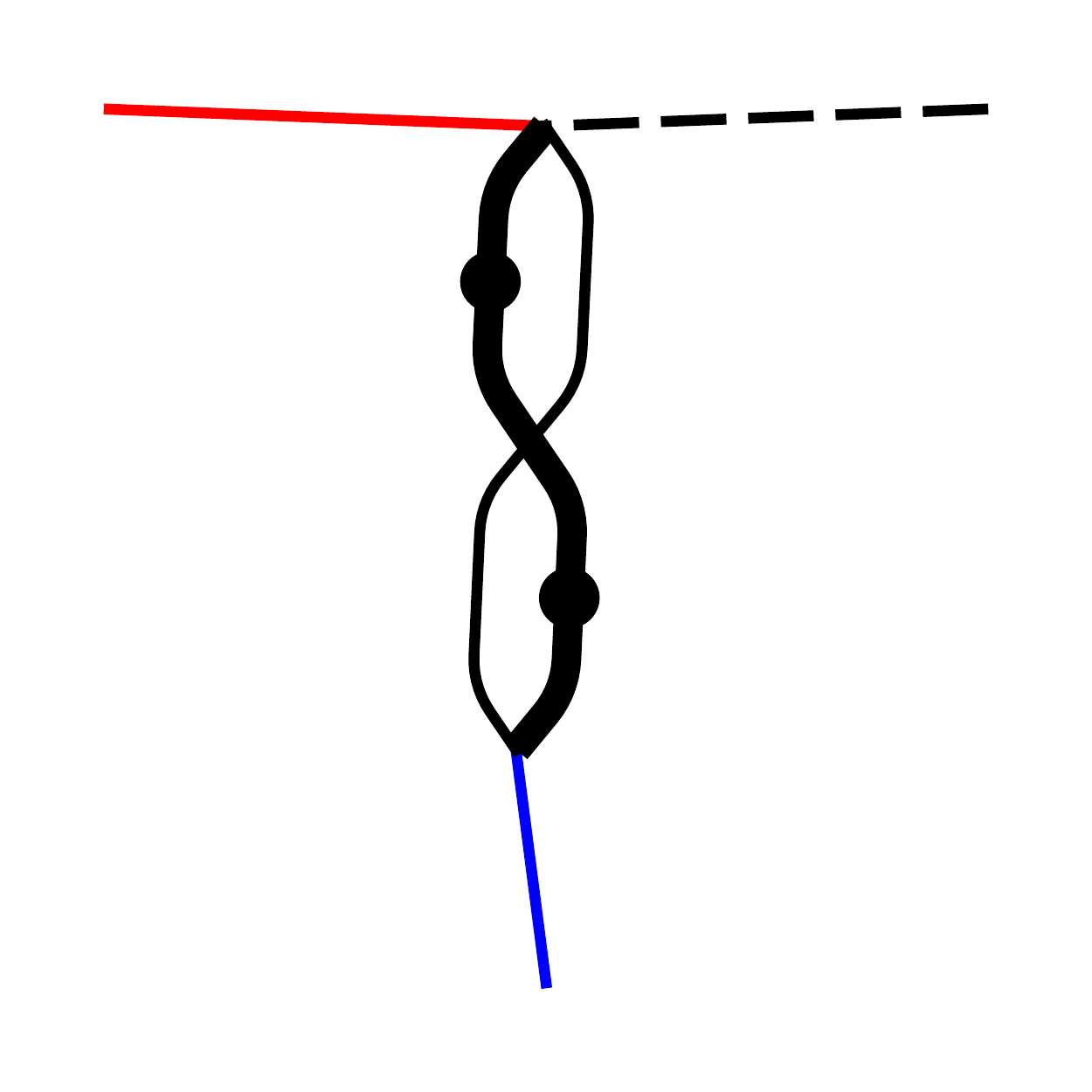}
  }
  \subfloat[$\mathcal{T}_{12}$]{%
    \includegraphics[width=0.14\textwidth]{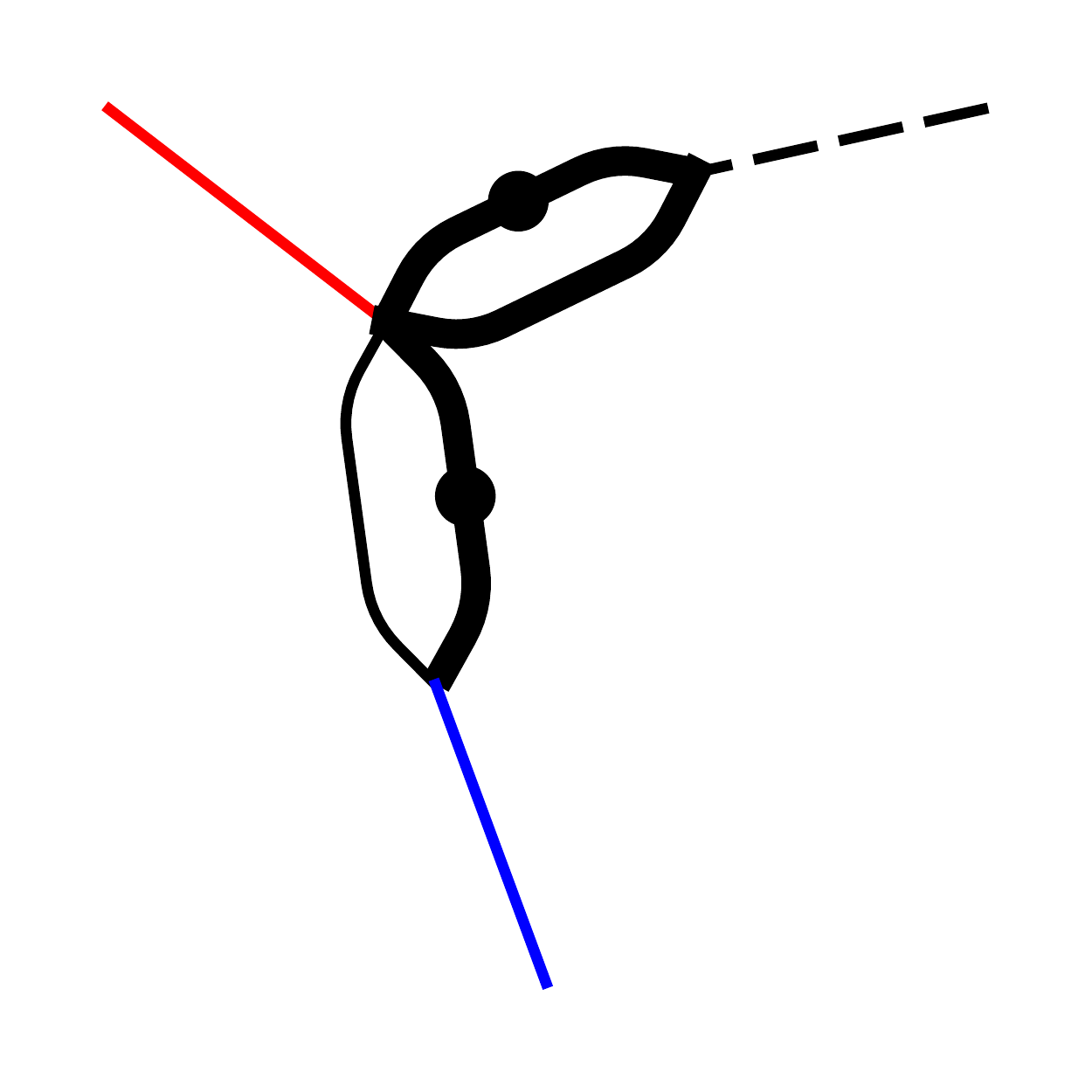}
  }
  \\
  \subfloat[$\mathcal{T}_{13}$]{%
    \includegraphics[width=0.14\textwidth]{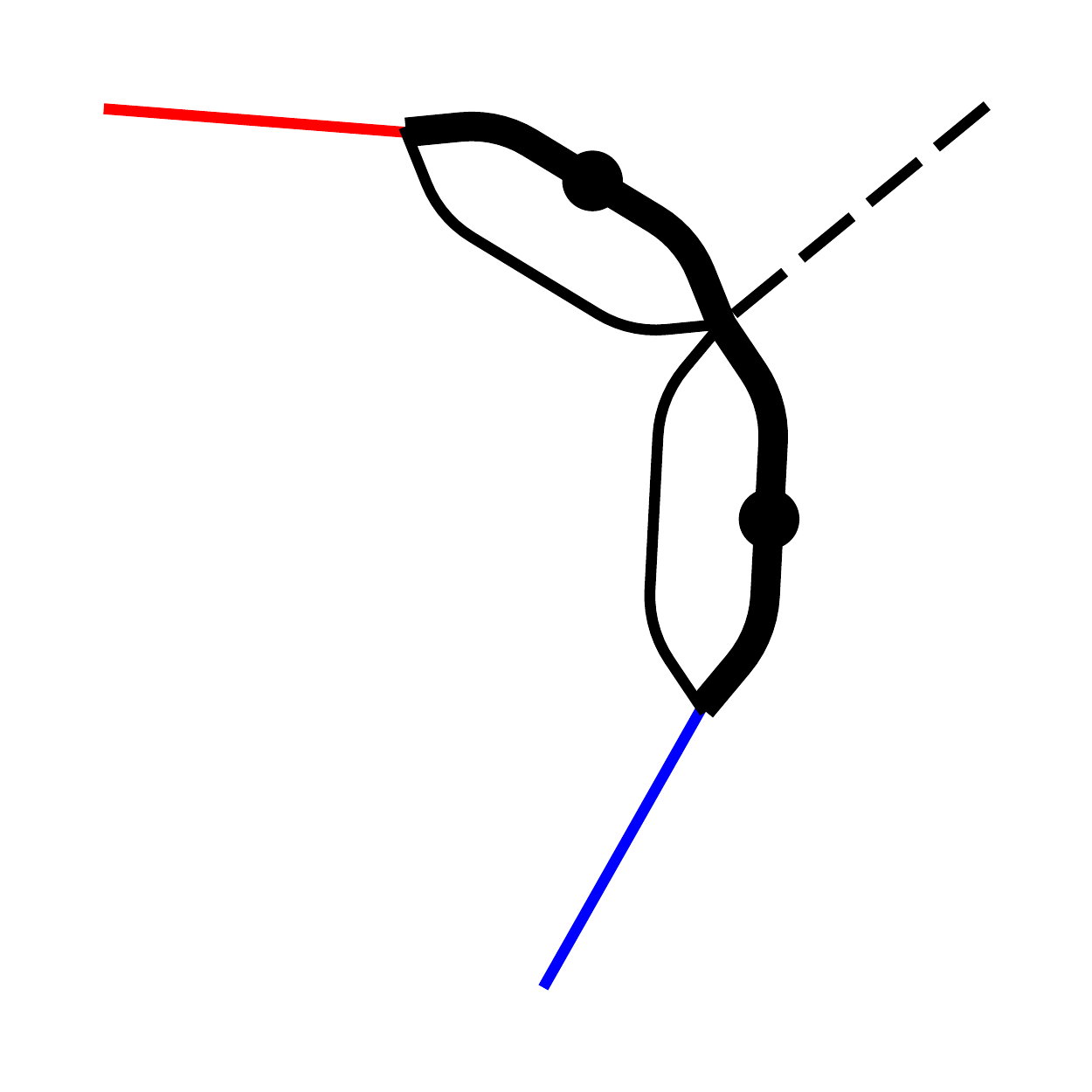}
  }
  \subfloat[$\mathcal{T}_{14}$]{%
    \includegraphics[width=0.14\textwidth]{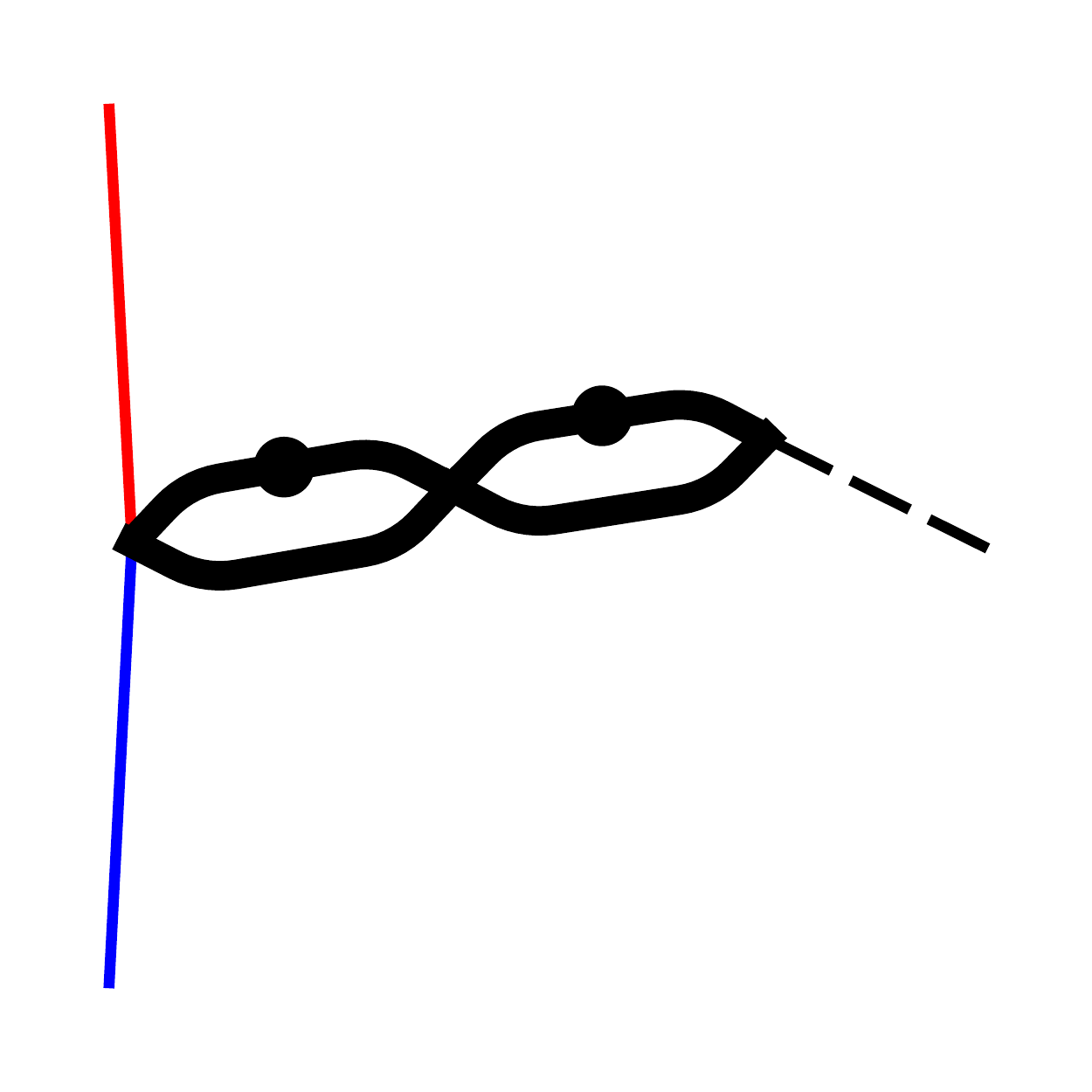}
  }
  \subfloat[$\mathcal{T}_{15}$]{%
    \includegraphics[width=0.14\textwidth]{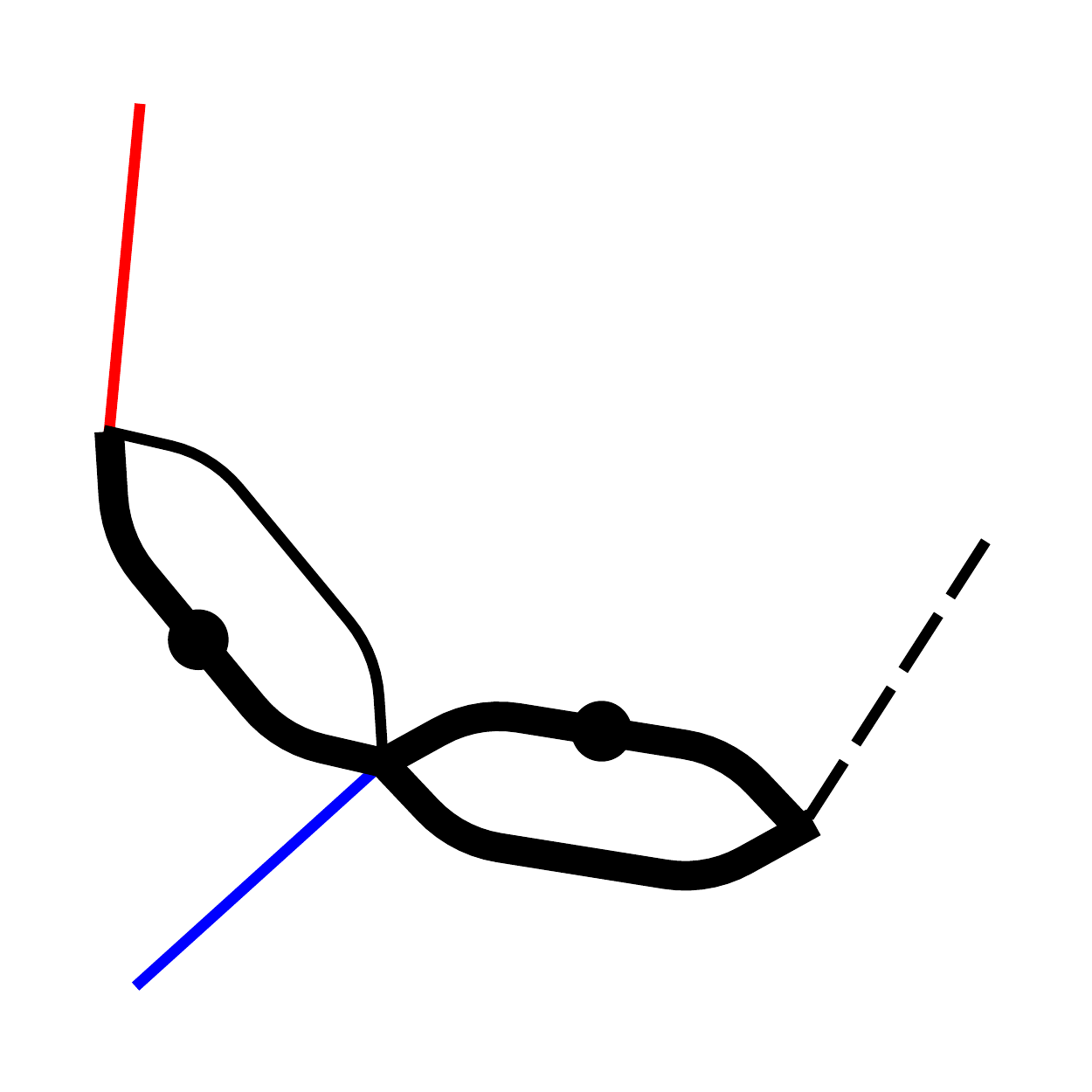}
  }
  \subfloat[$\mathcal{T}_{16}$]{%
    \includegraphics[width=0.14\textwidth]{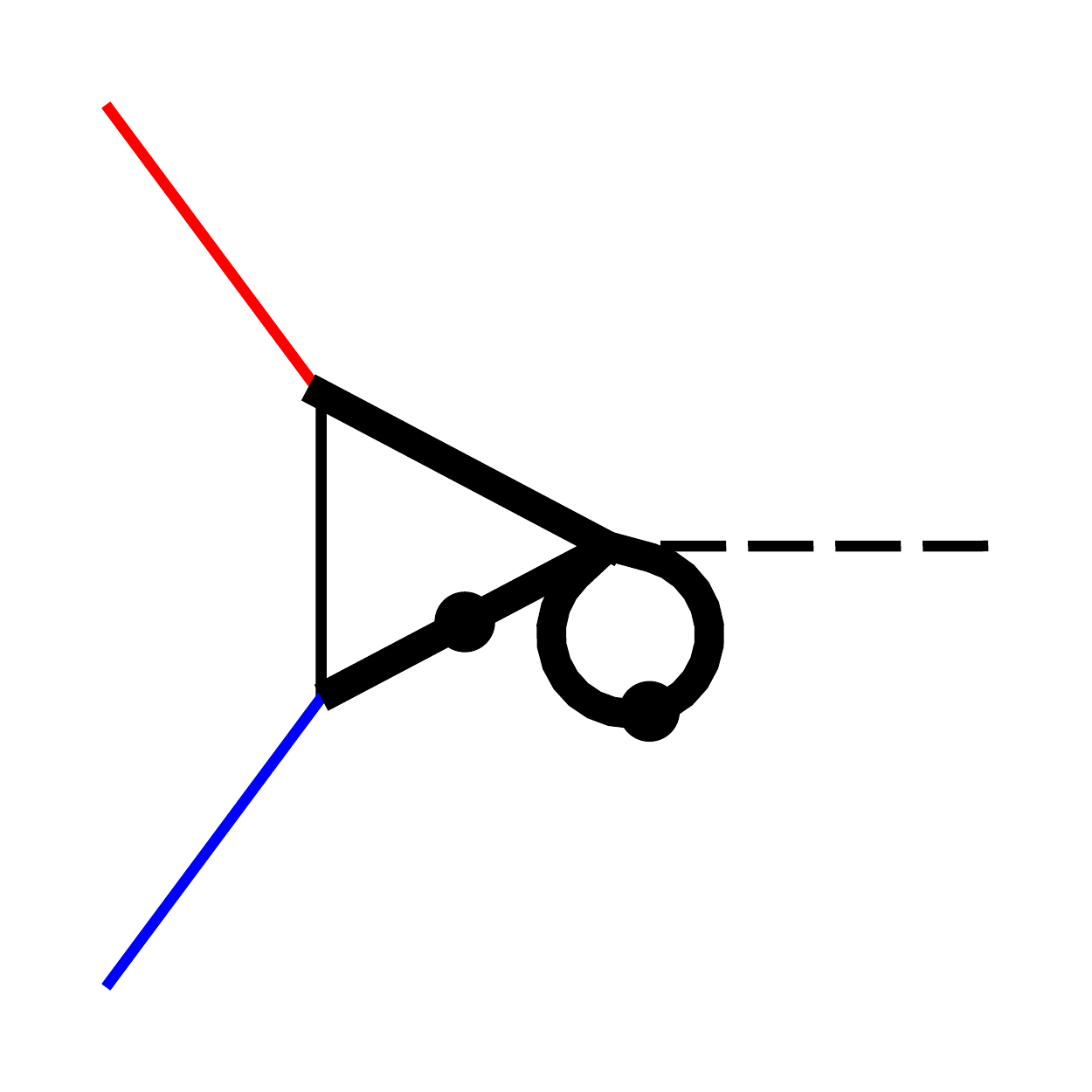}
  }
  \subfloat[$\mathcal{T}_{17}$]{%
    \includegraphics[width=0.14\textwidth]{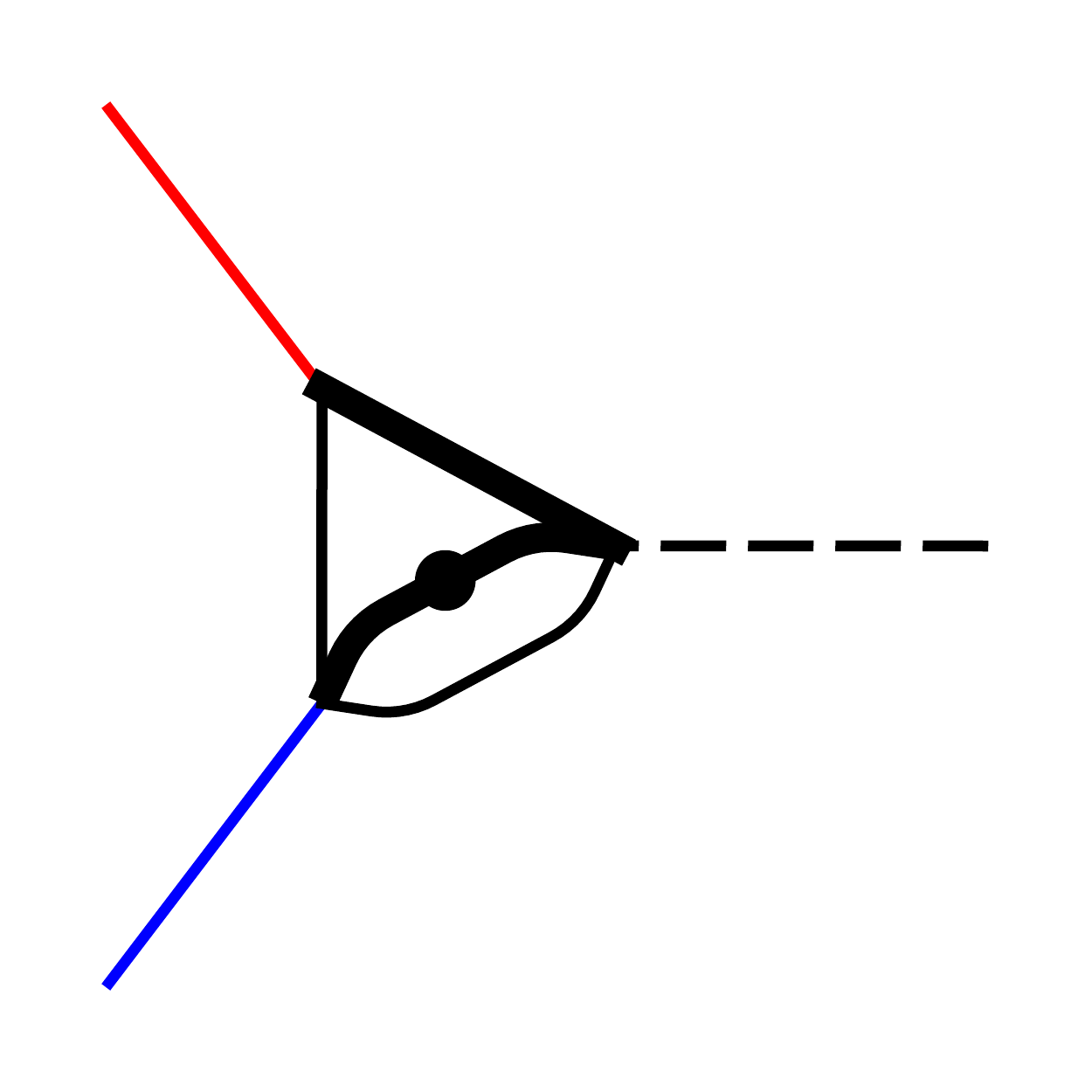}
  }
  \subfloat[$\mathcal{T}_{18}$]{%
    \includegraphics[width=0.14\textwidth]{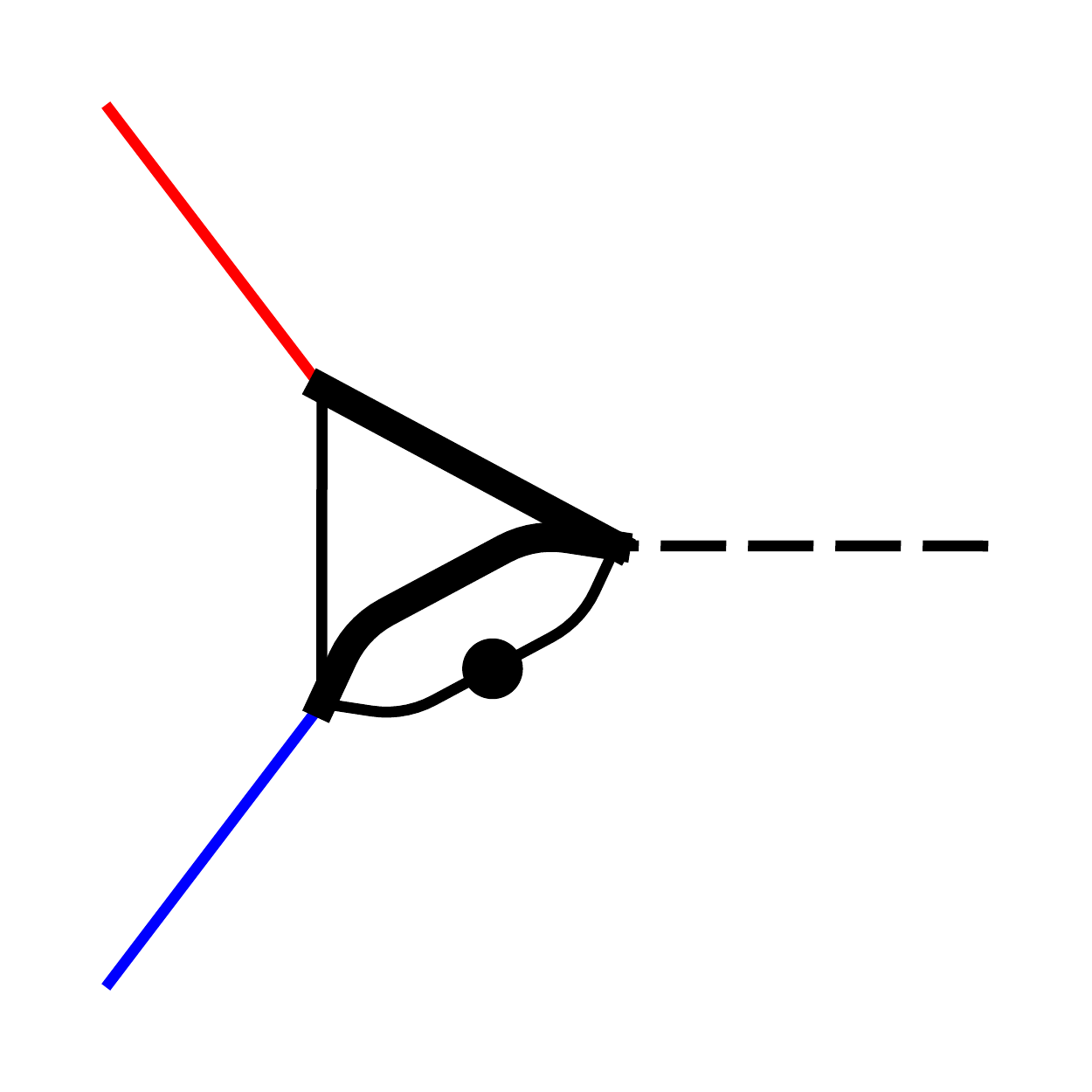}
  }
  \\
  \subfloat[$\mathcal{T}_{19}$]{%
    \includegraphics[width=0.14\textwidth]{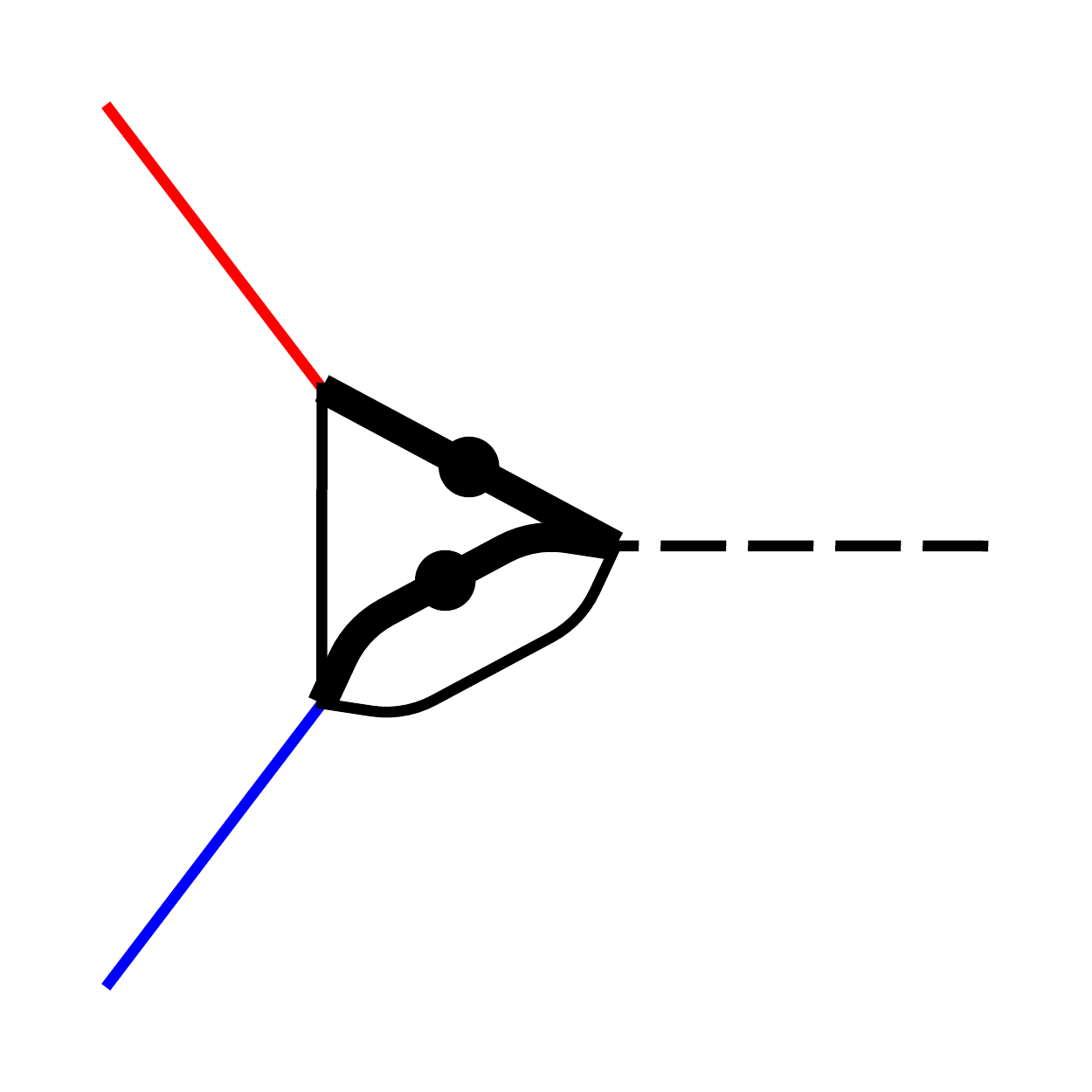}
  }
  \subfloat[$\mathcal{T}_{20}$]{%
    \includegraphics[width=0.14\textwidth]{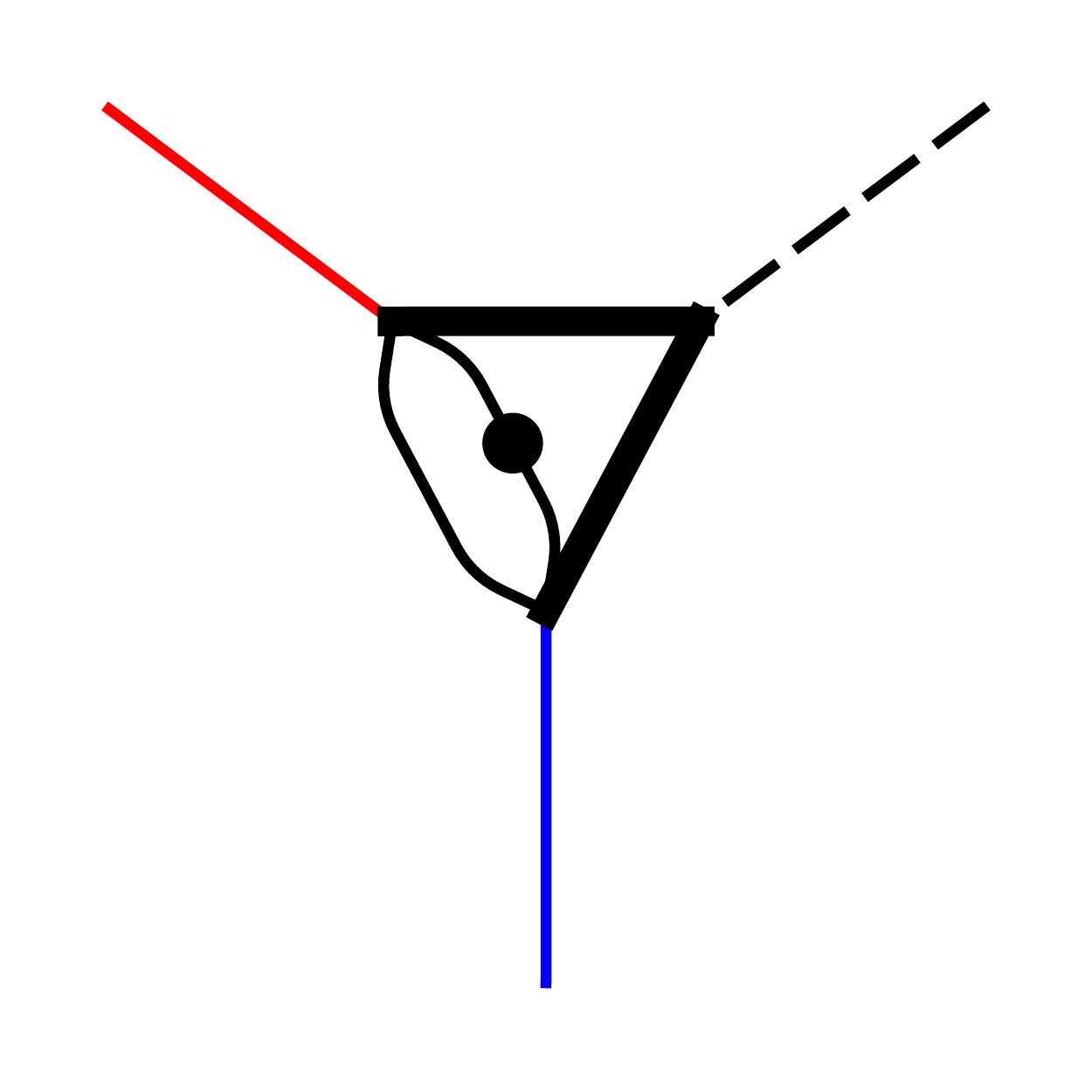}
  }
  \subfloat[$\mathcal{T}_{21}$]{%
    \includegraphics[width=0.14\textwidth]{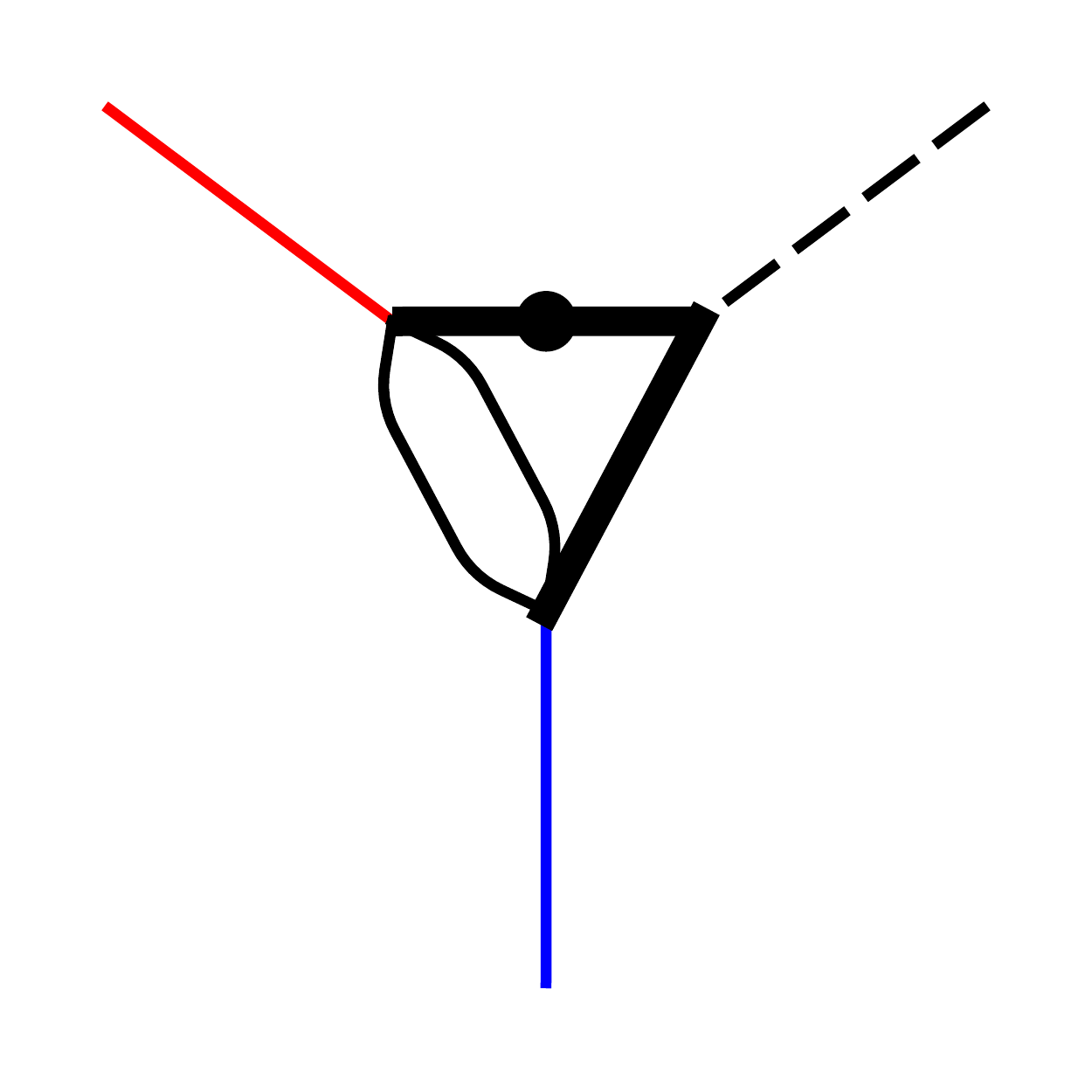}
  }
  \subfloat[$\mathcal{T}_{22}$]{%
    \includegraphics[width=0.14\textwidth]{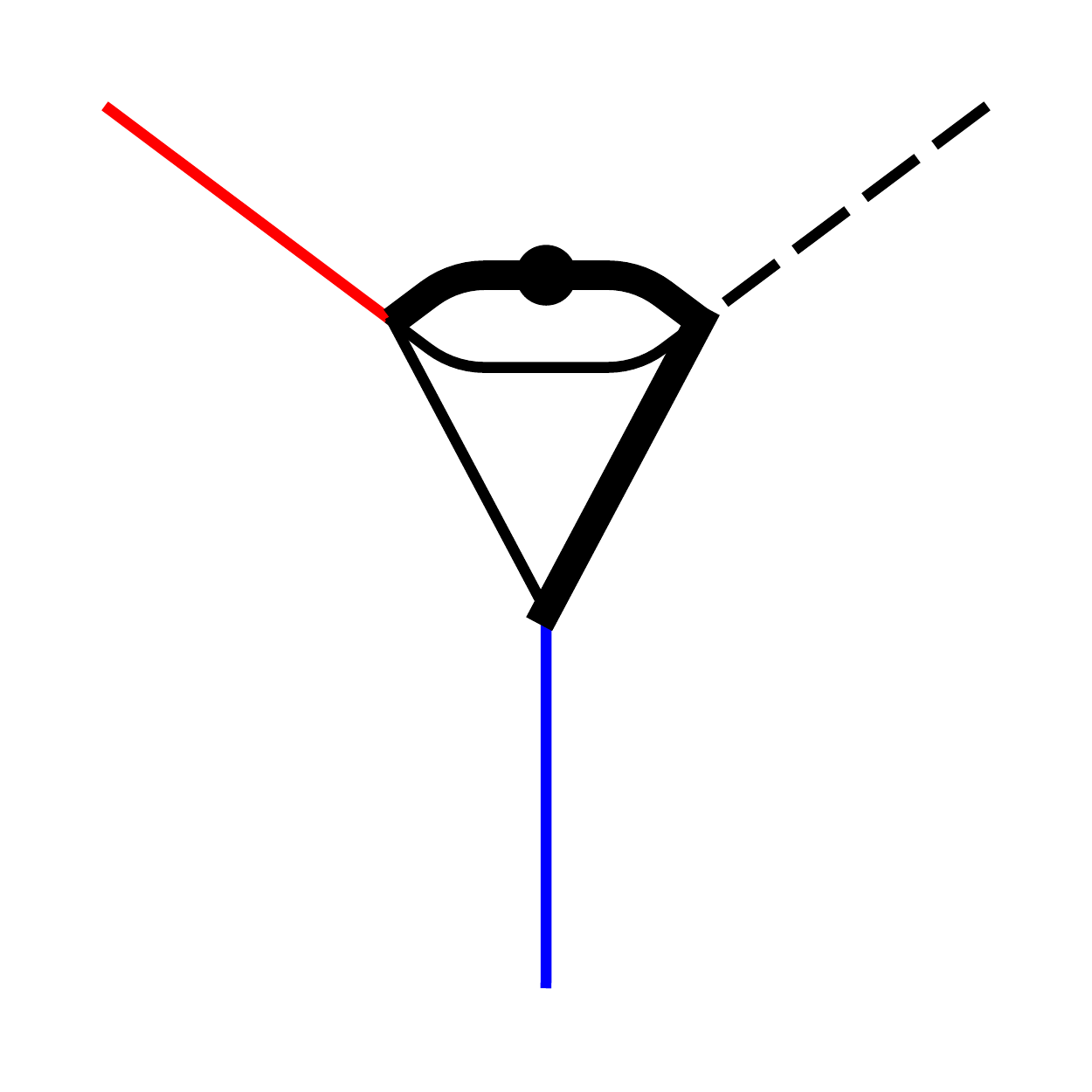}
  }
  \subfloat[$\mathcal{T}_{23}$]{%
    \includegraphics[width=0.14\textwidth]{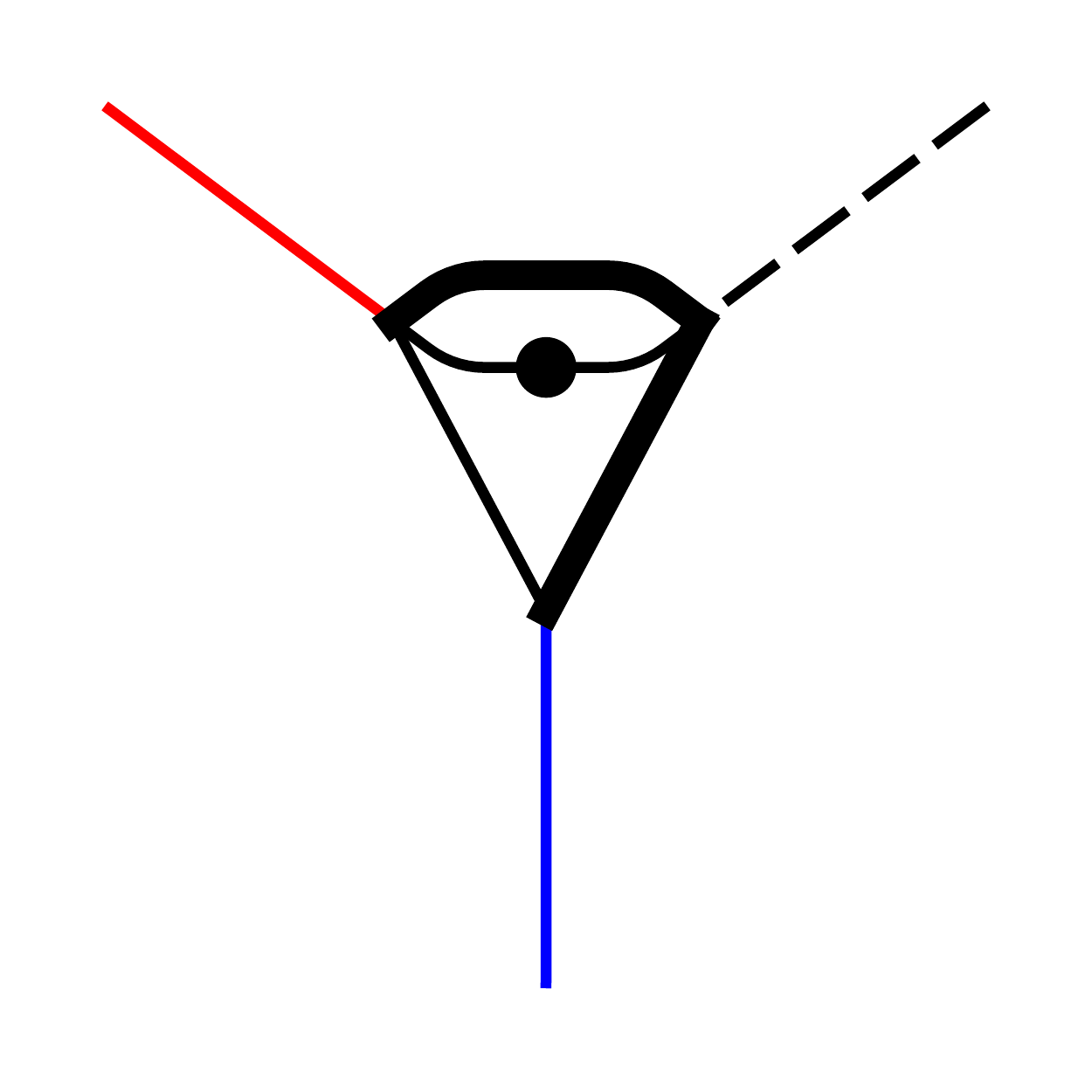}
  }
  \subfloat[$\mathcal{T}_{24}$]{%
    \includegraphics[width=0.14\textwidth]{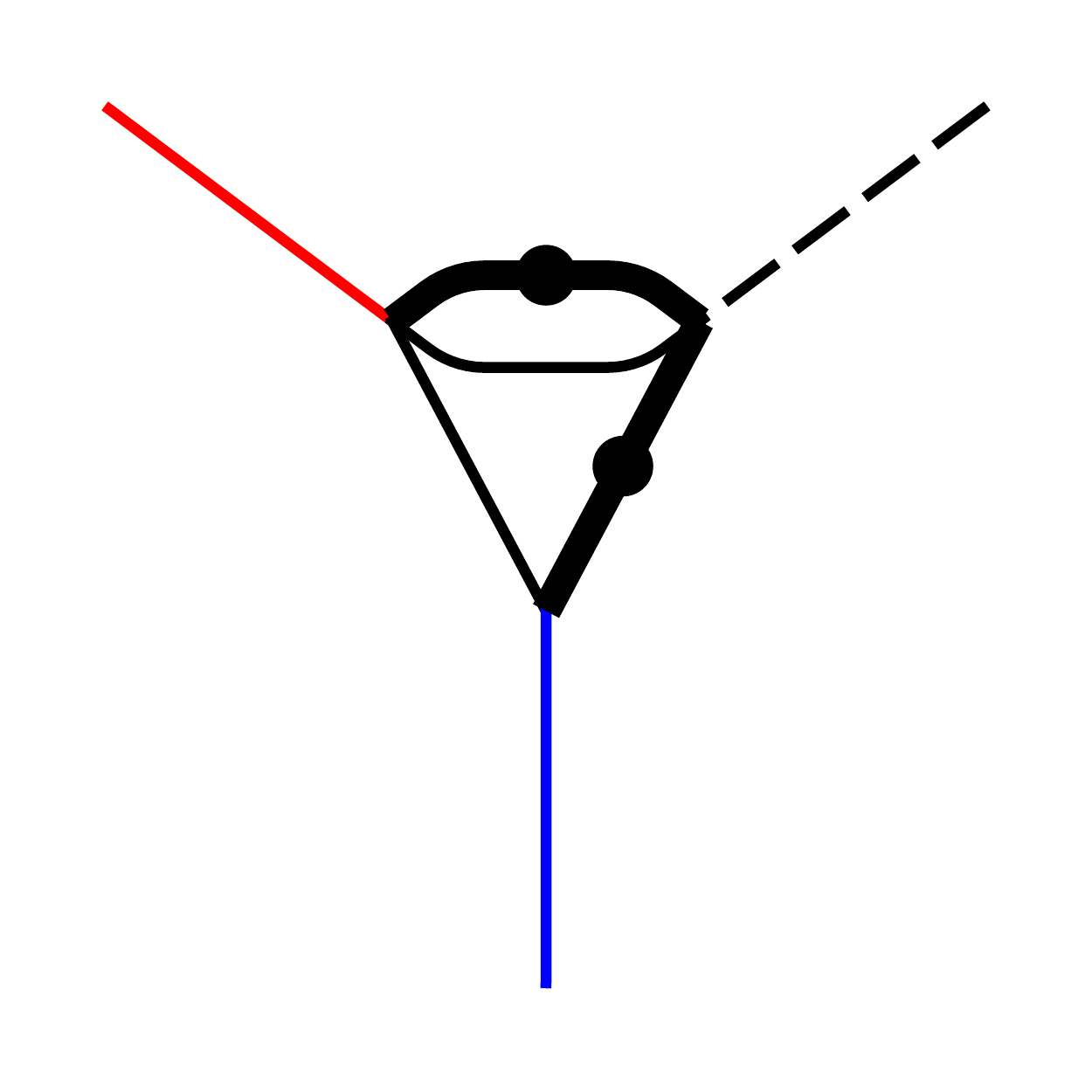}
  }
  \\
  \subfloat[$\mathcal{T}_{25}$]{%
    \includegraphics[width=0.14\textwidth]{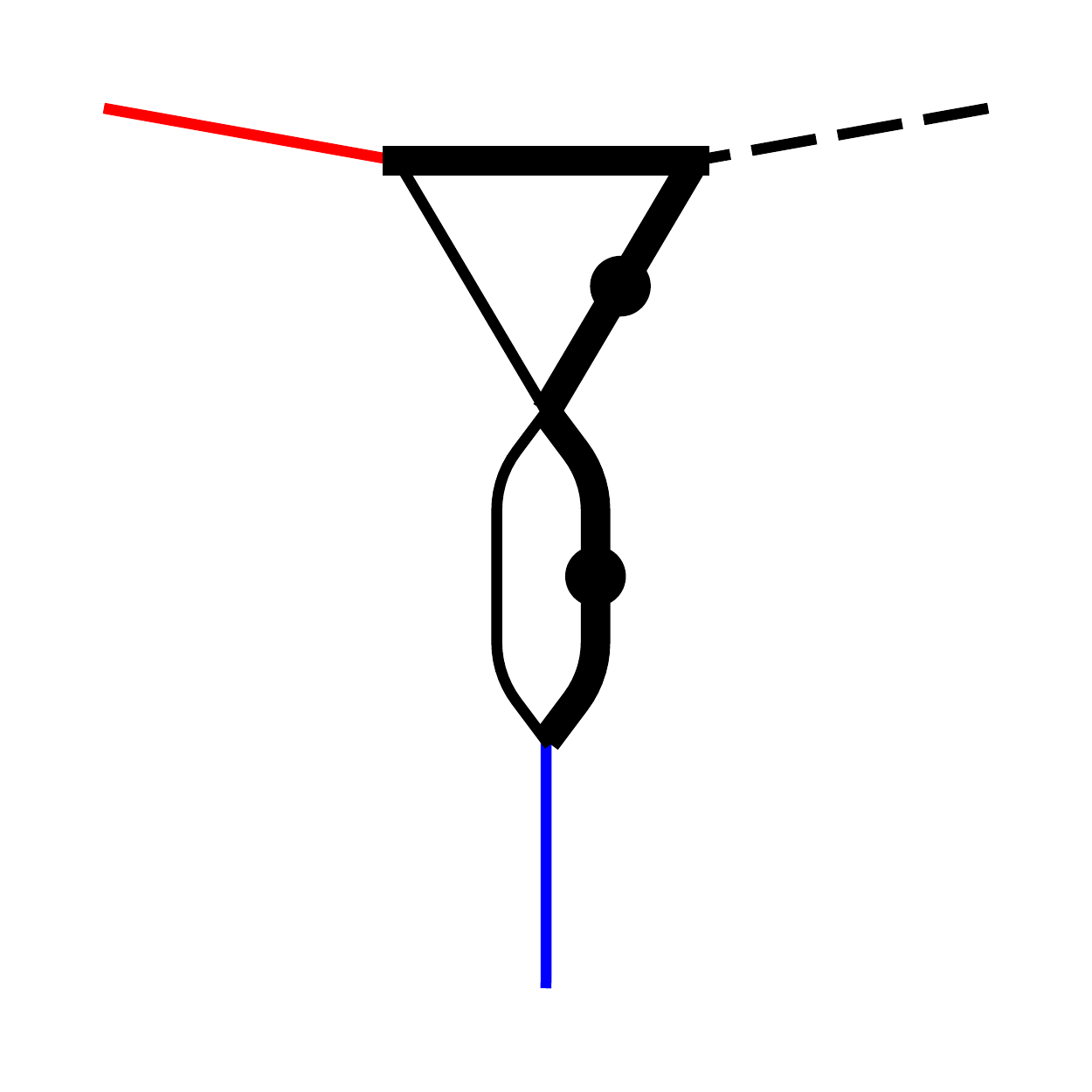}
  }
  \subfloat[$\mathcal{T}_{26}$]{%
    \includegraphics[width=0.14\textwidth]{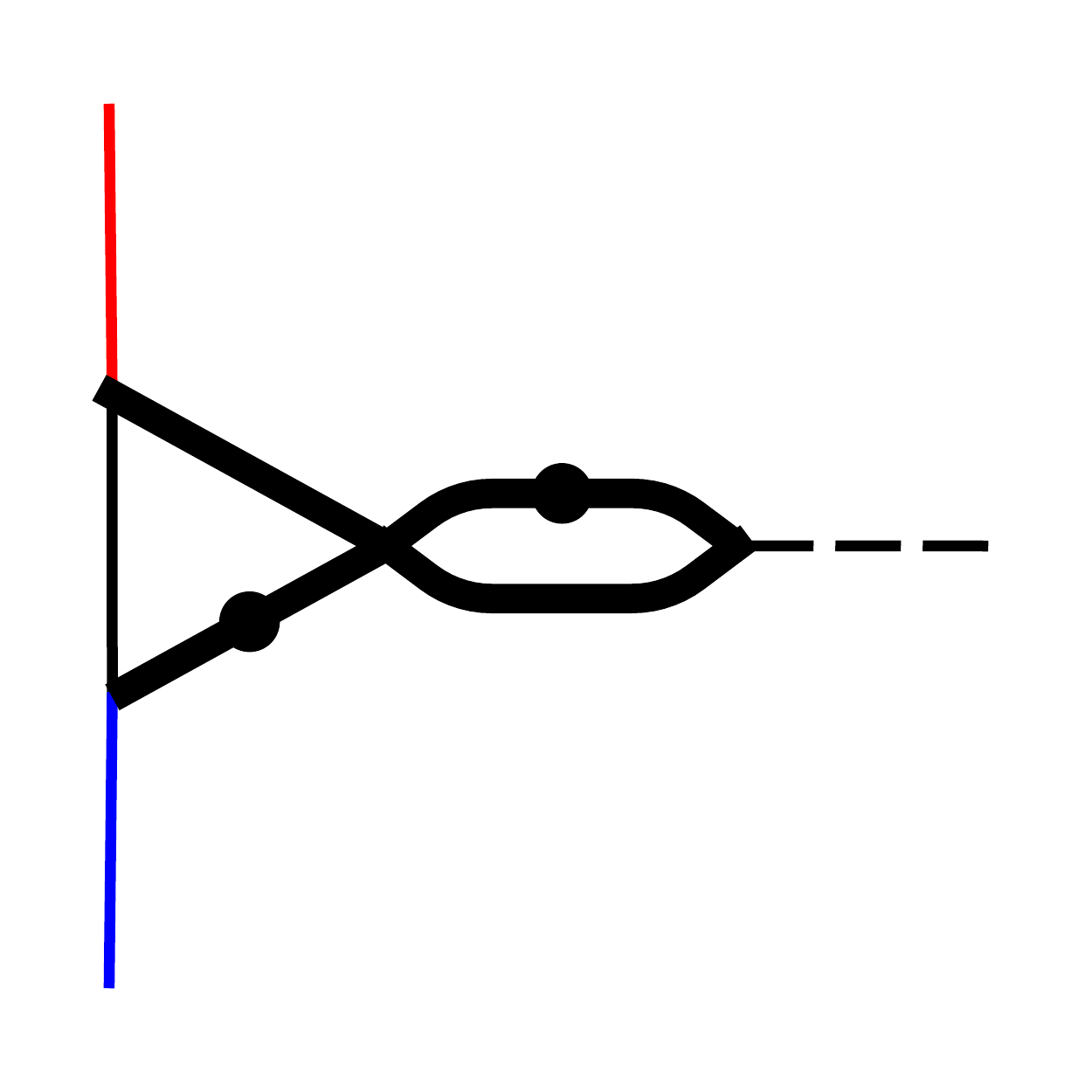}
  }
  \subfloat[$\mathcal{T}_{27}$]{%
    \includegraphics[width=0.14\textwidth]{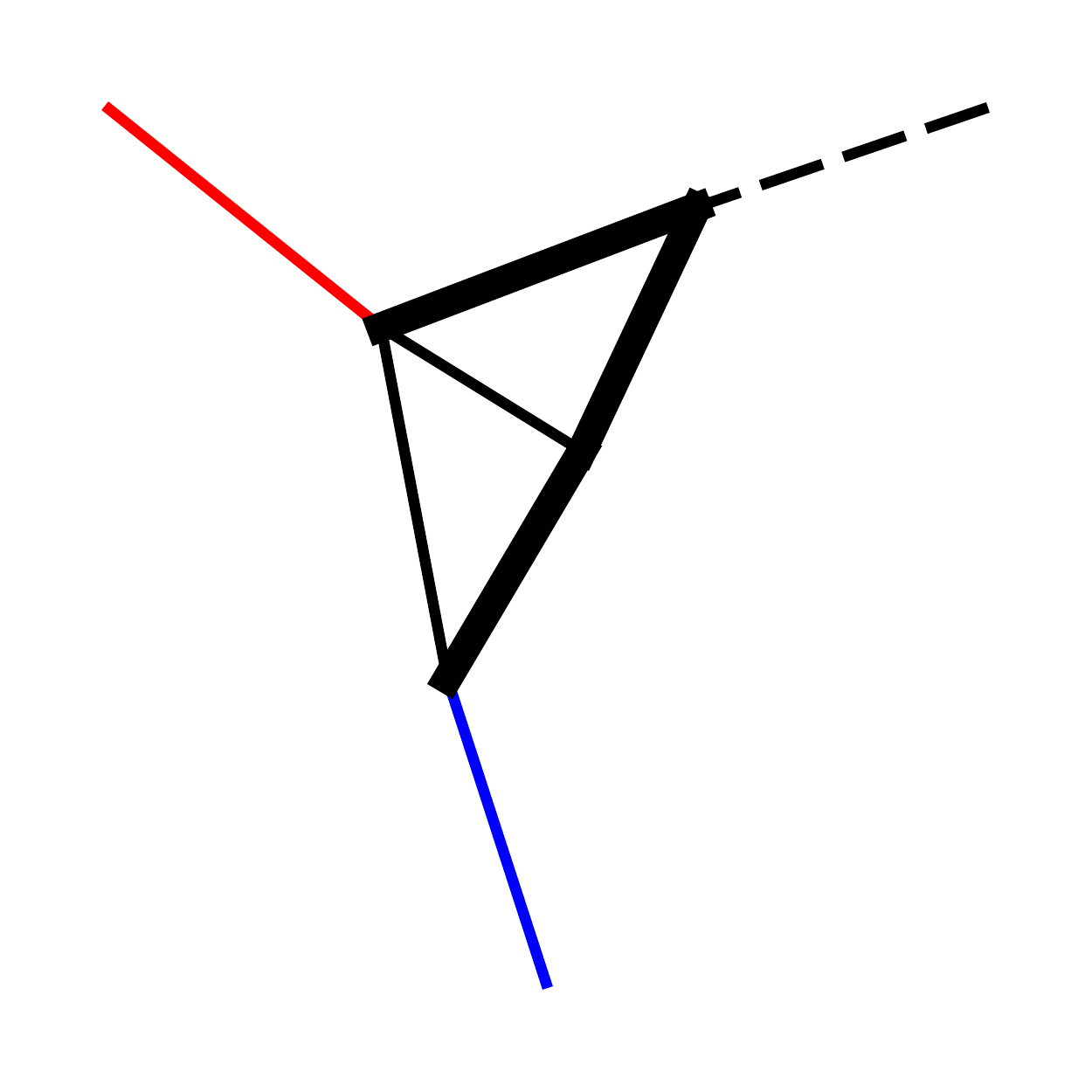}
  }
  \subfloat[$\mathcal{T}_{28}$]{%
    \includegraphics[width=0.14\textwidth]{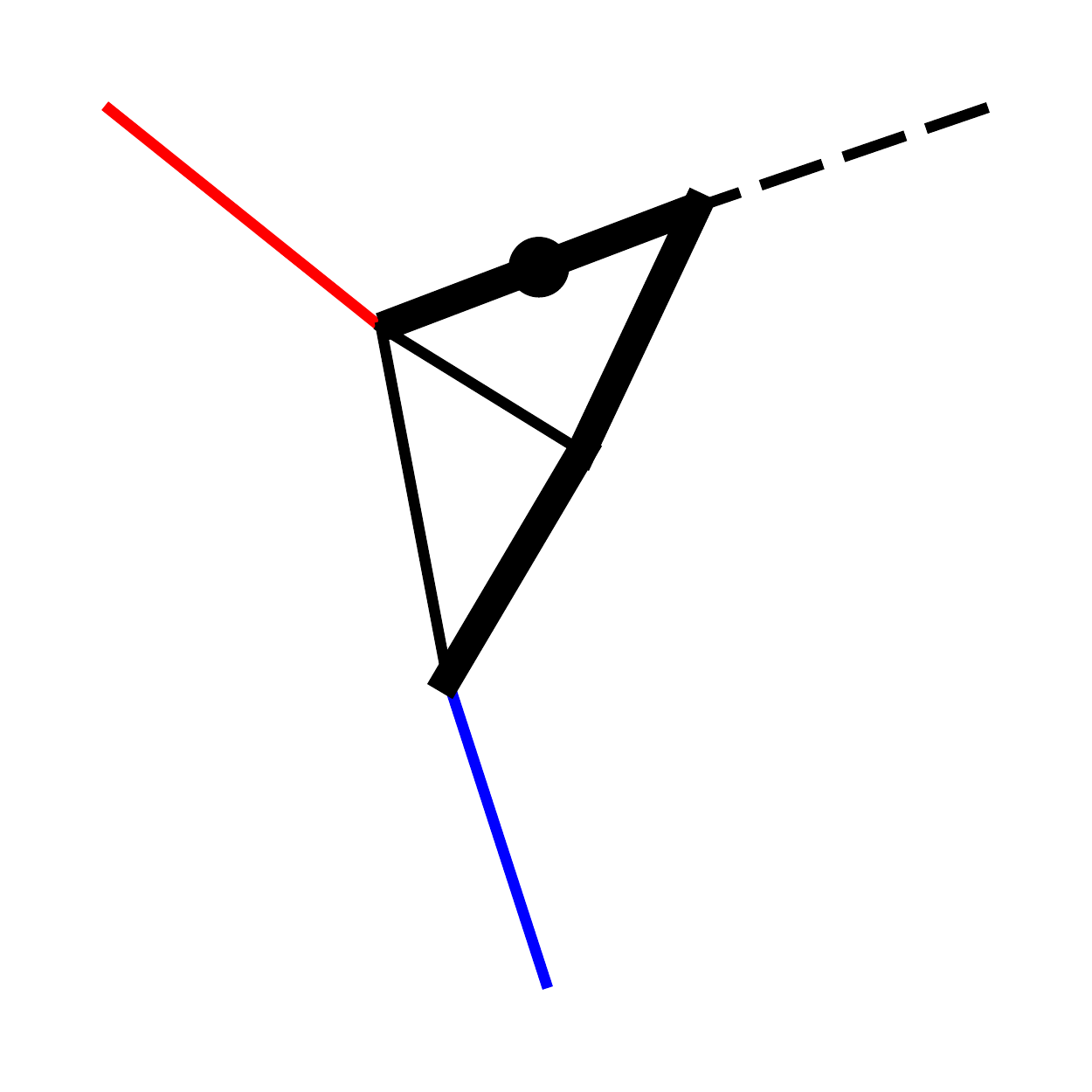}
  }
  \subfloat[$\mathcal{T}_{29}$]{%
    \includegraphics[width=0.14\textwidth]{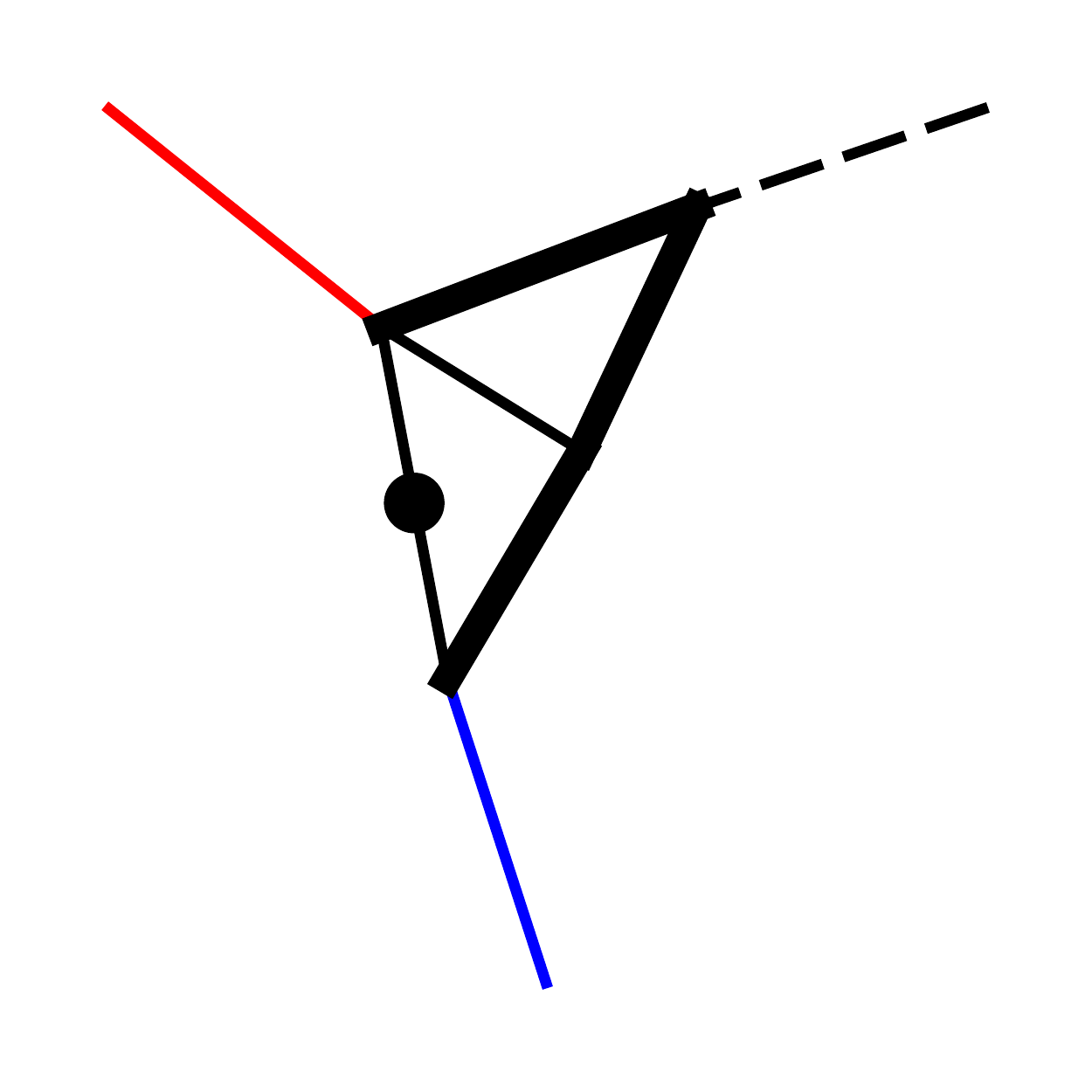}
  }
  \subfloat[$\mathcal{T}_{30}$]{%
    \includegraphics[width=0.14\textwidth]{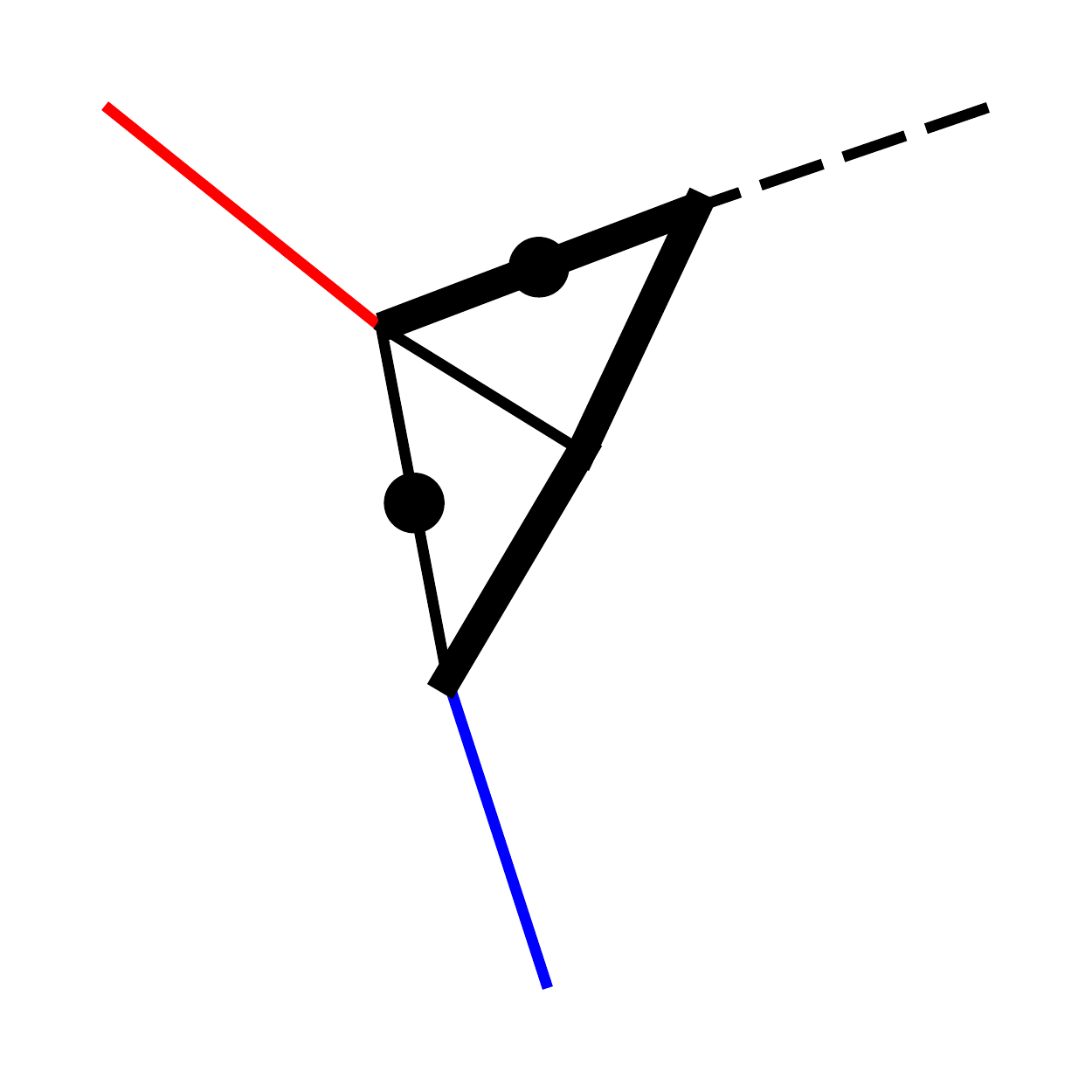}
  }
  \\
  \subfloat[$\mathcal{T}_{31}$]{%
    \includegraphics[width=0.14\textwidth]{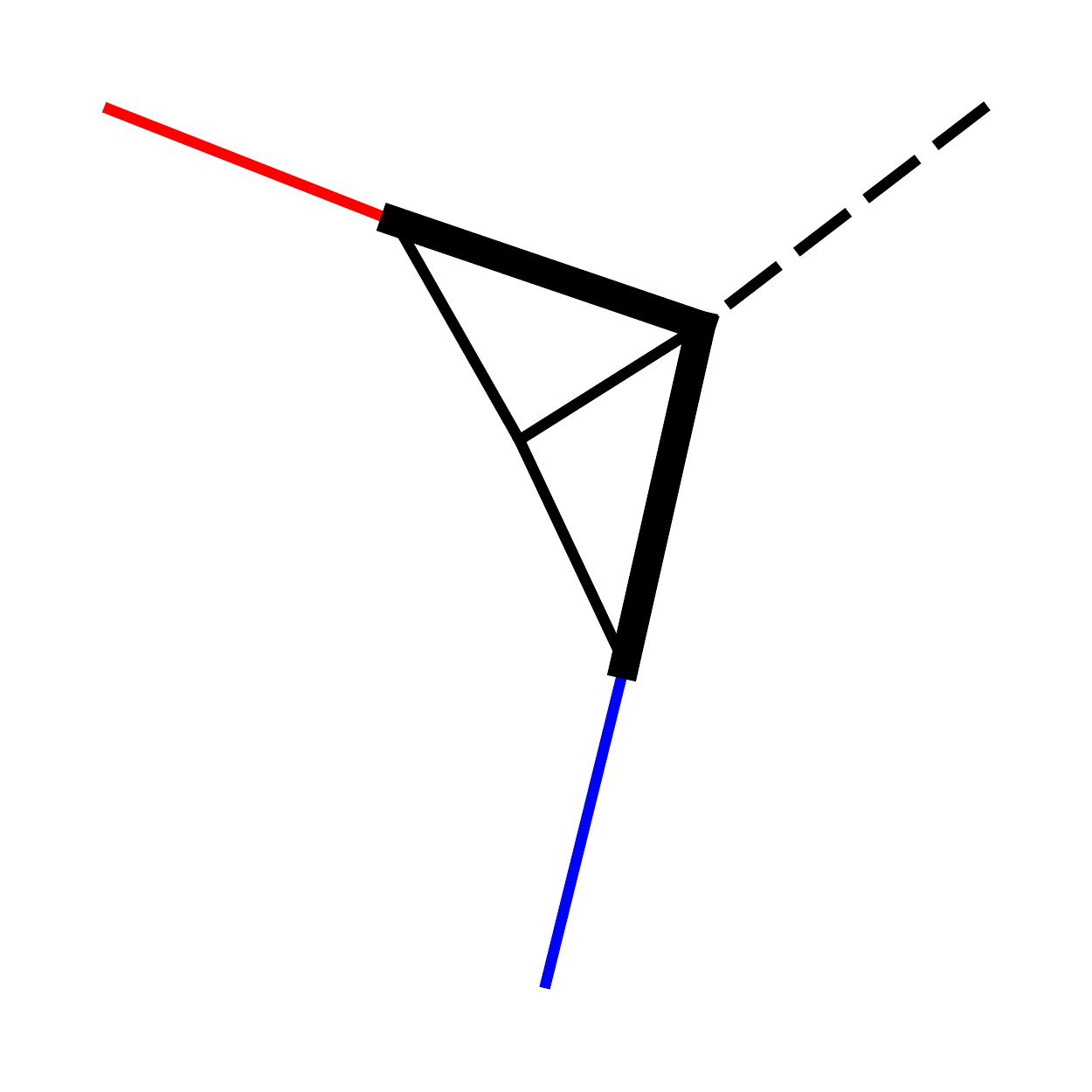}
  }
  \caption{Two-loop MIs $\mathcal{T}_{1,\ldots,31}$ for the topologies
    \protect\subref{fig:FigTop1WW}-\protect\subref{fig:FigTop2WW}. Graphical
    conventions are the same as in figure~\ref{fig:TopoWW}. Dots
    indicate squared propagators.}
  \label{fig:WWMIstt}
\end{figure}
The following set of MIs obeys a system of DEs which is linear in
$\eps$ :
\begin{align*}
\FF_{1}&=\eps^2 \, \top{1}\,,  &
\FF_{2}&=\eps^2 \, \top{2}\,,  &
\FF_{3}&=\eps^2 \, \top{3}\,,  \\
\FF_{4}&=\eps^2 \, \top{4}\,,  &
\FF_{5}&=\eps^2 \, \top{5}\,,  &
\FF_{6}&=\eps^2 \, \top{6}\,,  \\
\FF_{7}&=\eps^2 \, \top{7}\,,  &
\FF_{8}&=\eps^2 \, \top{8}\,,  &
\FF_{9}&=\eps^2 \, \top{9}\,,  \\
\FF_{10}&=\eps^2 \, \top{10}\,,  &
\FF_{11}&=\eps^2 \, \top{11}\,,  &
\FF_{12}&=\eps^2 \, \top{12}\,,  \\
\FF_{13}&=\eps^2 \, \top{13}\,,  &
\FF_{14}&=\eps^2 \, \top{14}\,,  &
\FF_{15}&=\eps^2 \, \top{15}\,,  \\
\FF_{16}&=\eps^2\, \top{16}\,,  &
\FF_{17}&=\eps^3\, \top{17}\,,  &
\FF_{18}&=\eps^3 \, \top{18}\,,  \\
\FF_{19}&=\eps^2 \, \top{19}\,,  &
\FF_{20}&=\eps^3 \, \top{20}\,,  &
\FF_{21}&=-\eps^2(1-2\eps) \, \top{21}\,,  \\
\FF_{22}&=\eps^3 \, \top{22}\,,  &
\FF_{23}&=\eps^3\, \top{34}\,,  &
\FF_{24}&=\eps^2 \, \top{24}\,, \\
\FF_{25}&=\eps^2 \, \top{25}\,,  &
\FF_{26}&=\eps^2\, \top{26}\,,  &
\FF_{27}&=\eps^4 \, \top{27}\,, \\
\FF_{28}&=\eps^3 \, \top{28}\,,  &
\FF_{29}&=\eps^3\, \top{29}\,,  &
\FF_{30}&=\eps^2 \, \top{30}\,, \\
\FF_{31}&=\eps^4 \, \top{31}\,,\stepcounter{equation}\tag{\theequation}
\label{def:1M1LBasisMIs}
\end{align*}
where the $\mathcal{T}_i$ are depicted in figure~\ref{fig:WWMIstt}. As
for the case of topologies
\protect\subref{fig:FigTop1WWbb}-\protect\subref{fig:FigTop2WWbb},
some of integrals $\mathcal{T}_i$ are related by
$p_1^2 \leftrightarrow p_2^2$,
\begin{align}
 \top{4} \leftrightarrow  \top{2}\,,\quad  \top{9} \leftrightarrow  \top{5}\,, \quad \top{10} \leftrightarrow  \top{6}\,,\quad  \top{15} \leftrightarrow  \top{12}\,, \quad  \top{22} \leftrightarrow  \top{17}\,, \quad  \top{23} \leftrightarrow  \top{18}\,, \quad  \top{24} \leftrightarrow  \top{19}\,,
\end{align}
so that the total number of independent integrals is 24. However, as
discussed already after eq.~\eqref{eq:Tabsymm}, we work with the
complete set of integrals given in eq.~\eqref{def:1M1LBasisMIs}.

The Magnus exponential allows us to obtain a set of canonical
MIs obeying a system of equations of the form
\eqref{eq:canonicalDE}, \raggedbottom
\begin{gather}
  \begin{alignedat}{2}
    \GG_{1}&=   \FF_1\,, & \qquad
    \GG_{2}&= -p_2^2  \,  \FF_2\,,\nn
    \GG_{3}&= \rho\, \FF_3\,,  & \qquad
    \GG_{4}&= -p_1^2 \FF_4 \,,\nn
    \GG_{5}&= (m^2-p_2^2) \,  \FF_5+2m^2\,\FF_6\,,  & \qquad
    \GG_{6}&= -p_2^2 \, \FF_6\, \nn
    \GG_{7}&= \rho \, \FF_7+\frac{1}{2}(\rho-s)\, \FF_8\,, & \qquad
    \GG_{8}&= -s\FF_8\,,\nn
    \GG_{9}&=-p_1^2\,\FF_9\,, & \qquad
    \GG_{10}&= 2m^2\,\FF_{9}+(m^2-p_1^2)\,\FF_{10}\,,\nn
    \GG_{11}&= p_2^4\, \FF_{11}\,,  & \qquad  
    \GG_{12}&= -p_2^2  \rho\, \FF_{12}\,, \nn
    \GG_{13}&= p_1^2p_2^2 \, \FF_{13}\, & \qquad  
    \GG_{14}&=\rho^2 \, \FF_{14}\,, \nn
    \GG_{15}&= -p_1^2 \rho\, \FF_{15}\,,
  \end{alignedat}
  \nn
  \begin{alignedat}{1}
    \GG_{16}&=c_{2,\,16}\, \FF_{2}+c_{3,\,16}\, \FF_{3}+c_{4,\,16}\, \FF_{4}+c_{16,\,16}\, \FF_{16}\,,          \end{alignedat}
  \nn
  \begin{alignedat}{2}
    \GG_{17}&= -\sqrt{\lambda} \, \FF_{17}\,,&\qquad
    \GG_{18}&= -\sqrt{\lambda} \, \FF_{18}\,, \nn
    \GG_{19}&=c_{17,\,19}\, \FF_{17}+c_{18,\,19}\, \FF_{18}+c_{19,\,19}\, \FF_{19}\,, &\qquad
    \GG_{20}&= -\sqrt{\lambda} \, \FF_{20}\,, \nn
  \end{alignedat}
  \nn
  \begin{alignedat}{1}
    \GG_{21}&=c_{5,\,21}\, \FF_{5}+c_{6,\,21}\, \FF_{6}+c_{9,\,21}\, \FF_{9}+c_{10,\,21}\, \FF_{10}+c_{20,\,21}\, \FF_{20}+c_{21,\,21}\, \FF_{21}\,, &\qquad
  \end{alignedat}
  \nn
  \begin{alignedat}{2}
    \GG_{22}&=  -\sqrt{\lambda} \, \FF_{22}\,,& \qquad
    \GG_{23}&= -\sqrt{\lambda} \, \FF_{23}\,, \nn
  \end{alignedat}
  \nn
  \begin{alignedat}{1}
    \GG_{24}&= c_{22,\,24}\, \FF_{22}+c_{23,\,24}\, \FF_{23}+c_{24,\,24}\, \FF_{24}\,,\nn
    \GG_{25}&= c_{11,\,25}\, \FF_{11}+c_{12,\,25}\, \FF_{12}+c_{13,\,25}\, \FF_{13}+c_{25,\,25}\, \FF_{25}\ \,, \nn
    \GG_{26}&= c_{12,\,26}\, \FF_{12}+c_{14,\,26}\, \FF_{14}+c_{15,\,26}\, \FF_{15}+c_{26,\,26}\, \FF_{26}\,, \nn
  \end{alignedat}
  \nn
  \begin{alignedat}{2}
    \GG_{27}&= -\sqrt{\lambda} \, \FF_{27}\,, &\qquad
    \GG_{28}&= -\rho\sqrt{\lambda} \, \FF_{28}\,, \nn
    \GG_{29}&= (p_2^2-m^2)\sqrt{\lambda}\, \FF_{29}\,,& \qquad
  \end{alignedat}
  \nn
  \begin{alignedat}{1}
    \GG_{30}&= c_{3,\,30}\, \FF_{3}+c_{12,\,30}\, \FF_{12}+c_{28,\,30}\, \FF_{28}+c_{29,\,30}\, \FF_{29}+c_{30,\,30}\, \FF_{30}\,, \nn
  \end{alignedat}
  \nn
  \begin{alignedat}{2}
    \GG_{31}&= -\sqrt{\lambda} \, \FF_{31}\,, 
    \label{def:WWCanonicalMIs}
  \end{alignedat}
\end{gather}
where $\rho\equiv\sqrt{-s}\sqrt{4m^2-s}$ and $\lambda$ is defined as in
eq.~\eqref{eq:lambda}. The expression of the coefficients $c_{i,\,j}$
is given in appendix~\ref{sec:gMIcoefft}.
The alphabet of the corresponding $\dlog$-form contains the following
18 letters
\begin{align}
  \eta_1 & =v\,,& \eta_2 & =1-v\,, \nn
  \eta_3 & =1+v\,, & \eta_4 & =\zz\,, \nn
  \eta_5 & =1-\zz\,,& \eta_6 &  =\zzb\,,\nn
  \eta_7 & =1-\zzb\,,& \eta_8 & =\zz-\zzb\,,\nn
  \eta_9 & =\zz+v(1-\zz)\,,& \eta_{10} & =1-\zz(1-v)\,,\nn
  \eta_{11} & =\zzb+v(1-\zzb)\,,& \eta_{12} & =1-\zzb(1-v)\,,\nn
  \eta_{13} &  =v+\zz\zzb(1-v)^2\,,& \eta_{14} &  =v+(1-\zz-\zzb+\zz\zzb)(1-v)^2\,,\nn
  \eta_{15} & =v+\zz(1-v)^2\,,&   \eta_{16} & =v+(1-\zz)(1-v)^2\,,\nn
  \eta_{17} & =v+\zzb(1-v)^2\,,& \eta_{18} & =v+(1-\zzb)(1-v)^2\,.
  \label{alphabet:MWW}
\end{align}
The coefficient matrices $\MM_i$ are collected in the
appendix~\ref{dlogWW2Lt}.  In this case, all the letters are real and
positive in the region
\begin{equation}
  \label{eq:positivitycdvzzb}
  0<v<1\,, \quad 0<z<1\,, \quad 0<\zzb<\zz\,.
\end{equation}
If one fixes $m^2>0$, this corresponds to a patch of the Euclidean
region, $s\,,p_1^2\,, p_2^2 < 0$, defined by the following
constraint
\begin{equation}
  \label{eq:positivitycdsp1p2}
  s<-\left(\sqrt{-p_1^2}+\sqrt{-p_2^2}\right)^2<0\,.
\end{equation}

The solution of the system of DEs is straightforwardly obtained in
terms of Chen's iterated integrals. Moreover, since the alphabet is
rational, the solution can be converted in terms of GPLs of argument
1, with kinematic-dependent weights, as we discuss in
appendix~\ref{sec:chen}. The prescriptions for the analytic
continuation to the other patches of the Euclidean region and to the
physical regions are given in section~\ref{sec:continuation}.

The boundary constants can be fixed by demanding the regularity
of the basis \eqref{def:1M1LBasisMIs} for vanishing external momenta,
$s=p_1^2=p_2^2=0$. In particular, if we choose as a base-point for the
integration
\begin{align}
  \xxi=(1,1,1), 
\end{align}
then the prefactors appearing in the definitions
\eqref{def:WWCanonicalMIs} of the canonical MIs $\mathbf{I}$ vanish,
with the only exceptions of $\GG_{1,5,10,19,21,24}$. Therefore the
boundaries of the former MIs are determined by demanding their
vanishing at $\vec{x}\to \xxi$,

\begin{align}
\GG_i(\eps,\vec{x}_0)=0, \qquad i\neq 1,5,10,19,21,24.
\end{align}
The boundary constants of integrals $\GG_{1, \, 5, \,10}$ can be taken
from the previous topologies in equations
(\ref{bound_i1_top},\ref{bound_i5_top}), whereas for integrals
$\GG_{19, \, 21, \,24}$ the boundary constants are fixed as follows:
\begin{itemize}
\item The boundary constants for $\GG_{19}$ and $\GG_{24}$ can be
  determined by imposing regularity at the pseudothresholds
  $ v \to 1 $ ($s=p_1^2=p_2^2=0$) and, respectively, $\zz\to 1$,
  $\zzb\to 1 $ (both corresponding to $p_2^2=0$),
  \begin{align}
    \GG_{19,\,24}(\eps,\vec{x}_0)=\frac{1}{6}\pi^2\eps^2-\zeta_3\eps^3+\frac{1}{20}\pi^4\eps^4+\mathcal{O}(\eps^5).
  \end{align}
\item Finally, the boundary constants for $\GG_{21}$ can be fixed by
  observing that, from \eqref{def:WWCanonicalMIs}, we can derive
  \begin{align}
    \FF_{21}(\eps,\,\xxi)=\lim_{\vec{x}\to\xxi}\frac{v}{m^2(1-v^2)}\GG_{21}(\eps,\,\xxi).
  \end{align}
  Therefore, in order for $\FF_{21}(\eps,\,\xxi)$ to be regular we
  must demand
  \begin{align}
    \GG_{21}(\eps,\vec{x}_0)=0,
  \end{align}
\end{itemize} 
All results have been numerically checked, in both the Euclidean and
the physical regions, with the help of the computer codes
\texttt{GiNaC} and \texttt{SecDec 3.0}, and the analytic expressions
of the MIs are given in electronic form in the ancillary files
attached to the \texttt{arXiv} version of the manuscript.


\section{Change of variables and analytic continuation}
\label{sec:continuation}

In this section we discuss in detail the variables used to parametrize
the dependence of the MIs on the kinematic invariants. In particular,
we elaborate on the prescriptions to analytically continue our results
to arbitrary values of $s,p_1^2,p_2^2$.
Both topologies
\protect\subref{fig:FigTop1WWbb}-\protect\subref{fig:FigTop2WWbb} and
\protect\subref{fig:FigTop1WW}-\protect\subref{fig:FigTop2WW} feature
two independent, kinematic structures:

\begin{enumerate}
\item the off-shell external legs are responsible for the presence in
  the DEs of the square root of the K\"all\'en function,
  $\sqrt{\lambda(s,p_1^2,p_2^2)}$;
\item the presence of massive internal lines can generate square roots
  in the DEs, as in the case of topologies
  \protect\subref{fig:FigTop1WW}-\protect\subref{fig:FigTop2WW} where
  one has also $\sqrt{-s}\sqrt{4m^2-s}$.
\end{enumerate}
In the following we separately discuss the variable changes that
rationalize the two types of square roots.

\subsection{Off-shell external legs: the $z,\zb$ variables}
\label{sec:zzb}
\begin{figure}
  \centering  \includegraphics[width=0.6\textwidth]{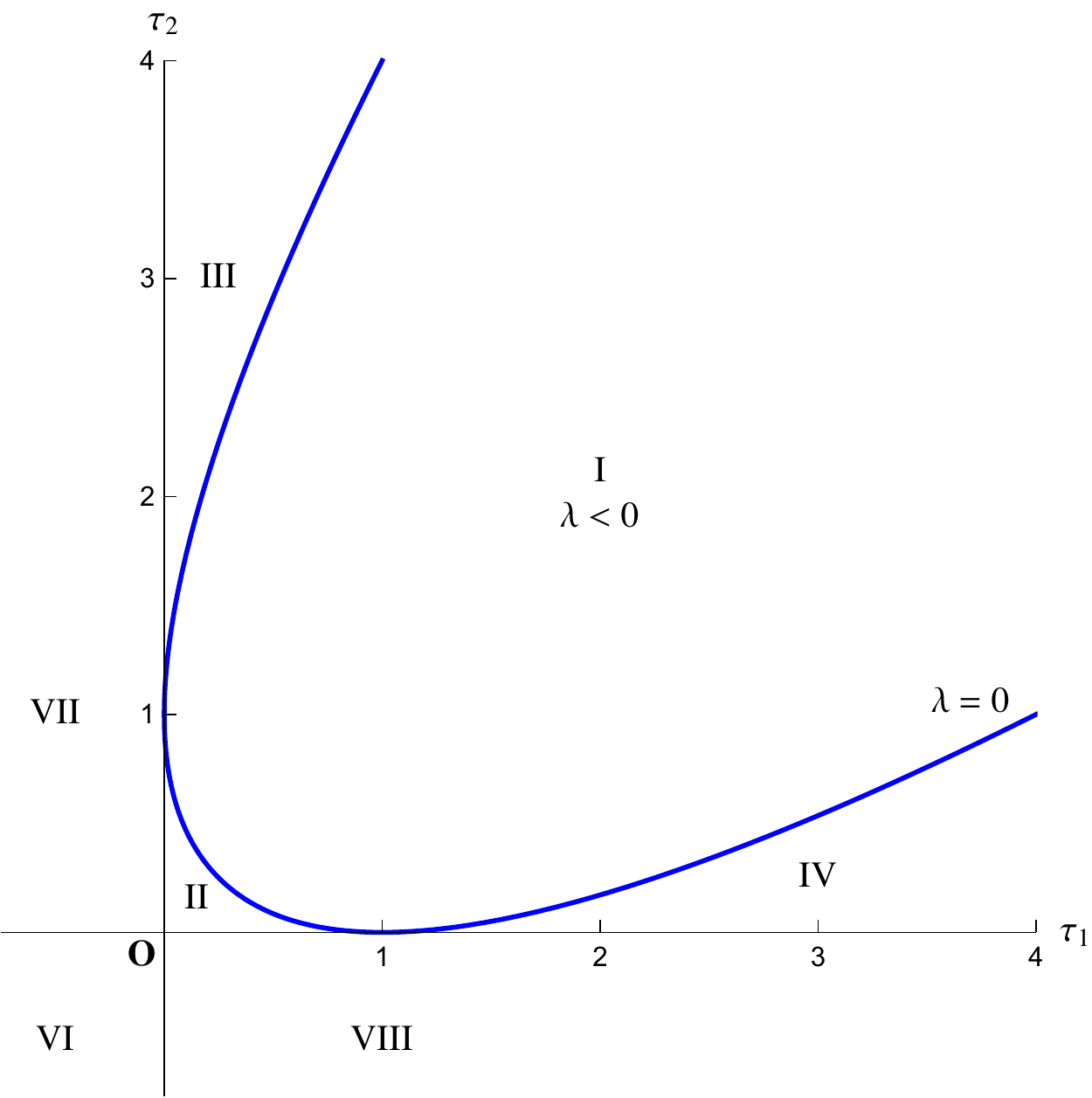}
  \caption{Regions of the $(\tau_1,\tau_2)$-plane classified in
    eq~\eqref{eq:zzbregions}. Region V, which is identified by the
    condition $\lambda(1,\tau_1,\tau_2)=0$, corresponds to the blue
    curve.}
  \label{fig:regions}
\end{figure}

To deal with the square root of the K\"all\'en function, we begin by
choosing one of the external legs as reference, $s$, and trading the
other squared momenta for dimensionless ratios
\begin{align}
  \tau_{1,2} = {} & \frac{p_{1,2}^2}{s}\,.
\end{align}
In the $(s,\tau_1,\tau_2)$ variables, the square root of the
K\"all\'en function is proportional to
\begin{align}
  \sqrt{\lambda(1,\tau_1,\tau_2)} = \sqrt{(1 - \tau_1 - \tau_2)^2 - 4 \tau_1 \tau_2}
\end{align}
and is rationalized by the following change of
variables~\cite{Chavez:2012kn}
\begin{align}
  \tau_1 = {} & z \zb\,, \\
  \tau_2 = {} & (1-z)(1-\zb)\,,
  \label{eq:tau}
\end{align}
(see eqs.~\eqref{eq:paramab} and~\eqref{eq:paramcd}), that leads to
\begin{align}
  \lambda\left(1,\tau_1(z,\zb),\tau_2(z,\zb)\right) = (z-\zb)^2\,.
\end{align}
Without loss of generality, we choose the following root of
eq.~\eqref{eq:tau}
\begin{align}
  z = {} & \frac12 \left( 1+\tau_1-\tau_2 + \sqrt{\lambda(1,\tau_1,\tau_2)} \right)\,, \\
  \zb = {} & \frac12 \left( 1+\tau_1-\tau_2 - \sqrt{\lambda(1,\tau_1,\tau_2)} \right)\,.
             \label{eq:zzbar}
\end{align}
Varying the pair $(\tau_1,\tau_2)$ in the real plane, we identify the
following possibilities for
$z,\zb$
\begin{equation}
  \left\{
  \begin{array}{lll}
    \zb = z^* & \lambda(1,\tau_1,\tau_2) < 0  \,,  \quad \tau_1,\tau_2 >0 & \text{(region I)}\\
    0<\zb<z<1 & \sqrt{\tau_1} + \sqrt{\tau_2} < 1\,, \quad 0<\tau_1\,,\tau_2 <1  \hphantom{\ldots}  & \text{(region II)} \\
    \zb<z < 0 & \sqrt{\tau_2}> 1+\sqrt{\tau_1} \,, \quad \tau_1>0 & \text{(region III)}\\
    z>\zb>1 & \sqrt{\tau_1} > 1+\sqrt{\tau_2} \,, \quad \tau_2>0 & \text{(region IV)} \\
    z=\zb=\pm \sqrt{\tau_1} & \tau_2 = \left(1 \pm \sqrt{\tau_1}\right)^2 \,,  \quad \tau_1,\tau_2 >0 & \text{(region V)} \\
    z>1\,, \quad \bar{z}<0 & \tau_1, \tau_2<0 & \text{(region VI)} \\
    0<z<1\,, \quad \bar{z}<0 \hphantom{\ldots} & \tau_1<0\,, \quad \tau_2>0 & \text{(region VII)} \\
    z>1\,, \quad 0<\bar{z}<1 & \tau_1>0\,, \quad \tau_2<0 & \text{(region VIII)} \\
  \end{array}
  \right.
  \label{eq:zzbregions}
\end{equation}
where the first five regions were discussed also
in~\cite{Chavez:2012kn}. A graphical representation of these eight
regions in the $(\tau_1,\tau_2)$-plane is shown if
fig~\ref{fig:regions}.

The variables $z,\zb$ are complex conjugates in region I, where
$\lambda(1,\tau_1,\tau_2)<0$, and real in all the other regions.
In regions I-V one has $\tau_{1,2} >0$, which requires that either
$s,p_1^2,p_2^2<0$ or $s,p_1^2,p_2^2>0$. The former case defines the
Euclidean region. The latter case, for $\lambda(1,\tau_1,\tau_2)>0$,
describes $1\to 2$ or $2\to 1$ processes involving
three timelike particles.
Region V is where $\lambda(1,\tau_1,\tau_2) = 0$, so that
$z=\zb$. Since our expressions are obtained in general for
$z\neq \zb$, the limit $\zb\to z$ has to be taken carefully.
Regions VI-VIII have at least one of the $\tau_i<0$, which requires
either two external legs to be spacelike and the remaining one to be
timelike, or vice versa. The former configuration, in the $2\to 1$
kinematics, describes the vertex entering the production of a timelike
particle via the ``fusion'' of two spacelike particles.

In regions other than II, the variables $z,\zb$ are not in the half of
the unit square where all the letters are real, therefore analytic
continuation is required.
A consistent physical prescription is inherited in regions VI-VIII
from the Feynman prescription on the kinematic invariants, and it is
naturally extended to the other regions, as we argue below. For the
moment we hold $s<0$, and we will discuss later the case $s>0$.

\noindent In region VI,  $s<0$ and $p_1^2,p_2^2>0$, then
\begin{equation}
  \tau_i \to -|\tau_i| - i \varepsilon  \,,
\end{equation}
so that the vanishing imaginary parts outside the square root
in~eq~\eqref{eq:zzbar} cancel against each other, and only the one
stemming from the square root is left:
\begin{equation}
  z \to z + i\varepsilon\,, \qquad     \zb \to \zb - i\varepsilon\,.
  \label{eq:zzboppositeVI}
\end{equation}
In region VII, $p_1^2>0$ and $s,p_2^2<0$, then
\begin{equation}
  \tau_1 \to -|\tau_1| - i\varepsilon\,, \qquad \tau_2 \to |\tau_2|\,,
\end{equation}
so that
\begin{align}
  z \to {} & z + i\, \frac{\varepsilon}{2} \left( \frac{1+|\tau_1|+\tau_2}{\sqrt{\lambda(1,-|\tau_1|,\tau_2)}} - 1 \right) \simeq z + i \varepsilon \,, \nn
  \zb \to {} & \zb - i\,\frac{\varepsilon}{2} \left( \frac{1+|\tau_1|+\tau_2}{\sqrt{\lambda(1,-|\tau_1|,\tau_2)}} + 1 \right) \simeq \zb - i \varepsilon \,,
  \label{eq:zzboppositeVII}
\end{align}
where the approximate equalities are allowed because the factor in the
bracket is always positive, and a redefinition of $\varepsilon$ is
understood.

\noindent
In region VIII, $p_2^2>0$ and $s,p_1^2<0$, then
\begin{equation}
  \tau_1 \to |\tau_1| \,, \qquad \tau_2 \to -|\tau_2| - i\varepsilon\,,
\end{equation}
so that
\begin{align}
  z \to {} & z + i\, \frac{\varepsilon}{2} \left( \frac{1+\tau_1+|\tau_2|}{\sqrt{\lambda(1,\tau_1,-|\tau_2|)}} + 1 \right) \simeq z + i\varepsilon \,, \nn
  \zb \to {} & \zb - i\,\frac{\varepsilon}{2} \left( \frac{1+\tau_1+|\tau_2|}{\sqrt{\lambda(1,\tau_1,-|\tau_2|)}} - 1 \right) \simeq  \zb - i\varepsilon \,,
  \label{eq:zzboppositeVIII}
\end{align}
where again the approximate equalities are allowed because the factor
in the bracket is always positive, and a redefinition of $\varepsilon$
is understood.

\noindent
We have so far only discussed the case in which $s<0$. It is easy to
see that, if instead $s>0$, the prescription on $z,\zb$ is the
opposite.

In regions I-V there is no physical prescription for the analytic
continuation of $z,\zb$.  Indeed, if $s,p_1^2,p_2^2>0$, then the
vanishing imaginary parts of the Feynman prescription cancel out in
the ratios $\tau_1$ and $\tau_2$:
\begin{equation}
  \tau_i \to \frac{p_i^2\,(1+i\varepsilon)}{s\,(1+i\varepsilon)} = \tau_i\,.
\end{equation}
This cancellation affects also region I, where
$\sqrt{\lambda(1,\tau_1,\tau_2)}<0$ and $z^*=\zb$. Indeed, while this
condition fixes the relative sign of their imaginary parts, the sign
of $\text{Im}\,z$ depends on the choice of the branch of the square
root in eq.~\eqref{eq:zzbar}, which is not fixed. This last
statement holds true also in the Euclidean region.

This ambiguity is resolved by the definite $i\varepsilon$ prescription
in regions VI-VIII discussed above.
In order to have a smooth analytic continuation in the Euclidean, in
region I, one chooses the branch of the square root that gives
$\text{Im}\sqrt{\lambda(1,\tau_1,\tau_2)}>0$, and in regions III-IV,
one assigns vanishing imaginary parts for $z,\zb$ according to the
previous discussion.  The opposite prescription should be used if the
three external legs are timelike.

Summarizing, according to the sign of $s$, we choose the following
analytic continuation prescriptions for $z,\zb$ in the whole real
$(p_1^2,p_2^2)$ plane
\begin{align}
  z \to z + i \varepsilon\,, \quad \zb \to \zb - i \varepsilon \quad s<0\,,
  \label{eq:zzbsneg} \\
  z \to z - i \varepsilon\,, \quad \zb \to \zb + i \varepsilon \quad s>0\,.
  \label{eq:zzbspos}
\end{align}

\subsection{Internal massive lines: the $u,v$ variables}
\label{sec:uv}
For topologies
\protect\subref{fig:FigTop1WWbb}-\protect\subref{fig:FigTop2WWbb}, the
change of variables eq.~\eqref{eq:tau} is actually enough to
rationalize the DEs completely. In eq.~\eqref{eq:paramab} we simply
rescale $s$ by the internal mass ($m^2>0$), $-s/m^2=u$, to deal with a
dimensionless variable. If $s<0$, $u>0$. If $s>0$, the Feynman
prescription $s\to s +i\varepsilon$ fixes the analytic continuation
for $u$
\begin{equation}
  \label{eq:uprescr}
  u \to - u' - i\varepsilon\,, \quad \text{with } u'>0\,.
\end{equation}

In the case of topologies
\protect\subref{fig:FigTop1WW}-\protect\subref{fig:FigTop2WW}, the DEs
still contain the square roots related to the $s$-channel threshold at
$s=4m^2$. They are rationalized by the usual variable change (see
eq.~\eqref{eq:paramcd}),
\begin{equation}
  \label{eq:landau}
  -\frac{s}{m^2} = \frac{(1-v)^2}{v}\,,
\end{equation}
of which we choose the following root
\begin{equation}
  \label{eq:landauroot}
  v = \frac{\sqrt{4m^2-s}-\sqrt{-s}}{\sqrt{4m^2-s}+\sqrt{-s}}\,.
\end{equation}
For completeness, we discuss how $v$ varies with $s$. Holding $m^2>0$,
and keeping in mind the Feynman prescription for $s>0$, one finds the
following cases
\begin{itemize}
\item For $s<0$, $v$ is on the unit interval, $0\leq v \leq 1$;
\item For $0\leq s \leq 4m^2$, $v$ is a pure phase, $v=e^{i \phi}$, with
  $0<\phi<\pi$;
\item For $s>4m^2$, $v$ is on the negative unit interval, and one must
  replace
  \begin{equation}
    \label{eq:vprescr}
    v\to -v'+i\varepsilon\,, \quad 0\leq v' \leq 1\,.
  \end{equation}
\end{itemize}

\subsection{Analytic continuation of the master integrals}
As discussed in section~\ref{sec:WWMIs}, for all the topologies we
start in the patch of the Euclidean region where the alphabet is real
and positive (see eqs.~\eqref{eq:positivityabuzzb}
and~\eqref{eq:positivitycdvzzb}), and we solve the DEs there.  As far
as the variables $z,\zb$ are concerned, the conditions of positivity
of the alphabet are the same for all our topologies,
\begin{equation}
  0<z<1\,, \quad 0<\zzb<\zz\,,
  \label{eq:positivityzzb}
\end{equation}
\ie we start from region II (see eq.~\eqref{eq:zzbregions}).
Regarding the condition on the variables associated to $s$, \ie $u$
and $v$ (respectively for topologies
\protect\subref{fig:FigTop1WWbb}-\protect\subref{fig:FigTop2WWbb} and
topologies
\protect\subref{fig:FigTop1WW}-\protect\subref{fig:FigTop2WW}), we
require
\begin{equation}
  0 < u < \frac{1}{z(1-\zb)}, \quad 0<v<1\,,
  \label{eq:positivityuv}
\end{equation}
It is clear from eq.~\eqref{eq:zzbregions} that, if these conditions
are satisfied, one does not have access even to the full Euclidean
region.  Results in the remaining patches of the latter, as well as in
the physical regions, are obtained by analytic continuation using the
prescriptions described in sections~\ref{sec:zzb}
and~\ref{sec:zzb}.

In the present work we performed the analytic continuation
numerically, \ie we assigned to $u,v,z,\zb$ the vanishing imaginary
parts discussed above choosing sufficiently small numerical values.
For convenience, we summarize the analytic continuation prescription
for the physically interesting cases.
\begin{itemize}
\item $\bm{\X\to W W}$: In this region a particle of mass $s>0$ decays in
  two (possibly off-shell) particles with invariant masses $p_1^2>0$
  and $p_2^2>0$, so that
  \begin{align}
    \sqrt{s} \geq {} & \sqrt{p_1^2} + \sqrt{p_2^2}\,.
  \end{align}
  Regarding $z,\zb$, this corresponds to region II (see
  eq.~\eqref{eq:zzbregions}), therefore no analytic continuation is
  needed. Furthermore, for topologies
  \protect\subref{fig:FigTop1WWbb}-\protect\subref{fig:FigTop2WWbb}
  one must replace
  $$u\to -u' - i\varepsilon$$ irrespectively of the value of
  $s$, according to eq.~\eqref{eq:uprescr}. Instead, for topologies
  \protect\subref{fig:FigTop1WW}-\protect\subref{fig:FigTop2WW}, if
  $0<s<4m^2$ then $v$ is on the unit circle in the complex plane,
  while if $s>4m^2$, one has to replace $$v\to -v+i\varepsilon\,,$$
  according to eq.~\eqref{eq:vprescr}.

\item $\bm{W\to W\X}$: This is again a $1\to 2$ process involving
  timelike particles, the only difference being that now
  \begin{align}
    \sqrt{p_1^2} \geq \sqrt{s} + \sqrt{p_2^2} > 0\,,
  \end{align}
  or
  \begin{align}
    \sqrt{p_2^2} \geq \sqrt{s} + \sqrt{p_1^2} > 0\,.
  \end{align}
  The former case corresponds to region IV, the latter to region III
  (see eq.~\eqref{eq:zzbregions}). Therefore, in addition to the
  analytic continuation in $u,v$ discussed already for the $\X$-decay,
  one must further use the replacement~\eqref{eq:zzbspos}
  \begin{equation}
    z \to z - i \varepsilon\,, \quad \zb \to \zb + i \varepsilon\,.
  \end{equation}
  
\item $\bm{WW\to\X}$: Here
  \begin{equation}
    p_1^2,p_2^2<0\,, \quad s>0\,,
  \end{equation}
  corresponding to region VI, so that $z,\zb$ inherit the analytic
  continuation prescription eq.~\eqref{eq:zzbspos} from
  $s\to s+i\varepsilon$
  \begin{equation*}
    z \to z - i \varepsilon\,, \quad \zb \to \zb + i \varepsilon\,.
  \end{equation*}
  Concerning $u,v$, the discussion is the same as for $\X$-decay.

\item $\bm{\X\,W\to W}$: Here
  \begin{equation}
    p_1^2,s<0\,, \quad p_2^2>0\,,
  \end{equation}
  or
  \begin{equation}
    p_2^2,s<0\,, \quad p_1^2>0\,,
  \end{equation}
  corresponding to region VII and VIII respectively, so that $z,\zb$
  inherit from $p_i^2\to p_i^2+i\varepsilon$ the
  prescription~\eqref{eq:zzbsneg}
  \begin{equation*}
    z \to z + i \varepsilon\,, \quad \zb \to \zb - i \varepsilon\,.
  \end{equation*}
  Since $s<0$, no continuation is due on $u,v$.

\end{itemize}


\section{Conclusions}
\label{sec:conclusions}

  In this paper we have computed the two-loop master integrals required for the
  leading QCD corrections to the interaction vertex between a massive
  neutral boson $\X$, such as $H,Z$ or $\gamma^{*}$, and pair of $W$
  bosons, mediated by a $SU(2)_L$ quark doublet composed of one
  massive and one massless flavor.  We considered external legs with arbitrary invariant
  masses. The master integrals were computed by means of the
  differential equation method. After identifying a set of master
  integrals obeying a system of equations which depends linearly on the
  space-time dimension $d$, we used 
  the Magnus exponential in order to to find a novel set of 
  integrals that, around $d=4$ dimensions, obey a canonical
  system of differential equations.  The canonical master integrals
  were finally obtained as a Taylor series in $\epsilon = (4-d)/2$, up to order
  four, with coefficients written as combination of Goncharov
  polylogarithms, respectively up to weight four.

  In the context of the Standard Model, our results are relevant for
  the mixed EW-QCD corrections to the Higgs decay to a $W$ pair, as
  well as the production channels obtained by crossing, and to the
  triple gauge boson vertices $ZWW$ and $\gamma^*WW$.


\section*{Acknowledgments}
We acknowledge clarifying discussions with R.~Bonciani, G.~Degrassi,
A.~Ferroglia, K.~Melnikov, P.~Paradisi, G.~Passarino, F.~Petriello,
F.~Tramontano, A.~Vicini and A.~Wulzer.  A.~P.\ wishes to thank the
Institute for Theoretical Particle Physics of the Karlsruhe Institute
of Technology for the kind hospitality during the completion of this
work. U.~S.\ is supported by the DOE contract DE-AC02-06CH11357.

\appendix
\section{Properties of Chen's iterated integrals}
\label{sec:chen}
In this appendix we recall the main properties of Chen's iterated integrals~\cite{Chen:1977oja}. We closely follow the notation of~\cite{Bonciani:2016ypc}. Chen's iterated integrals are defined by
\begin{align}
\label{ap:chendlog}
\chen{i_k,\ldots,i_1}{\gamma} = \int_\gamma \dlog \eta_{i_k} \ldots \dlog \eta_{i_1}= {} & \int_{0\leq t_1 \leq \ldots \leq t_k \leq 1} g^\gamma_{i_k}(t_k) \ldots g^\gamma_{i_1}(t_1) \,dt_1 \ldots \,dt_k\,,
\end{align}
where
$\gamma$ is a piecewise-smooth path connecting $\vec{x}_0$ to
$\vec{x}$,
\begin{align}
\begin{cases}
\gammaAB{\xxi}{\xxf}:[0,1]\ni t \mapsto  \gamma(t) = (\gamma^1(t),\gamma^2(t),\gamma^3(t))\\
\gamma(0)=\xxi \\
\gamma(1)=\xxf \, .
\end{cases}
\label{eq:gamma}
\end{align}
and
\begin{align}
\label{ap:dlogweight}
g^\gamma_i(t) = {} & \frac{d}{dt} \log \eta_i(\gamma(t))\,.
\end{align}
\begin{itemize}
\item \textit{Invariance under path reparametrization.}  The integral
  $\chen{i_k,\ldots,i_1}{\gamma}$ does not depend on the way one chooses
  to parametrize the path $\gamma$.
\item \textit{Reverse path formula.}  If the path $\gamma^{-1}$ is the
  path $\gamma$ traversed in the opposite direction, then
  \begin{align}
    \label{eq:reverse}
    \chen{i_k,\ldots,i_1}{\gamma^{-1}} = (-1)^k \chen{i_k,\ldots,i_1}{\gamma}\,.
  \end{align}
\item\textit{Recursive structure.}  From \eqref{ap:chendlog}
  and~\eqref{ap:dlogweight} it follows that the line integral of one
  $\dlog$ is defined, as usual, by
  \begin{align}
    \label{eq:lineintegral}
    \int_{\gamma} \dlog \eta = {} & \int_{0\leq t \leq 1}  \frac{\dlog \eta(\gamma(t))}{d t} dt = \log{\eta(\xxf)} - \log \eta(\xxi)\,,
  \end{align}
  and only depends on the endpoints $\xxi,\xxf$.\\
  
  It is convenient to introduce the path integral ``up to some point
  along $\gamma$'': given some path $\gamma$ and a parameter $s\in[0,1]$,
  one can define the 1-parameter family of paths
  \begin{align}
    \label{eq:gammas}
    \gammas:[0,1]\ni t \mapsto \xx = (\gamma^1(s\, t),\gamma^2(s\, t),\gamma^3(s\, t))\,.
  \end{align}
  If $s=1$, then trivially $\gammas=\gamma$ whereas if $s=0$ the image of
  the interval $[0,1]$ is just $\{\xxi\}$.  If $s\in(0,1)$, then the
  curve $\gammas([0,1])$ starts at $\gamma(0)=\xxi$ and overlaps with
  the curve $\gamma([0,1])$ up to the point $\gamma(s)$, where it
  ends.  It can be easily shown that the path integral along
  $\gamma_s$ can be written as
  \begin{align}
    \label{eq:chens}
    \chen{i_k,\ldots,i_1}{\gammas} = \int_{0\leq t_1 \leq \ldots \leq t_k \leq s} g^\gamma_{i_k}(t_k) \ldots g^\gamma_{i_1}(t_1) \,dt_1 \ldots \,dt_k\,,
  \end{align}
  which differs from eq.~\eqref{eq:chendlog} by the fact that the
  outer integration (\ie the one in $dt_k$) is performed over $[0,s]$
  instead of $[0,1]$.
  Having introduced $\gammas$, we can rewrite~\eqref{eq:chendlog} in a
  recursive manner:
  \begin{align}
    \label{eq:chenrecursive}
    \chen{i_k,\ldots,i_1}{\gamma} = {} & \int_0^1 g^\gamma_{i_k}(s) \, \chen{i_{k-1},\ldots,i_1}{\gammas} ds\,.
  \end{align}
  In addition, eq.~\eqref{eq:chens} can be used in order to derive the identity
  \begin{align}
    \label{eq:chenderivative}
    \frac{d}{ds}\, \chen{i_k,\ldots,i_1}{\gammas} = g^\gamma_{i_k}(s) \, \chen{i_{k-1},\ldots,i_1}{\gammas}\,.
  \end{align}
  
\item\textit{Shuffle algebra.}  Chen's iterated integrals fulfill
  shuffle algebra relations: if $\vec{m}=(m_M,\ldots,m_1)$ and
  $\vec{n}=(n_N,\ldots,n_1)$, with $M$ and $N$ natural numbers, one has
  \begin{align}
    \chen{\vec{m}}{\gamma} \, \chen{\vec{n}}{\gamma} = {} &
    \sum_{\text{shuffles }\sigma} \chen{\sigma(m_M),\ldots,\sigma(m_1),\sigma(n_N),\ldots,\sigma(n_1)}{\gamma}\,,
                                                            \label{chen:shuffle}
  \end{align}
  where the sum runs over all the permutations $\sigma$ that preserve
  the relative order of $\vec{m}$ and $\vec{n}$.
    
\item\textit{Path composition formula.}  If
  $\alpha,\beta:[0,1]\to\mathcal{M}$ are two paths such that
  $\alpha(0)=\xxi$, $\alpha(1)=\beta(0)$, and $\beta(1)=\xxf$, then
  the composed path $\gamma = \alpha\beta$ is obtained by first
  traversing $\alpha$ and then $\beta$. One can prove that the
  integral over such a composed path satisfies
  \begin{align}
    \label{eq:composed}
    \chen{i_k,\ldots,i_1}{\alpha\beta} = \sum_{p=0}^k \chen{i_k,\ldots,i_{p+1}}{\beta} \, \chen{i_p,\ldots,i_1}{\alpha}\,.
\end{align}

\item\textit{Integration-by-parts formula.} The computation of
  eq.~\eqref{ap:chendlog} requires, in principle, the evaluation of
  $k$ nested integrals. Nevertheless, we observe that the innermost
  integration is always reduced to \eqref{eq:lineintegral}, so that
  one has $k-1$ actual integrations to perform. For instance, at
  weight $k=2$, we have
\begin{align}
\label{eq:innerinteg}
  \chen{m,n}{\gamma} = {} & \int_0^1 g_{m}(t) \,\chen{n}{\gammat} \,dt \nonumber\\
  = {} & \int_0^1 g_{m}(t) (\logxt{n}{t} - \logxi{n}) \,dt\,
\end{align}
and one is left with a single integral to be evaluated, either analytically or numerically.

\noindent
Moreover, one can show that the integration involving the outermost
weight $g_k$ can be performed by parts, returning
\begin{align}
  \label{eq:chenibpk}
  \chen{i_k,\ldots,i_1}{\gamma} = {} & \logxf{i_k}\, \chen{i_{k-1},\ldots,i_1}{\gamma} - \int_0^1  \logxt{i_k}{t} \, g_{i_{k-1}}(t) \, \chen{i_{k-2},\ldots,i_1}{\gammat} dt\,.
\end{align}
The combined use of eqs.~\eqref{eq:innerinteg} and \eqref{eq:chenibpk}
allows, for instance, a remarkable simplification in the numerical
evaluation of weight $k\geq 3$ iterated integrals, since the analytic
calculation of the inner- and outermost integrals leaves only $k-2$
integrations to be performed via numerical methods.

\item\textit{Conversion to} GPLs \textit{formula}. If all letters
  appearing in a Chen's iterated integral are rational functions with algebraic roots, then the iterated integral can be converted in terms of GPLs. \noindent

Suppose we connect the endpoints $\vec{x}_0=(a_1,\, a_2,\, a_3)$ and
$\vec{x}_1=(b_1,\, b_2,\, b_3)$ through a piecewise
path of the type
\begin{align}
  \begin{cases}
    \gamma_1(t): \; \left(a_1+t(b_1-a_1),\, a_2,\, a_3\right)\\
    \gamma_2(t): \; \left(b_1,\, a_2+t(b_2-a_2),\, a_3\right)\\
    \gamma_3(t): \; \left(b_1,\, b_2,\, a_3+t(b_3-a_3)\right) \, .
  \end{cases}
\end{align}
the conversion of the iterated integral to GPLs can be achieved by factorizing all letters such that the dependence on the varied variable $x_i$, becomes linear. Any weight-$k$ integral can then be transformed by
\begin{align}
\label{eq:GPLconv}
\int_{\gamma_i}\; d\log(x_i-w_k)\dots d\log(x_i-w_1)=G\left(\frac{a_i-w_{k}}{a_i-b_i},\dots,\frac{a_i-w_{1}}{a_i-b_i};1\right),
\end{align}
where $w_{i}$ are weights which may depend on the constant variables along the path $\gamma_i$.

\end{itemize}


\section{Canonical master integrals for $\X\, W^{+}W^{-}$}
\label{sec:gMIcoeff}
In this appendix we give the explicit expression of the kinematic coefficients appearing in the definition of canonical master integrals defined by eq.~\eqref{def:WWCanonicalMIsbb} and eq.~\eqref{def:WWCanonicalMIs}.
 \subsection{Topologies (a)-(b)}
The coefficients of the canonical MIs listed in eq.~\eqref{def:WWCanonicalMIsbb} are given by
\label{sec:gMIcoeffb}
 \begin{align}
  c_{16,\,17}={}&\frac{3}{2}\left(\sqrt{\lambda}+s-p_1^2-p_2^2+2m^2\right)\,,\\
  c_{17,\,17}={}&p_1^2p_2^2-(p_1^2+p_2^2)m^2+m^4+m^2s\,,\\
  c_{21,\,21}={}&\frac{3}{2}\left(\sqrt{\lambda}+s-p_1^2-p_2^2+2m^2\right)\,,\\
  c_{22,\,22}={}&p_1^2p_2^2-(p_1^2+p_2^2)m^2+m^4+m^2s\,,\\
  c_{1,\,29}={}&\frac{ m^2s}{(p_1^2-m^2)(p_2^2-m^2)}\,,\\
  c_{2,\,29}={}&-\frac{m^2sp_2^2}{(p_1^2-m^2)(p_2^2-m^2)}\,,\\
  c_{4,\,29}={}&-\frac{ m^2sp_1^2}{(p_1^2-m^2)(p_2^2-m^2)}\,,\\
  c_{27,\,29}={}&p_1^2 \left(p_2^2-m^2\right)+p_2^2 \left(\sqrt{\lambda} +m^2+s-p_2^2\right)+m^2 \left(s-\sqrt{\lambda} \right)\,,\\
   c_{28,\,29}={}&p_1^2 \left(p_2^2-\sqrt{\lambda} +m^2+s-p_1^2\right)+m^2 (\sqrt{\lambda}+s-p_2^2)\,,\\
  c_{29,\,29}={}&-s\left(p_1^2(p_2^2-m^2)+m^2(s+m^2-p_2^2)\right)\,,
  \end{align}
where the K\"all\'en function $\lambda$ is defined in eq.~\eqref{eq:lambda}.

\subsection{Topologies (c)-(d)}
The coefficients of the canonical MIs listed in eq.~\eqref{def:WWCanonicalMIs} are given by
\label{sec:gMIcoefft}
\begin{align}
  c_{2,\,16}={}&-\frac{p_2^2\left(-m^2\left(p_2^2-s+\sqrt{\lambda}\right)-p_1^2\left(p_2^2-p_1^2+s-m^2-\sqrt{\lambda}\right)\right)}{(p_1^2-p_2^2)(p_1^2+m^2)-s(p_1^2-m^2)},\,\\
  c_{3,\,16}={}&\frac{\sqrt{\lambda}\left(p_1^2(2m^2-s)-m^2(2p_2^2-s)\right)}{(p_1^2-p_2^2)(p_1^2+m^2)-(p_1^2-m^2)s}- \rho,\,\\
  c_{4,\,16}={}&-\frac{p_1^2\left(m^2\left(p_2^2-s+\sqrt{\lambda}\right)+p_1^2\left(p_2^2-p_1^2+s-m^2-\sqrt{\lambda}\right)\right)}{(p_1^2-p_2^2)(p_1^2+m^2)-(p_1^2-m^2)s},\,\\
  c_{16,\,16}={}&\frac{\sqrt{\lambda}\left(m^2(p_1^2-p_2^2)^2+s(p_1^2-m^2)(p_2^2-m^2)\right)}{(p_1^2-p_2^2)(p_1^2+m^2)-(p_1^2-m^2)s},\, \\
  c_{17,\,19}={}&\left(p_2^2-p_1^2+2m^2-s-\sqrt{\lambda}\right),\\
  c_{18,\,19}={}&\frac{1}{2}\left(p_2^2-p_1^2+2m^2-s-\sqrt{\lambda}\right),\\
  c_{19,\,19}={}&m^2(p_2^2+m^2-s)+p_1^2(s-m^2),\\
  c_{5,\,21}={}&\frac{m^2(p_2^2-m^2)}{s+\rho},\\
  c_{6,\,21}={}&-\frac{2m^2(p_2^2+m^2)}{s+\rho},\\
  c_{9,\,21}={}&\frac{2m^2(p_1^2+m^2)}{s+\rho},\\
  c_{10,\,21}={}&-\frac{m^2(p_1^2-m^2)}{s+\rho},\\
  c_{20,\,21}={}&-\frac{1}{2}\sqrt{\lambda}+\frac{1}{2s}(s+p_2^2-p_1^2)\rho,{}\\
  c_{21,\,21}={}&\rho,\\
  c_{22,\,24}={}&p_1^2-p_2^2-s+2m^2+\sqrt{\lambda},\\
  c_{23,\,24}={}&\frac{1}{2}(p_1^2-p_2^2-s+2m^2+\sqrt{\lambda}),\\
  c_{24,\,24}={}&m^2p_1^2-(p_2^2-m^2)(m^2-s),\\
  c_{11,\,25}={}&-\frac{p_2^2\left(m^2\left(p_2^2-s+\sqrt{\lambda}\right)+p_1^2\left(p_2^2-p_1^2+s-m^2-\sqrt{\lambda}\right)\right)}{(p_1^2-p_2^2)(p_1^2+m^2)-s(p_1^2-m^2)},\,\\
  c_{12,\,25}={}&\frac{p_2^2\sqrt{\lambda}\left(s(p_1^2-m^2)-2m^2(p_1^2-p_2^2)\right)}{(p_1^2-p_2^2)(p_1^2+m^2)-(p_1^2-m^2)s}+p_2^2\rho,\,\\
  c_{13,\,25}={}&\frac{p_1^2p_2^2\left(p_1^2\left(p_2^2-p_1^2+s-m^2-\sqrt{\lambda}\right)+m^2\left(p_2^2-s+\sqrt{\lambda}\right)\right)}{(p_1^2-p_2^2)(p_1^2+m^2)-(p_1^2-m^2)s},\,\\
  c_{25,\,25}={}&-\frac{p_2^2\sqrt{\lambda}\left(m^2(p_1^2-p_2^2)^2+s(p_1^2-m^2)(p_2^2-m^2)\right)}{(p_1^2-p_2^2)(p_1^2+m^2)-(p_1^2-m^2)s},\\
  c_{12,\,26}={}&\frac{p_2^2\rho\left(p_1^2\left(p_2^2-p_1^2+s-m^2-\sqrt{\lambda}\right)+m^2\left(p_2^2-s+\sqrt{\lambda}\right)\right)}{(p_1^2-p_2^2)(p_1^2+m^2)-(p_1^2-m^2)s},\,\\
  c_{14,\,26}={}&\rho^2+\frac{\rho\sqrt{\lambda}\left(p_1^2\left(2m^2-s\right)+m^2\left(s-2p_2^2\right)\right)}{(p_1^2-p_2^2)(p_1^2+m^2)-(p_1^2-m^2)s},\,\\
  c_{15,\,26}={}&-\frac{p_1^2\rho\left(p_1^2\left(p_2^2-p_1^2-s+\sqrt{\lambda}\right)+m^2\left(p_2^2-s+\sqrt{\lambda}\right)\right)}{(p_1^2-p_2^2)(p_1^2+m^2)-(p_1^2-m^2)s},\,\\
  c_{26,\,26}={}&\frac{8\sqrt{\lambda}\left(m^2(p_1^2-p_2^2)^2+s(p_1^2-m^2)(p_2^2-m^2)\right)}{\left((p_1^2-p_2^2)(p_1^2+m^2)-(p_1^2-m^2)s\right)(s-2m^2+\rho)^4}\times\nn
               &\times \bigg(s^4\left(s+\rho\right)+2m^8\left(8s+\rho\right)-m^2s^3\left(5s+4\rho\right)-4m^6s\left(11s+4\rho\right)\nn
               &+2m^4s^2\left(17s+10\rho\right)\bigg),\\
  c_{3,\,30}={}&2(p_1^2-p_2^2)-\frac{2p_2^2(p_1^2-p_2^2)}{p_2^1-m^2}-s+\rho\,,\\
  c_{12,\,30}={}&2p_2^2\left(-2(p_1^2-p_2^2)+\frac{2p_2^2(p_1^2-p_2^2)}{p_2^2-m^2}+s-\rho\right),\\
  c_{28,\,30}={}&s(p_1^2+p_2^2+2m^2-s)-\rho\sqrt{\lambda},\\
  c_{29,\,30}={}&\left(p_2^2(s-m^2-p_2^2)+p_1^2(p_2^2-m^2)-m^2s\right)+(p_2^2-m^2)\sqrt{\lambda}\,,\\
  c_{30,\,30}={}&-m^2(p_1^2-p_2^2)^2-(p_1^2-m^2)(p_2^2-m^2)s\,,
\end{align}
where $\rho$ is defined after eq.~\eqref{def:WWCanonicalMIs} and the
K\"all\'en function $\lambda$ is given in eq.~\eqref{eq:lambda}.


\section[dlog-forms]{$\dlog$-forms}
\label{sec:dlog-form}
In this appendix we collect the coefficient matrices of the
$\dlog$-forms for the master integrals \eqref{def:WWCanonicalMIsbb} and \eqref{def:WWCanonicalMIs}. 
\subsection{Topologies  (a)-(b)}
\label{dlogWW2Lbb}
For the two-loop integrals discussed in section~\ref{sec:WWMIsb} we have
\begin{align}
  \dA = {} & \MM_1 \, \dlog(u) + \MM_2 \, \dlog(\zz) + \MM_3 \, \dlog(1-\zz) \nn
  &+ \MM_4 \, \dlog(\zzb) + \MM_5 \, \dlog(1-\zzb)  + \MM_6\, \dlog(\zz-\zzb) \nn
  &+ \MM_7\, \dlog(1+u\,\zz\,\zzb) + \MM_8 \, \dlog\left(1-u\,\zz (1-\zzb)\right) \nn
  &+ \MM_9\, \dlog\left(1-u\,\zzb(1-\zz)\right) + \MM_{10} \, \dlog\left(1+u(1-\zz)(1-\zzb)\right)\,,
\end{align}
with 

\begin{align}
\MM_1 = \scalemath{0.53}{
\left(
\begin{array}{ccccccccccccccccccccccccccccc}
 0 & \pminus 0 & \pminus 0 & \pminus 0 & \pminus 0 & \pminus 0 & \pminus 0 & \pminus 0 & \pminus 0 & \pminus 0 & \pminus 0 & \pminus 0 & \pminus 0 & \pminus 0 & \pminus 0 & \pminus 0 & \pminus 0 & \pminus 0 & \pminus 0 & \pminus 0 & \pminus 0 & \pminus 0 & \pminus 0 & \pminus 0 & \pminus 0 & \pminus 0 & \pminus 0 & \pminus 0 & \pminus 0 \\
 0 & \pminus 1 & \pminus 0 & \pminus 0 & \pminus 0 & \pminus 0 & \pminus 0 & \pminus 0 & \pminus 0 & \pminus 0 & \pminus 0 & \pminus 0 & \pminus 0 & \pminus 0 & \pminus 0 & \pminus 0 & \pminus 0 & \pminus 0 & \pminus 0 & \pminus 0 & \pminus 0 & \pminus 0 & \pminus 0 & \pminus 0 & \pminus 0 & \pminus 0 & \pminus 0 & \pminus 0 & \pminus 0 \\
 0 & \pminus 0 & -1 & \pminus 0 & \pminus 0 & \pminus 0 & \pminus 0 & \pminus 0 & \pminus 0 & \pminus 0 & \pminus 0 & \pminus 0 & \pminus 0 & \pminus 0 & \pminus 0 & \pminus 0 & \pminus 0 & \pminus 0 & \pminus 0 & \pminus 0 & \pminus 0 & \pminus 0 & \pminus 0 & \pminus 0 & \pminus 0 & \pminus 0 & \pminus 0 & \pminus 0 & \pminus 0 \\
 0 & \pminus 0 & \pminus 0 & \pminus 1 & \pminus 0 & \pminus 0 & \pminus 0 & \pminus 0 & \pminus 0 & \pminus 0 & \pminus 0 & \pminus 0 & \pminus 0 & \pminus 0 & \pminus 0 & \pminus 0 & \pminus 0 & \pminus 0 & \pminus 0 & \pminus 0 & \pminus 0 & \pminus 0 & \pminus 0 & \pminus 0 & \pminus 0 & \pminus 0 & \pminus 0 & \pminus 0 & \pminus 0 \\
 0 & \pminus 0 & \pminus 0 & \pminus 0 & \pminus 1 & \pminus 0 & \pminus 0 & \pminus 0 & \pminus 0 & \pminus 0 & \pminus 0 & \pminus 0 & \pminus 0 & \pminus 0 & \pminus 0 & \pminus 0 & \pminus 0 & \pminus 0 & \pminus 0 & \pminus 0 & \pminus 0 & \pminus 0 & \pminus 0 & \pminus 0 & \pminus 0 & \pminus 0 & \pminus 0 & \pminus 0 & \pminus 0 \\
\end{array}
\right)
  }\,,
\end{align}

\begin{align}
\MM_2 = \scalemath{0.53}{
\left(
\begin{array}{ccccccccccccccccccccccccccccc}
 0 & \pminus 0 & \pminus 0 & \pminus 0 & \pminus 0 & \pminus 0 & \pminus 0 & \pminus 0 & \pminus 0 & \pminus 0 & \pminus 0 & \pminus 0 & \pminus 0 & \pminus 0 & \pminus 0 & \pminus 0 & \pminus 0 & \pminus 0 & \pminus 0 & \pminus 0 & \pminus 0 & \pminus 0 & \pminus 0 & \pminus 0 & \pminus 0 & \pminus 0 & \pminus 0 & \pminus 0 & \pminus 0 \\
 0 & \pminus 0 & \pminus 0 & \pminus 0 & \pminus 0 & \pminus 0 & \pminus 0 & \pminus 0 & \pminus 0 & \pminus 0 & \pminus 0 & \pminus 0 & \pminus 0 & \pminus 0 & \pminus 0 & \pminus 0 & \pminus 0 & \pminus 0 & \pminus 0 & \pminus 0 & \pminus 0 & \pminus 0 & \pminus 0 & \pminus 0 & \pminus 0 & \pminus 0 & \pminus 0 & \pminus 0 & \pminus 0 \\
 0 & \pminus 0 & \pminus 0 & \pminus 0 & \pminus 0 & \pminus 0 & \pminus 0 & \pminus 0 & \pminus 0 & \pminus 0 & \pminus 0 & \pminus 0 & \pminus 0 & \pminus 0 & \pminus 0 & \pminus 0 & \pminus 0 & \pminus 0 & \pminus 0 & \pminus 0 & \pminus 0 & \pminus 0 & \pminus 0 & \pminus 0 & \pminus 0 & \pminus 0 & \pminus 0 & \pminus 0 & \pminus 0 \\
 0 & \pminus 0 & \pminus 0 & \pminus 1 & \pminus 0 & \pminus 0 & \pminus 0 & \pminus 0 & \pminus 0 & \pminus 0 & \pminus 0 & \pminus 0 & \pminus 0 & \pminus 0 & \pminus 0 & \pminus 0 & \pminus 0 & \pminus 0 & \pminus 0 & \pminus 0 & \pminus 0 & \pminus 0 & \pminus 0 & \pminus 0 & \pminus 0 & \pminus 0 & \pminus 0 & \pminus 0 & \pminus 0 \\
 0 & \pminus 0 & \pminus 0 & \pminus 0 & \pminus 0 & \pminus 0 & \pminus 0 & \pminus 0 & \pminus 0 & \pminus 0 & \pminus 0 & \pminus 0 & \pminus 0 & \pminus 0 & \pminus 0 & \pminus 0 & \pminus 0 & \pminus 0 & \pminus 0 & \pminus 0 & \pminus 0 & \pminus 0 & \pminus 0 & \pminus 0 & \pminus 0 & \pminus 0 & \pminus 0 & \pminus 0 & \pminus 0 \\
\end{array}
\right)
  }\,,
\end{align}

\begin{align}
\MM_3 = \scalemath{0.53}{
\left(
\begin{array}{ccccccccccccccccccccccccccccc}
 0 & \pminus 0 & \pminus 0 & \pminus 0 & \pminus 0 & \pminus 0 & \pminus 0 & \pminus 0 & \pminus 0 & \pminus 0 & \pminus 0 & \pminus 0 & \pminus 0 & \pminus 0 & \pminus 0 & \pminus 0 & \pminus 0 & \pminus 0 & \pminus 0 & \pminus 0 & \pminus 0 & \pminus 0 & \pminus 0 & \pminus 0 & \pminus 0 & \pminus 0 & \pminus 0 & \pminus 0 & \pminus 0 \\
 0 & \pminus 1 & \pminus 0 & \pminus 0 & \pminus 0 & \pminus 0 & \pminus 0 & \pminus 0 & \pminus 0 & \pminus 0 & \pminus 0 & \pminus 0 & \pminus 0 & \pminus 0 & \pminus 0 & \pminus 0 & \pminus 0 & \pminus 0 & \pminus 0 & \pminus 0 & \pminus 0 & \pminus 0 & \pminus 0 & \pminus 0 & \pminus 0 & \pminus 0 & \pminus 0 & \pminus 0 & \pminus 0 \\
 0 & \pminus 0 & \pminus 0 & \pminus 0 & \pminus 0 & \pminus 0 & \pminus 0 & \pminus 0 & \pminus 0 & \pminus 0 & \pminus 0 & \pminus 0 & \pminus 0 & \pminus 0 & \pminus 0 & \pminus 0 & \pminus 0 & \pminus 0 & \pminus 0 & \pminus 0 & \pminus 0 & \pminus 0 & \pminus 0 & \pminus 0 & \pminus 0 & \pminus 0 & \pminus 0 & \pminus 0 & \pminus 0 \\
 0 & \pminus 0 & \pminus 0 & \pminus 0 & \pminus 0 & \pminus 0 & \pminus 0 & \pminus 0 & \pminus 0 & \pminus 0 & \pminus 0 & \pminus 0 & \pminus 0 & \pminus 0 & \pminus 0 & \pminus 0 & \pminus 0 & \pminus 0 & \pminus 0 & \pminus 0 & \pminus 0 & \pminus 0 & \pminus 0 & \pminus 0 & \pminus 0 & \pminus 0 & \pminus 0 & \pminus 0 & \pminus 0 \\
 0 & \pminus 0 & \pminus 0 & \pminus 0 & \pminus 1 & \pminus 0 & \pminus 0 & \pminus 0 & \pminus 0 & \pminus 0 & \pminus 0 & \pminus 0 & \pminus 0 & \pminus 0 & \pminus 0 & \pminus 0 & \pminus 0 & \pminus 0 & \pminus 0 & \pminus 0 & \pminus 0 & \pminus 0 & \pminus 0 & \pminus 0 & \pminus 0 & \pminus 0 & \pminus 0 & \pminus 0 & \pminus 0 \\
\end{array}
\right)
  }\,,
\end{align}

\begin{align}
\MM_4 = \scalemath{0.53}{
\left(
\begin{array}{ccccccccccccccccccccccccccccc}
 0 & \pminus0 & \pminus0 & \pminus0 & \pminus0 & \pminus0 & \pminus0 & \pminus0 & \pminus0 & \pminus0 & \pminus0 & \pminus0 & \pminus0 & \pminus0 & \pminus0 & \pminus0 & \pminus0 & \pminus0 & \pminus0 & \pminus0 & \pminus0 & \pminus0 & \pminus0 & \pminus0 & \pminus0 & \pminus0 & \pminus0 & \pminus0 & \pminus0 \\
 0 & \pminus0 & \pminus0 & \pminus0 & \pminus0 & \pminus0 & \pminus0 & \pminus0 & \pminus0 & \pminus0 & \pminus0 & \pminus0 & \pminus0 & \pminus0 & \pminus0 & \pminus0 & \pminus0 & \pminus0 & \pminus0 & \pminus0 & \pminus0 & \pminus0 & \pminus0 & \pminus0 & \pminus0 & \pminus0 & \pminus0 & \pminus0 & \pminus0 \\
 0 & \pminus0 & \pminus0 & \pminus0 & \pminus0 & \pminus0 & \pminus0 & \pminus0 & \pminus0 & \pminus0 & \pminus0 & \pminus0 & \pminus0 & \pminus0 & \pminus0 & \pminus0 & \pminus0 & \pminus0 & \pminus0 & \pminus0 & \pminus0 & \pminus0 & \pminus0 & \pminus0 & \pminus0 & \pminus0 & \pminus0 & \pminus0 & \pminus0 \\
 0 & \pminus0 & \pminus0 & \pminus1 & \pminus0 & \pminus0 & \pminus0 & \pminus0 & \pminus0 & \pminus0 & \pminus0 & \pminus0 & \pminus0 & \pminus0 & \pminus0 & \pminus0 & \pminus0 & \pminus0 & \pminus0 & \pminus0 & \pminus0 & \pminus0 & \pminus0 & \pminus0 & \pminus0 & \pminus0 & \pminus0 & \pminus0 & \pminus0 \\
 0 & \pminus0 & \pminus0 & \pminus0 & \pminus0 & \pminus0 & \pminus0 & \pminus0 & \pminus0 & \pminus0 & \pminus0 & \pminus0 & \pminus0 & \pminus0 & \pminus0 & \pminus0 & \pminus0 & \pminus0 & \pminus0 & \pminus0 & \pminus0 & \pminus0 & \pminus0 & \pminus0 & \pminus0 & \pminus0 & \pminus0 & \pminus0 & \pminus0 \\
 0 & \pminus0 & \pminus0 & \pminus0 & \pminus0 & \pminus0 & \pminus0 & \pminus0 & \pminus0 & \pminus0 & \pminus0 & \pminus0 & \pminus0 & \pminus0 & \pminus0 & \pminus0 & \pminus0 & \pminus0 & \pminus0 & \pminus0 & \pminus0 & \pminus0 & \pminus0 & \pminus0 & \pminus0 & \pminus0 & \pminus0 & \pminus0 & \pminus0 \\
\end{array}
\right)
  }\,,
\end{align}

\begin{align}
\MM_5 = \scalemath{0.53}{
\left(
\begin{array}{ccccccccccccccccccccccccccccc}
 0 & \pminus 0 & \pminus 0 & \pminus 0 & \pminus 0 & \pminus 0 & \pminus 0 & \pminus 0 & \pminus 0 & \pminus 0 & \pminus 0 & \pminus 0 & \pminus 0 & \pminus 0 & \pminus 0 & \pminus 0 & \pminus 0 & \pminus 0 & \pminus 0 & \pminus 0 & \pminus 0 & \pminus 0 & \pminus 0 & \pminus 0 & \pminus 0 & \pminus 0 & \pminus 0 & \pminus 0 & \pminus 0 \\
 0 & \pminus 1 & \pminus 0 & \pminus 0 & \pminus 0 & \pminus 0 & \pminus 0 & \pminus 0 & \pminus 0 & \pminus 0 & \pminus 0 & \pminus 0 & \pminus 0 & \pminus 0 & \pminus 0 & \pminus 0 & \pminus 0 & \pminus 0 & \pminus 0 & \pminus 0 & \pminus 0 & \pminus 0 & \pminus 0 & \pminus 0 & \pminus 0 & \pminus 0 & \pminus 0 & \pminus 0 & \pminus 0 \\
 0 & \pminus 0 & \pminus 0 & \pminus 0 & \pminus 0 & \pminus 0 & \pminus 0 & \pminus 0 & \pminus 0 & \pminus 0 & \pminus 0 & \pminus 0 & \pminus 0 & \pminus 0 & \pminus 0 & \pminus 0 & \pminus 0 & \pminus 0 & \pminus 0 & \pminus 0 & \pminus 0 & \pminus 0 & \pminus 0 & \pminus 0 & \pminus 0 & \pminus 0 & \pminus 0 & \pminus 0 & \pminus 0 \\
 0 & \pminus 0 & \pminus 0 & \pminus 0 & \pminus 0 & \pminus 0 & \pminus 0 & \pminus 0 & \pminus 0 & \pminus 0 & \pminus 0 & \pminus 0 & \pminus 0 & \pminus 0 & \pminus 0 & \pminus 0 & \pminus 0 & \pminus 0 & \pminus 0 & \pminus 0 & \pminus 0 & \pminus 0 & \pminus 0 & \pminus 0 & \pminus 0 & \pminus 0 & \pminus 0 & \pminus 0 & \pminus 0 \\
 0 & \pminus 0 & \pminus 0 & \pminus 0 & \pminus 1 & \pminus 0 & \pminus 0 & \pminus 0 & \pminus 0 & \pminus 0 & \pminus 0 & \pminus 0 & \pminus 0 & \pminus 0 & \pminus 0 & \pminus 0 & \pminus 0 & \pminus 0 & \pminus 0 & \pminus 0 & \pminus 0 & \pminus 0 & \pminus 0 & \pminus 0 & \pminus 0 & \pminus 0 & \pminus 0 & \pminus 0 & \pminus 0 \\
\end{array}
\right)
  }\,,
\end{align}

\begin{align}
\MM_6 = \scalemath{0.53}{
\left(
\begin{array}{ccccccccccccccccccccccccccccc}
 0 & \pminus 0 & \pminus 0 & \pminus 0 & \pminus 0 & \pminus 0 & \pminus 0 & \pminus 0 & \pminus 0 & \pminus 0 & \pminus 0 & \pminus 0 & \pminus 0 & \pminus 0 & \pminus 0 & \pminus 0 & \pminus 0 & \pminus 0 & \pminus 0 & \pminus 0 & \pminus 0 & \pminus 0 & \pminus 0 & \pminus 0 & \pminus 0 & \pminus 0 & \pminus 0 & \pminus 0 & \pminus 0 \\
 0 & \pminus 0 & \pminus 0 & \pminus 0 & \pminus 0 & \pminus 0 & \pminus 0 & \pminus 0 & \pminus 0 & \pminus 0 & \pminus 0 & \pminus 0 & \pminus 0 & \pminus 0 & \pminus 0 & \pminus 0 & \pminus 0 & \pminus 0 & \pminus 0 & \pminus 0 & \pminus 0 & \pminus 0 & \pminus 0 & \pminus 0 & \pminus 0 & \pminus 0 & \pminus 0 & \pminus 0 & \pminus 0 \\
 0 & \pminus 0 & \pminus 0 & \pminus 0 & \pminus 0 & \pminus 0 & \pminus 0 & \pminus 0 & \pminus 0 & \pminus 0 & \pminus 0 & \pminus 0 & \pminus 0 & \pminus 0 & \pminus 0 & \pminus 0 & \pminus 0 & \pminus 0 & \pminus 0 & \pminus 0 & \pminus 0 & \pminus 0 & \pminus 0 & \pminus 0 & \pminus 0 & \pminus 0 & \pminus 0 & \pminus 0 & \pminus 0 \\
 0 & \pminus 0 & \pminus 0 & \pminus 0 & \pminus 0 & \pminus 0 & \pminus 0 & \pminus 0 & \pminus 0 & \pminus 0 & \pminus 0 & \pminus 0 & \pminus 0 & \pminus 0 & \pminus 0 & \pminus 0 & \pminus 0 & \pminus 0 & \pminus 0 & \pminus 0 & \pminus 0 & \pminus 0 & \pminus 0 & \pminus 0 & \pminus 0 & \pminus 0 & \pminus 0 & \pminus 0 & \pminus 0 \\
 0 & \pminus 0 & \pminus 0 & \pminus 0 & \pminus 0 & \pminus 0 & \pminus 0 & \pminus 0 & \pminus 0 & \pminus 0 & \pminus 0 & \pminus 0 & \pminus 0 & \pminus 0 & \pminus 0 & \pminus 0 & \pminus 0 & \pminus 0 & \pminus 0 & \pminus 0 & \pminus 0 & \pminus 0 & \pminus 0 & \pminus 0 & \pminus 0 & \pminus 0 & \pminus 0 & \pminus 0 & \pminus 0 \\
\end{array}
\right)
  }\,,
\end{align}

\begin{align}
\MM_7 = \scalemath{0.53}{
\left(
\begin{array}{ccccccccccccccccccccccccccccc}
 0 & \pminus 0 & \pminus 0 & \pminus 0 & \pminus 0 & \pminus 0 & \pminus 0 & \pminus 0 & \pminus 0 & \pminus 0 & \pminus 0 & \pminus 0 & \pminus 0 & \pminus 0 & \pminus 0 & \pminus 0 & \pminus 0 & \pminus 0 & \pminus 0 & \pminus 0 & \pminus 0 & \pminus 0 & \pminus 0 & \pminus 0 & \pminus 0 & \pminus 0 & \pminus 0 & \pminus 0 & \pminus 0 \\
 0 & \pminus 0 & \pminus 0 & \pminus 0 & \pminus 0 & \pminus 0 & \pminus 0 & \pminus 0 & \pminus 0 & \pminus 0 & \pminus 0 & \pminus 0 & \pminus 0 & \pminus 0 & \pminus 0 & \pminus 0 & \pminus 0 & \pminus 0 & \pminus 0 & \pminus 0 & \pminus 0 & \pminus 0 & \pminus 0 & \pminus 0 & \pminus 0 & \pminus 0 & \pminus 0 & \pminus 0 & \pminus 0 \\
 0 & \pminus 0 & \pminus 0 & \pminus 0 & \pminus 0 & \pminus 0 & \pminus 0 & \pminus 0 & \pminus 0 & \pminus 0 & \pminus 0 & \pminus 0 & \pminus 0 & \pminus 0 & \pminus 0 & \pminus 0 & \pminus 0 & \pminus 0 & \pminus 0 & \pminus 0 & \pminus 0 & \pminus 0 & \pminus 0 & \pminus 0 & \pminus 0 & \pminus 0 & \pminus 0 & \pminus 0 & \pminus 0 \\
 -1 & \pminus 0 & \pminus 0 & -2 & \pminus 0 & \pminus 0 & \pminus 0 & \pminus 0 & \pminus 0 & \pminus 0 & \pminus 0 & \pminus 0 & \pminus 0 & \pminus 0 & \pminus 0 & \pminus 0 & \pminus 0 & \pminus 0 & \pminus 0 & \pminus 0 & \pminus 0 & \pminus 0 & \pminus 0 & \pminus 0 & \pminus 0 & \pminus 0 & \pminus 0 & \pminus 0 & \pminus 0 \\
 0 & \pminus 0 & \pminus 0 & \pminus 0 & \pminus 0 & \pminus 0 & \pminus 0 & \pminus 0 & \pminus 0 & \pminus 0 & \pminus 0 & \pminus 0 & \pminus 0 & \pminus 0 & \pminus 0 & \pminus 0 & \pminus 0 & \pminus 0 & \pminus 0 & \pminus 0 & \pminus 0 & \pminus 0 & \pminus 0 & \pminus 0 & \pminus 0 & \pminus 0 & \pminus 0 & \pminus 0 & \pminus 0 \\
\end{array}
\right)
}\,,
\end{align}

\begin{align}
\MM_8 = \scalemath{0.53}{
\left(
\begin{array}{ccccccccccccccccccccccccccccc}
 0 & \pminus 0 & \pminus 0 & \pminus 0 & \pminus 0 & \pminus 0 & \pminus 0 & \pminus 0 & \pminus 0 & \pminus 0 & \pminus 0 & \pminus 0 & \pminus 0 & \pminus 0 & \pminus 0 & \pminus 0 & \pminus 0 & \pminus 0 & \pminus 0 & \pminus 0 & \pminus 0 & \pminus 0 & \pminus 0 & \pminus 0 & \pminus 0 & \pminus 0 & \pminus 0 & \pminus 0 & \pminus 0 \\
 0 & \pminus 0 & \pminus 0 & \pminus 0 & \pminus 0 & \pminus 0 & \pminus 0 & \pminus 0 & \pminus 0 & \pminus 0 & \pminus 0 & \pminus 0 & \pminus 0 & \pminus 0 & \pminus 0 & \pminus 0 & \pminus 0 & \pminus 0 & \pminus 0 & \pminus 0 & \pminus 0 & \pminus 0 & \pminus 0 & \pminus 0 & \pminus 0 & \pminus 0 & \pminus 0 & \pminus 0 & \pminus 0 \\
 0 & \pminus 0 & \pminus 0 & \pminus 0 & \pminus 0 & \pminus 0 & \pminus 0 & \pminus 0 & \pminus 0 & \pminus 0 & \pminus 0 & \pminus 0 & \pminus 0 & \pminus 0 & \pminus 0 & \pminus 0 & \pminus 0 & \pminus 0 & \pminus 0 & \pminus 0 & \pminus 0 & \pminus 0 & \pminus 0 & \pminus 0 & \pminus 0 & \pminus 0 & \pminus 0 & \pminus 0 & \pminus 0 \\
 0 & \pminus 0 & \pminus 0 & \pminus 0 & \pminus 0 & \pminus 0 & \pminus 0 & \pminus 0 & \pminus 0 & \pminus 0 & \pminus 0 & \pminus 0 & \pminus 0 & \pminus 0 & \pminus 0 & \pminus 0 & \pminus 0 & \pminus 0 & \pminus 0 & \pminus 0 & \pminus 0 & \pminus 0 & \pminus 0 & \pminus 0 & \pminus 0 & \pminus 0 & \pminus 0 & \pminus 0 & \pminus 0 \\
 0 & \pminus 0 & \pminus 0 & \pminus 0 & \pminus 0 & \pminus 0 & \pminus 0 & \pminus 0 & \pminus 0 & \pminus 0 & \pminus 0 & \pminus 0 & \pminus 0 & \pminus 0 & \pminus 0 & \pminus 0 & \pminus 0 & \pminus 0 & \pminus 0 & \pminus 0 & \pminus 0 & \pminus 0 & \pminus 0 & \pminus 0 & \pminus 0 & \pminus 0 & \pminus 0 & \pminus 0 & \pminus 0 \\
\end{array}
\right)
}\,,
\end{align}

\begin{align}
\MM_9 = \scalemath{0.53}{
\left(
\begin{array}{ccccccccccccccccccccccccccccc}
 0 & \pminus 0 & \pminus 0 & \pminus 0 & \pminus 0 & \pminus 0 & \pminus 0 & \pminus 0 & \pminus 0 & \pminus 0 & \pminus 0 & \pminus 0 & \pminus 0 & \pminus 0 & \pminus 0 & \pminus 0 & \pminus 0 & \pminus 0 & \pminus 0 & \pminus 0 & \pminus 0 & \pminus 0 & \pminus 0 & \pminus 0 & \pminus 0 & \pminus 0 & \pminus 0 & \pminus 0 & \pminus 0 \\
 0 & \pminus 0 & \pminus 0 & \pminus 0 & \pminus 0 & \pminus 0 & \pminus 0 & \pminus 0 & \pminus 0 & \pminus 0 & \pminus 0 & \pminus 0 & \pminus 0 & \pminus 0 & \pminus 0 & \pminus 0 & \pminus 0 & \pminus 0 & \pminus 0 & \pminus 0 & \pminus 0 & \pminus 0 & \pminus 0 & \pminus 0 & \pminus 0 & \pminus 0 & \pminus 0 & \pminus 0 & \pminus 0 \\
 0 & \pminus 0 & \pminus 0 & \pminus 0 & \pminus 0 & \pminus 0 & \pminus 0 & \pminus 0 & \pminus 0 & \pminus 0 & \pminus 0 & \pminus 0 & \pminus 0 & \pminus 0 & \pminus 0 & \pminus 0 & \pminus 0 & \pminus 0 & \pminus 0 & \pminus 0 & \pminus 0 & \pminus 0 & \pminus 0 & \pminus 0 & \pminus 0 & \pminus 0 & \pminus 0 & \pminus 0 & \pminus 0 \\
 0 & \pminus 0 & \pminus 0 & \pminus 0 & \pminus 0 & \pminus 0 & \pminus 0 & \pminus 0 & \pminus 0 & \pminus 0 & \pminus 0 & \pminus 0 & \pminus 0 & \pminus 0 & \pminus 0 & \pminus 0 & \pminus 0 & \pminus 0 & \pminus 0 & \pminus 0 & \pminus 0 & \pminus 0 & \pminus 0 & \pminus 0 & \pminus 0 & \pminus 0 & \pminus 0 & \pminus 0 & \pminus 0 \\
 0 & \pminus 0 & \pminus 0 & \pminus 0 & \pminus 0 & \pminus 0 & \pminus 0 & \pminus 0 & \pminus 0 & \pminus 0 & \pminus 0 & \pminus 0 & \pminus 0 & \pminus 0 & \pminus 0 & \pminus 0 & \pminus 0 & \pminus 0 & \pminus 0 & \pminus 0 & \pminus 0 & \pminus 0 & \pminus 0 & \pminus 0 & \pminus 0 & \pminus 0 & \pminus 0 & \pminus 0 & \pminus 0 \\
\end{array}
\right)
}\,,
\end{align}

\begin{align}
\MM_{10} = \scalemath{0.53}{
\left(
\begin{array}{ccccccccccccccccccccccccccccc}
 0 & \pminus 0 & \pminus 0 & \pminus 0 & \pminus 0 & \pminus 0 & \pminus 0 & \pminus 0 & \pminus 0 & \pminus 0 & \pminus 0 & \pminus 0 & \pminus 0 & \pminus 0 & \pminus 0 & \pminus 0 & \pminus 0 & \pminus 0 & \pminus 0 & \pminus 0 & \pminus 0 & \pminus 0 & \pminus 0 & \pminus 0 & \pminus 0 & \pminus 0 & \pminus 0 & \pminus 0 & \pminus 0 \\
 -1 & -2 & \pminus 0 & \pminus 0 & \pminus 0 & \pminus 0 & \pminus 0 & \pminus 0 & \pminus 0 & \pminus 0 & \pminus 0 & \pminus 0 & \pminus 0 & \pminus 0 & \pminus 0 & \pminus 0 & \pminus 0 & \pminus 0 & \pminus 0 & \pminus 0 & \pminus 0 & \pminus 0 & \pminus 0 & \pminus 0 & \pminus 0 & \pminus 0 & \pminus 0 & \pminus 0 & \pminus 0 \\
 0 & \pminus 0 & \pminus 0 & \pminus 0 & \pminus 0 & \pminus 0 & \pminus 0 & \pminus 0 & \pminus 0 & \pminus 0 & \pminus 0 & \pminus 0 & \pminus 0 & \pminus 0 & \pminus 0 & \pminus 0 & \pminus 0 & \pminus 0 & \pminus 0 & \pminus 0 & \pminus 0 & \pminus 0 & \pminus 0 & \pminus 0 & \pminus 0 & \pminus 0 & \pminus 0 & \pminus 0 & \pminus 0 \\
 0 & \pminus 0 & \pminus 0 & \pminus 0 & \pminus 0 & \pminus 0 & \pminus 0 & \pminus 0 & \pminus 0 & \pminus 0 & \pminus 0 & \pminus 0 & \pminus 0 & \pminus 0 & \pminus 0 & \pminus 0 & \pminus 0 & \pminus 0 & \pminus 0 & \pminus 0 & \pminus 0 & \pminus 0 & \pminus 0 & \pminus 0 & \pminus 0 & \pminus 0 & \pminus 0 & \pminus 0 & \pminus 0 \\
 0 & \pminus 0 & \pminus 0 & \pminus 0 & -2 & -1 & \pminus 0 & \pminus 0 & \pminus 0 & \pminus 0 & \pminus 0 & \pminus 0 & \pminus 0 & \pminus 0 & \pminus 0 & \pminus 0 & \pminus 0 & \pminus 0 & \pminus 0 & \pminus 0 & \pminus 0 & \pminus 0 & \pminus 0 & \pminus 0 & \pminus 0 & \pminus 0 & \pminus 0 & \pminus 0 & \pminus 0 \\
\end{array}
\right)
}\,.
\end{align}
\subsection{Topologies (c)-(d)}
\label{dlogWW2Lt}
For the two-loop integrals discussed in section~\ref{sec:WWMIst} we have 
\begin{align}
\dA = {} & \MM_1 \, \dlog(v) + \MM_2 \, \dlog(1-v) + \MM_3 \, \dlog(1+v) + \MM_4 \, \dlog(\zz)\nn
 &+ \MM_5 \, \dlog(1-\zz)  + \MM_6\, \dlog(\zzb)+ \MM_7\, \dlog(1-\zzb)\nn
  &+ \MM_8 \, \dlog(\zz-\zzb)+ \MM_9\, \dlog\left(\zz+v(1-\zz)\right) + \MM_{10} \, \dlog\left(1-\zz(1-v)\right)\nn
  &+ \MM_{11}\, \dlog\left(\zzb+v(1-\zzb)\right)+ \MM_{12} \, \dlog\left(1-\zzb(1-v)\right)\nn
  &+ \MM_{13} \, \dlog\left(v+\zz\zzb(1-v)^2\right)+ \MM_{14}\, \dlog\left(v+(1-\zz-\zzb+\zz\zzb)(1-v)^2\right)\nn
  &+ \MM_{15} \, \dlog\left(v+\zz(1-v)^2\right)  + \MM_{16}\, \dlog\left(v+(1-\zz)(1-v)^2\right)\nn
  &+ \MM_{17}\, \dlog\left(v+\zzb(1-v)^2\right) + \MM_{18} \, \dlog\left(v+(1-\zzb)(1-v)^2\right)\,,
\end{align}
with 

\begin{align}
\MM_1 = \scalemath{0.53}{
\left(
\begin{array}{ccccccccccccccccccccccccccccccc}
 0 & \pminus 0 & \pminus 0 & \pminus 0 & \pminus 0 & \pminus 0 & \pminus 0 & \pminus 0 & \pminus 0 & \pminus 0 & \pminus 0 & \pminus 0 & \pminus 0 & \pminus 0 & \pminus 0 & \pminus 0 & \pminus 0 & \pminus 0 & \pminus 0 & \pminus
   0 & \pminus 0 & \pminus 0 & \pminus 0 & \pminus 0 & \pminus 0 & \pminus 0 & \pminus 0 & \pminus 0 & \pminus 0 & \pminus 0 & \pminus 0 \\
 1 & \pminus 1 & \pminus 0 & \pminus 0 & \pminus 0 & \pminus 0 & \pminus 0 & \pminus 0 & \pminus 0 & \pminus 0 & \pminus 0 & \pminus 0 & \pminus 0 & \pminus 0 & \pminus 0 & \pminus 0 & \pminus 0 & \pminus 0 & \pminus 0 & \pminus
   0 & \pminus 0 & \pminus 0 & \pminus 0 & \pminus 0 & \pminus 0 & \pminus 0 & \pminus 0 & \pminus 0 & \pminus 0 & \pminus 0 & \pminus 0 \\
 1 & \pminus 0 & \pminus 1 & \pminus 0 & \pminus 0 & \pminus 0 & \pminus 0 & \pminus 0 & \pminus 0 & \pminus 0 & \pminus 0 & \pminus 0 & \pminus 0 & \pminus 0 & \pminus 0 & \pminus 0 & \pminus 0 & \pminus 0 & \pminus 0 & \pminus
   0 & \pminus 0 & \pminus 0 & \pminus 0 & \pminus 0 & \pminus 0 & \pminus 0 & \pminus 0 & \pminus 0 & \pminus 0 & \pminus 0 & \pminus 0 \\
 1 & \pminus 0 & \pminus 0 & \pminus 1 & \pminus 0 & \pminus 0 & \pminus 0 & \pminus 0 & \pminus 0 & \pminus 0 & \pminus 0 & \pminus 0 & \pminus 0 & \pminus 0 & \pminus 0 & \pminus 0 & \pminus 0 & \pminus 0 & \pminus 0 & \pminus
   0 & \pminus 0 & \pminus 0 & \pminus 0 & \pminus 0 & \pminus 0 & \pminus 0 & \pminus 0 & \pminus 0 & \pminus 0 & \pminus 0 & \pminus 0 \\
 0 & \pminus 0 & \pminus 0 & \pminus 0 & \pminus 2 & \pminus 0 & \pminus 0 & \pminus 0 & \pminus 0 & \pminus 0 & \pminus 0 & \pminus 0 & \pminus 0 & \pminus 0 & \pminus 0 & \pminus 0 & \pminus 0 & \pminus 0 & \pminus 0 & \pminus
   0 & \pminus 0 & \pminus 0 & \pminus 0 & \pminus 0 & \pminus 0 & \pminus 0 & \pminus 0 & \pminus 0 & \pminus 0 & \pminus 0 & \pminus 0 \\
\end{array}
\right)
  }\,,
\end{align}

\begin{align}
\MM_2 = \scalemath{0.53}{
\left(
\begin{array}{ccccccccccccccccccccccccccccccc}
 0 & \pminus 0 & \pminus 0 & \pminus 0 & \pminus 0 & \pminus 0 & \pminus 0 & \pminus 0 & \pminus 0 & \pminus 0 & \pminus 0 & \pminus 0 & \pminus 0 & \pminus 0 & \pminus 0 & \pminus 0 & \pminus 0 & \pminus 0 & \pminus 0 & \pminus
   0 & \pminus 0 & \pminus 0 & \pminus 0 & \pminus 0 & \pminus 0 & \pminus 0 & \pminus 0 & \pminus 0 & \pminus 0 & \pminus 0 & \pminus 0 \\
 0 & \pminus 2 & \pminus 0 & \pminus 0 & \pminus 0 & \pminus 0 & \pminus 0 & \pminus 0 & \pminus 0 & \pminus 0 & \pminus 0 & \pminus 0 & \pminus 0 & \pminus 0 & \pminus 0 & \pminus 0 & \pminus 0 & \pminus 0 & \pminus 0 & \pminus
   0 & \pminus 0 & \pminus 0 & \pminus 0 & \pminus 0 & \pminus 0 & \pminus 0 & \pminus 0 & \pminus 0 & \pminus 0 & \pminus 0 & \pminus 0 \\
 0 & \pminus 0 & \pminus 0 & \pminus 0 & \pminus 0 & \pminus 0 & \pminus 0 & \pminus 0 & \pminus 0 & \pminus 0 & \pminus 0 & \pminus 0 & \pminus 0 & \pminus 0 & \pminus 0 & \pminus 0 & \pminus 0 & \pminus 0 & \pminus 0 & \pminus
   0 & \pminus 0 & \pminus 0 & \pminus 0 & \pminus 0 & \pminus 0 & \pminus 0 & \pminus 0 & \pminus 0 & \pminus 0 & \pminus 0 & \pminus 0 \\
 0 & \pminus 0 & \pminus 0 & \pminus 2 & \pminus 0 & \pminus 0 & \pminus 0 & \pminus 0 & \pminus 0 & \pminus 0 & \pminus 0 & \pminus 0 & \pminus 0 & \pminus 0 & \pminus 0 & \pminus 0 & \pminus 0 & \pminus 0 & \pminus 0 & \pminus
   0 & \pminus 0 & \pminus 0 & \pminus 0 & \pminus 0 & \pminus 0 & \pminus 0 & \pminus 0 & \pminus 0 & \pminus 0 & \pminus 0 & \pminus 0 \\
 0 & \pminus 0 & \pminus 0 & \pminus 0 & \pminus 0 & \pminus 8 & \pminus 0 & \pminus 0 & \pminus 0 & \pminus 0 & \pminus 0 & \pminus 0 & \pminus 0 & \pminus 0 & \pminus 0 & \pminus 0 & \pminus 0 & \pminus 0 & \pminus 0 & \pminus
   0 & \pminus 0 & \pminus 0 & \pminus 0 & \pminus 0 & \pminus 0 & \pminus 0 & \pminus 0 & \pminus 0 & \pminus 0 & \pminus 0 & \pminus 0 \\
\end{array}
\right)
  }\,,
\end{align}

\begin{align}
\MM_3 = \scalemath{0.53}{
\left(
\begin{array}{ccccccccccccccccccccccccccccccc}
 0 & \pminus 0 & \pminus 0 & \pminus 0 & \pminus 0 & \pminus 0 & \pminus 0 & \pminus 0 & \pminus 0 & \pminus 0 & \pminus 0 & \pminus 0 & \pminus 0 & \pminus 0 & \pminus 0 & \pminus 0 & \pminus 0 & \pminus 0 & \pminus 0 & \pminus
   0 & \pminus 0 & \pminus 0 & \pminus 0 & \pminus 0 & \pminus 0 & \pminus 0 & \pminus 0 & \pminus 0 & \pminus 0 & \pminus 0 & \pminus 0 \\
 0 & \pminus 0 & \pminus 0 & \pminus 0 & \pminus 0 & \pminus 0 & \pminus 0 & \pminus 0 & \pminus 0 & \pminus 0 & \pminus 0 & \pminus 0 & \pminus 0 & \pminus 0 & \pminus 0 & \pminus 0 & \pminus 0 & \pminus 0 & \pminus 0 & \pminus
   0 & \pminus 0 & \pminus 0 & \pminus 0 & \pminus 0 & \pminus 0 & \pminus 0 & \pminus 0 & \pminus 0 & \pminus 0 & \pminus 0 & \pminus 0 \\
 0 & \pminus 0 & -2 & \pminus 0 & \pminus 0 & \pminus 0 & \pminus 0 & \pminus 0 & \pminus 0 & \pminus 0 & \pminus 0 & \pminus 0 & \pminus 0 & \pminus 0 & \pminus 0 & \pminus 0 & \pminus 0 & \pminus 0 & \pminus 0 & \pminus
   0 & \pminus 0 & \pminus 0 & \pminus 0 & \pminus 0 & \pminus 0 & \pminus 0 & \pminus 0 & \pminus 0 & \pminus 0 & \pminus 0 & \pminus 0 \\
 0 & \pminus 0 & \pminus 0 & \pminus 0 & \pminus 0 & \pminus 0 & \pminus 0 & \pminus 0 & \pminus 0 & \pminus 0 & \pminus 0 & \pminus 0 & \pminus 0 & \pminus 0 & \pminus 0 & \pminus 0 & \pminus 0 & \pminus 0 & \pminus 0 & \pminus
   0 & \pminus 0 & \pminus 0 & \pminus 0 & \pminus 0 & \pminus 0 & \pminus 0 & \pminus 0 & \pminus 0 & \pminus 0 & \pminus 0 & \pminus 0 \\
 0 & \pminus 0 & \pminus 0 & \pminus 0 & \pminus 0 & \pminus 0 & \pminus 0 & \pminus 0 & \pminus 0 & \pminus 0 & \pminus 0 & \pminus 0 & \pminus 0 & \pminus 0 & \pminus 0 & \pminus 0 & \pminus 0 & \pminus 0 & \pminus 0 & \pminus
   0 & \pminus 0 & \pminus 0 & \pminus 0 & \pminus 0 & \pminus 0 & \pminus 0 & \pminus 0 & \pminus 0 & \pminus 0 & \pminus 0 & \pminus 0 \\
\end{array}
\right)
  }\,,
\end{align}

\begin{align}
\MM_4 = \scalemath{0.53}{
\left(
\begin{array}{ccccccccccccccccccccccccccccccc}
 0 & \pminus 0 & \pminus 0 & \pminus 0 & \pminus 0 & \pminus 0 & \pminus 0 & \pminus 0 & \pminus 0 & \pminus 0 & \pminus 0 & \pminus 0 & \pminus 0 & \pminus 0 & \pminus 0 & \pminus 0 & \pminus 0 & \pminus 0 & \pminus 0 & \pminus
   0 & \pminus 0 & \pminus 0 & \pminus 0 & \pminus 0 & \pminus 0 & \pminus 0 & \pminus 0 & \pminus 0 & \pminus 0 & \pminus 0 & \pminus 0 \\
 0 & \pminus 0 & \pminus 0 & \pminus 0 & \pminus 0 & \pminus 0 & \pminus 0 & \pminus 0 & \pminus 0 & \pminus 0 & \pminus 0 & \pminus 0 & \pminus 0 & \pminus 0 & \pminus 0 & \pminus 0 & \pminus 0 & \pminus 0 & \pminus 0 & \pminus
   0 & \pminus 0 & \pminus 0 & \pminus 0 & \pminus 0 & \pminus 0 & \pminus 0 & \pminus 0 & \pminus 0 & \pminus 0 & \pminus 0 & \pminus 0 \\
 0 & \pminus 0 & \pminus 0 & \pminus 0 & \pminus 0 & \pminus 0 & \pminus 0 & \pminus 0 & \pminus 0 & \pminus 0 & \pminus 0 & \pminus 0 & \pminus 0 & \pminus 0 & \pminus 0 & \pminus 0 & \pminus 0 & \pminus 0 & \pminus 0 & \pminus
   0 & \pminus 0 & \pminus 0 & \pminus 0 & \pminus 0 & \pminus 0 & \pminus 0 & \pminus 0 & \pminus 0 & \pminus 0 & \pminus 0 & \pminus 0 \\
 0 & \pminus 0 & \pminus 0 & \pminus 1 & \pminus 0 & \pminus 0 & \pminus 0 & \pminus 0 & \pminus 0 & \pminus 0 & \pminus 0 & \pminus 0 & \pminus 0 & \pminus 0 & \pminus 0 & \pminus 0 & \pminus 0 & \pminus 0 & \pminus 0 & \pminus
   0 & \pminus 0 & \pminus 0 & \pminus 0 & \pminus 0 & \pminus 0 & \pminus 0 & \pminus 0 & \pminus 0 & \pminus 0 & \pminus 0 & \pminus 0 \\
 0 & \pminus 0 & \pminus 0 & \pminus 0 & \pminus 0 & \pminus 0 & \pminus 0 & \pminus 0 & \pminus 0 & \pminus 0 & \pminus 0 & \pminus 0 & \pminus 0 & \pminus 0 & \pminus 0 & \pminus 0 & \pminus 0 & \pminus 0 & \pminus 0 & \pminus
   0 & \pminus 0 & \pminus 0 & \pminus 0 & \pminus 0 & \pminus 0 & \pminus 0 & \pminus 0 & \pminus 0 & \pminus 0 & \pminus 0 & \pminus 0 \\
\end{array}
\right)
  }\,,
\end{align}

\begin{align}
\MM_5 = \scalemath{0.53}{
\left(
\begin{array}{ccccccccccccccccccccccccccccccc}
 0 & \pminus 0 & \pminus 0 & \pminus 0 & \pminus 0 & \pminus 0 & \pminus 0 & \pminus 0 & \pminus 0 & \pminus 0 & \pminus 0 & \pminus 0 & \pminus 0 & \pminus 0 & \pminus 0 & \pminus 0 & \pminus 0 & \pminus 0 & \pminus 0 & \pminus
   0 & \pminus 0 & \pminus 0 & \pminus 0 & \pminus 0 & \pminus 0 & \pminus 0 & \pminus 0 & \pminus 0 & \pminus 0 & \pminus 0 & \pminus 0 \\
 0 & \pminus 1 & \pminus 0 & \pminus 0 & \pminus 0 & \pminus 0 & \pminus 0 & \pminus 0 & \pminus 0 & \pminus 0 & \pminus 0 & \pminus 0 & \pminus 0 & \pminus 0 & \pminus 0 & \pminus 0 & \pminus 0 & \pminus 0 & \pminus 0 & \pminus
   0 & \pminus 0 & \pminus 0 & \pminus 0 & \pminus 0 & \pminus 0 & \pminus 0 & \pminus 0 & \pminus 0 & \pminus 0 & \pminus 0 & \pminus 0 \\
 0 & \pminus 0 & \pminus 0 & \pminus 0 & \pminus 0 & \pminus 0 & \pminus 0 & \pminus 0 & \pminus 0 & \pminus 0 & \pminus 0 & \pminus 0 & \pminus 0 & \pminus 0 & \pminus 0 & \pminus 0 & \pminus 0 & \pminus 0 & \pminus 0 & \pminus
   0 & \pminus 0 & \pminus 0 & \pminus 0 & \pminus 0 & \pminus 0 & \pminus 0 & \pminus 0 & \pminus 0 & \pminus 0 & \pminus 0 & \pminus 0 \\
 0 & \pminus 0 & \pminus 0 & \pminus 0 & \pminus 0 & \pminus 0 & \pminus 0 & \pminus 0 & \pminus 0 & \pminus 0 & \pminus 0 & \pminus 0 & \pminus 0 & \pminus 0 & \pminus 0 & \pminus 0 & \pminus 0 & \pminus 0 & \pminus 0 & \pminus
   0 & \pminus 0 & \pminus 0 & \pminus 0 & \pminus 0 & \pminus 0 & \pminus 0 & \pminus 0 & \pminus 0 & \pminus 0 & \pminus 0 & \pminus 0 \\
 0 & \pminus 0 & \pminus 0 & \pminus 0 & \pminus 0 & \pminus 4 & \pminus 0 & \pminus 0 & \pminus 0 & \pminus 0 & \pminus 0 & \pminus 0 & \pminus 0 & \pminus 0 & \pminus 0 & \pminus 0 & \pminus 0 & \pminus 0 & \pminus 0 & \pminus
   0 & \pminus 0 & \pminus 0 & \pminus 0 & \pminus 0 & \pminus 0 & \pminus 0 & \pminus 0 & \pminus 0 & \pminus 0 & \pminus 0 & \pminus 0 \\
\end{array}
\right)
  }\,,
\end{align}

\begin{align}
\MM_6 = \scalemath{0.53}{
\left(
\begin{array}{ccccccccccccccccccccccccccccccc}
 0 & \pminus 0 & \pminus 0 & \pminus 0 & \pminus 0 & \pminus 0 & \pminus 0 & \pminus 0 & \pminus 0 & \pminus 0 & \pminus 0 & \pminus 0 & \pminus 0 & \pminus 0 & \pminus 0 & \pminus 0 & \pminus 0 & \pminus 0 & \pminus 0 & \pminus
   0 & \pminus 0 & \pminus 0 & \pminus 0 & \pminus 0 & \pminus 0 & \pminus 0 & \pminus 0 & \pminus 0 & \pminus 0 & \pminus 0 & \pminus 0 \\
 0 & \pminus 0 & \pminus 0 & \pminus 0 & \pminus 0 & \pminus 0 & \pminus 0 & \pminus 0 & \pminus 0 & \pminus 0 & \pminus 0 & \pminus 0 & \pminus 0 & \pminus 0 & \pminus 0 & \pminus 0 & \pminus 0 & \pminus 0 & \pminus 0 & \pminus
   0 & \pminus 0 & \pminus 0 & \pminus 0 & \pminus 0 & \pminus 0 & \pminus 0 & \pminus 0 & \pminus 0 & \pminus 0 & \pminus 0 & \pminus 0 \\
 0 & \pminus 0 & \pminus 0 & \pminus 0 & \pminus 0 & \pminus 0 & \pminus 0 & \pminus 0 & \pminus 0 & \pminus 0 & \pminus 0 & \pminus 0 & \pminus 0 & \pminus 0 & \pminus 0 & \pminus 0 & \pminus 0 & \pminus 0 & \pminus 0 & \pminus
   0 & \pminus 0 & \pminus 0 & \pminus 0 & \pminus 0 & \pminus 0 & \pminus 0 & \pminus 0 & \pminus 0 & \pminus 0 & \pminus 0 & \pminus 0 \\
 0 & \pminus 0 & \pminus 0 & \pminus 1 & \pminus 0 & \pminus 0 & \pminus 0 & \pminus 0 & \pminus 0 & \pminus 0 & \pminus 0 & \pminus 0 & \pminus 0 & \pminus 0 & \pminus 0 & \pminus 0 & \pminus 0 & \pminus 0 & \pminus 0 & \pminus
   0 & \pminus 0 & \pminus 0 & \pminus 0 & \pminus 0 & \pminus 0 & \pminus 0 & \pminus 0 & \pminus 0 & \pminus 0 & \pminus 0 & \pminus 0 \\
 0 & \pminus 0 & \pminus 0 & \pminus 0 & \pminus 0 & \pminus 0 & \pminus 0 & \pminus 0 & \pminus 0 & \pminus 0 & \pminus 0 & \pminus 0 & \pminus 0 & \pminus 0 & \pminus 0 & \pminus 0 & \pminus 0 & \pminus 0 & \pminus 0 & \pminus
   0 & \pminus 0 & \pminus 0 & \pminus 0 & \pminus 0 & \pminus 0 & \pminus 0 & \pminus 0 & \pminus 0 & \pminus 0 & \pminus 0 & \pminus 0 \\
\end{array}
\right)
  }\,,
\end{align}

\begin{align}
\MM_7 = \scalemath{0.53}{
\left(
\begin{array}{ccccccccccccccccccccccccccccccc}
 0 & \pminus 0 & \pminus 0 & \pminus 0 & \pminus 0 & \pminus 0 & \pminus 0 & \pminus 0 & \pminus 0 & \pminus 0 & \pminus 0 & \pminus 0 & \pminus 0 & \pminus 0 & \pminus 0 & \pminus 0 & \pminus 0 & \pminus 0 & \pminus 0 & \pminus
   0 & \pminus 0 & \pminus 0 & \pminus 0 & \pminus 0 & \pminus 0 & \pminus 0 & \pminus 0 & \pminus 0 & \pminus 0 & \pminus 0 & \pminus 0 \\
 0 & \pminus 1 & \pminus 0 & \pminus 0 & \pminus 0 & \pminus 0 & \pminus 0 & \pminus 0 & \pminus 0 & \pminus 0 & \pminus 0 & \pminus 0 & \pminus 0 & \pminus 0 & \pminus 0 & \pminus 0 & \pminus 0 & \pminus 0 & \pminus 0 & \pminus
   0 & \pminus 0 & \pminus 0 & \pminus 0 & \pminus 0 & \pminus 0 & \pminus 0 & \pminus 0 & \pminus 0 & \pminus 0 & \pminus 0 & \pminus 0 \\
 0 & \pminus 0 & \pminus 0 & \pminus 0 & \pminus 0 & \pminus 0 & \pminus 0 & \pminus 0 & \pminus 0 & \pminus 0 & \pminus 0 & \pminus 0 & \pminus 0 & \pminus 0 & \pminus 0 & \pminus 0 & \pminus 0 & \pminus 0 & \pminus 0 & \pminus
   0 & \pminus 0 & \pminus 0 & \pminus 0 & \pminus 0 & \pminus 0 & \pminus 0 & \pminus 0 & \pminus 0 & \pminus 0 & \pminus 0 & \pminus 0 \\
 0 & \pminus 0 & \pminus 0 & \pminus 0 & \pminus 0 & \pminus 0 & \pminus 0 & \pminus 0 & \pminus 0 & \pminus 0 & \pminus 0 & \pminus 0 & \pminus 0 & \pminus 0 & \pminus 0 & \pminus 0 & \pminus 0 & \pminus 0 & \pminus 0 & \pminus
   0 & \pminus 0 & \pminus 0 & \pminus 0 & \pminus 0 & \pminus 0 & \pminus 0 & \pminus 0 & \pminus 0 & \pminus 0 & \pminus 0 & \pminus 0 \\
 0 & \pminus 0 & \pminus 0 & \pminus 0 & \pminus 0 & \pminus 4 & \pminus 0 & \pminus 0 & \pminus 0 & \pminus 0 & \pminus 0 & \pminus 0 & \pminus 0 & \pminus 0 & \pminus 0 & \pminus 0 & \pminus 0 & \pminus 0 & \pminus 0 & \pminus
   0 & \pminus 0 & \pminus 0 & \pminus 0 & \pminus 0 & \pminus 0 & \pminus 0 & \pminus 0 & \pminus 0 & \pminus 0 & \pminus 0 & \pminus 0 \\
\end{array}
\right)
}\,,
\end{align}

\begin{align}
\MM_8 = \scalemath{0.53}{
\left(
\begin{array}{ccccccccccccccccccccccccccccccc}
 0 & \pminus0 & \pminus0 & \pminus0 & \pminus0 & \pminus0 & \pminus0 & \pminus0 & \pminus0 & \pminus0 & \pminus0 & \pminus0 & \pminus0 & \pminus0 & \pminus0 & \pminus0 & \pminus0 & \pminus0 & \pminus0 &
   0 & \pminus0 & \pminus0 & \pminus0 & \pminus0 & \pminus0 & \pminus0 & \pminus0 & \pminus0 & \pminus0 & \pminus0 & \pminus0 \\
 0 & \pminus0 & \pminus0 & \pminus0 & \pminus0 & \pminus0 & \pminus0 & \pminus0 & \pminus0 & \pminus0 & \pminus0 & \pminus0 & \pminus0 & \pminus0 & \pminus0 & \pminus0 & \pminus0 & \pminus0 & \pminus0 &
   0 & \pminus0 & \pminus0 & \pminus0 & \pminus0 & \pminus0 & \pminus0 & \pminus0 & \pminus0 & \pminus0 & \pminus0 & \pminus0 \\
 0 & \pminus0 & \pminus0 & \pminus0 & \pminus0 & \pminus0 & \pminus0 & \pminus0 & \pminus0 & \pminus0 & \pminus0 & \pminus0 & \pminus0 & \pminus0 & \pminus0 & \pminus0 & \pminus0 & \pminus0 & \pminus0 &
   0 & \pminus0 & \pminus0 & \pminus0 & \pminus0 & \pminus0 & \pminus0 & \pminus0 & \pminus0 & \pminus0 & \pminus0 & \pminus0 \\
 0 & \pminus0 & \pminus0 & \pminus0 & \pminus0 & \pminus0 & \pminus0 & \pminus0 & \pminus0 & \pminus0 & \pminus0 & \pminus0 & \pminus0 & \pminus0 & \pminus0 & \pminus0 & \pminus0 & \pminus0 & \pminus0 &
   0 & \pminus0 & \pminus0 & \pminus0 & \pminus0 & \pminus0 & \pminus0 & \pminus0 & \pminus0 & \pminus0 & \pminus0 & \pminus0 \\
 0 & \pminus0 & \pminus0 & \pminus0 & \pminus0 & \pminus0 & \pminus0 & \pminus0 & \pminus0 & \pminus0 & \pminus0 & \pminus0 & \pminus0 & \pminus0 & \pminus0 & \pminus0 & \pminus0 & \pminus0 & \pminus0 &
   0 & \pminus0 & \pminus0 & \pminus0 & \pminus0 & \pminus0 & \pminus0 & \pminus0 & \pminus0 & \pminus0 & \pminus0 & \pminus0 \\
\end{array}
\right)
}\,,
\end{align}

\begin{align}
\MM_9 = \scalemath{0.53}{
\left(
\begin{array}{ccccccccccccccccccccccccccccccc}
 0 & \pminus 0 & \pminus 0 & \pminus 0 & \pminus 0 & \pminus 0 & \pminus 0 & \pminus 0 & \pminus 0 & \pminus 0 & \pminus 0 & \pminus 0 & \pminus 0 & \pminus 0 & \pminus 0 & \pminus 0 & \pminus 0 & \pminus 0 & \pminus 0 & \pminus
   0 & \pminus 0 & \pminus 0 & \pminus 0 & \pminus 0 & \pminus 0 & \pminus 0 & \pminus 0 & \pminus 0 & \pminus 0 & \pminus 0 & \pminus 0 \\
 0 & \pminus 0 & \pminus 0 & \pminus 0 & \pminus 0 & \pminus 0 & \pminus 0 & \pminus 0 & \pminus 0 & \pminus 0 & \pminus 0 & \pminus 0 & \pminus 0 & \pminus 0 & \pminus 0 & \pminus 0 & \pminus 0 & \pminus 0 & \pminus 0 & \pminus
   0 & \pminus 0 & \pminus 0 & \pminus 0 & \pminus 0 & \pminus 0 & \pminus 0 & \pminus 0 & \pminus 0 & \pminus 0 & \pminus 0 & \pminus 0 \\
 0 & \pminus 0 & \pminus 0 & \pminus 0 & \pminus 0 & \pminus 0 & \pminus 0 & \pminus 0 & \pminus 0 & \pminus 0 & \pminus 0 & \pminus 0 & \pminus 0 & \pminus 0 & \pminus 0 & \pminus 0 & \pminus 0 & \pminus 0 & \pminus 0 & \pminus
   0 & \pminus 0 & \pminus 0 & \pminus 0 & \pminus 0 & \pminus 0 & \pminus 0 & \pminus 0 & \pminus 0 & \pminus 0 & \pminus 0 & \pminus 0 \\
 0 & \pminus 0 & \pminus 0 & \pminus 0 & \pminus 0 & \pminus 0 & \pminus 0 & \pminus 0 & \pminus 0 & \pminus 0 & \pminus 0 & \pminus 0 & \pminus 0 & \pminus 0 & \pminus 0 & \pminus 0 & \pminus 0 & \pminus 0 & \pminus 0 & \pminus
   0 & \pminus 0 & \pminus 0 & \pminus 0 & \pminus 0 & \pminus 0 & \pminus 0 & \pminus 0 & \pminus 0 & \pminus 0 & \pminus 0 & \pminus 0 \\
 0 & \pminus 0 & \pminus 0 & \pminus 0 & \pminus 0 & \pminus 0 & \pminus 0 & \pminus 0 & \pminus 0 & \pminus 0 & \pminus 0 & \pminus 0 & \pminus 0 & \pminus 0 & \pminus 0 & \pminus 0 & \pminus 0 & \pminus 0 & \pminus 0 & \pminus
   0 & \pminus 0 & \pminus 0 & \pminus 0 & \pminus 0 & \pminus 0 & \pminus 0 & \pminus 0 & \pminus 0 & \pminus 0 & \pminus 0 & \pminus 0 \\
\end{array}
\right)
}\,,
\end{align}

\begin{align}
\MM_{10} = \scalemath{0.53}{
\left(
\begin{array}{ccccccccccccccccccccccccccccccc}
 0 & \pminus 0 & \pminus 0 & \pminus 0 & \pminus 0 & \pminus 0 & \pminus 0 & \pminus 0 & \pminus 0 & \pminus 0 & \pminus 0 & \pminus 0 & \pminus 0 & \pminus 0 & \pminus 0 & \pminus 0 & \pminus 0 & \pminus 0 & \pminus 0 & \pminus
   0 & \pminus 0 & \pminus 0 & \pminus 0 & \pminus 0 & \pminus 0 & \pminus 0 & \pminus 0 & \pminus 0 & \pminus 0 & \pminus 0 & \pminus 0 \\
 0 & \pminus 0 & \pminus 0 & \pminus 0 & \pminus 0 & \pminus 0 & \pminus 0 & \pminus 0 & \pminus 0 & \pminus 0 & \pminus 0 & \pminus 0 & \pminus 0 & \pminus 0 & \pminus 0 & \pminus 0 & \pminus 0 & \pminus 0 & \pminus 0 & \pminus
   0 & \pminus 0 & \pminus 0 & \pminus 0 & \pminus 0 & \pminus 0 & \pminus 0 & \pminus 0 & \pminus 0 & \pminus 0 & \pminus 0 & \pminus 0 \\
 0 & \pminus 0 & \pminus 0 & \pminus 0 & \pminus 0 & \pminus 0 & \pminus 0 & \pminus 0 & \pminus 0 & \pminus 0 & \pminus 0 & \pminus 0 & \pminus 0 & \pminus 0 & \pminus 0 & \pminus 0 & \pminus 0 & \pminus 0 & \pminus 0 & \pminus
   0 & \pminus 0 & \pminus 0 & \pminus 0 & \pminus 0 & \pminus 0 & \pminus 0 & \pminus 0 & \pminus 0 & \pminus 0 & \pminus 0 & \pminus 0 \\
 0 & \pminus 0 & \pminus 0 & \pminus 0 & \pminus 0 & \pminus 0 & \pminus 0 & \pminus 0 & \pminus 0 & \pminus 0 & \pminus 0 & \pminus 0 & \pminus 0 & \pminus 0 & \pminus 0 & \pminus 0 & \pminus 0 & \pminus 0 & \pminus 0 & \pminus
   0 & \pminus 0 & \pminus 0 & \pminus 0 & \pminus 0 & \pminus 0 & \pminus 0 & \pminus 0 & \pminus 0 & \pminus 0 & \pminus 0 & \pminus 0 \\
 0 & \pminus 0 & \pminus 0 & \pminus 0 & \pminus 0 & \pminus 0 & \pminus 0 & \pminus 0 & \pminus 0 & \pminus 0 & \pminus 0 & \pminus 0 & \pminus 0 & \pminus 0 & \pminus 0 & \pminus 0 & \pminus 0 & \pminus 0 & \pminus 0 & \pminus
   0 & \pminus 0 & \pminus 0 & \pminus 0 & \pminus 0 & \pminus 0 & \pminus 0 & \pminus 0 & \pminus 0 & \pminus 0 & \pminus 0 & \pminus 0 \\
\end{array}
\right)
}\,,
\end{align}

\begin{align}
\MM_{11} = \scalemath{0.53}{
\left(
\begin{array}{ccccccccccccccccccccccccccccccc}
 0 & \pminus 0 & \pminus 0 & \pminus 0 & \pminus 0 & \pminus 0 & \pminus 0 & \pminus 0 & \pminus 0 & \pminus 0 & \pminus 0 & \pminus 0 & \pminus 0 & \pminus 0 & \pminus 0 & \pminus 0 & \pminus 0 & \pminus 0 & \pminus 0 & \pminus
   0 & \pminus 0 & \pminus 0 & \pminus 0 & \pminus 0 & \pminus 0 & \pminus 0 & \pminus 0 & \pminus 0 & \pminus 0 & \pminus 0 & \pminus 0 \\
 0 & \pminus 0 & \pminus 0 & \pminus 0 & \pminus 0 & \pminus 0 & \pminus 0 & \pminus 0 & \pminus 0 & \pminus 0 & \pminus 0 & \pminus 0 & \pminus 0 & \pminus 0 & \pminus 0 & \pminus 0 & \pminus 0 & \pminus 0 & \pminus 0 & \pminus
   0 & \pminus 0 & \pminus 0 & \pminus 0 & \pminus 0 & \pminus 0 & \pminus 0 & \pminus 0 & \pminus 0 & \pminus 0 & \pminus 0 & \pminus 0 \\
 0 & \pminus 0 & \pminus 0 & \pminus 0 & \pminus 0 & \pminus 0 & \pminus 0 & \pminus 0 & \pminus 0 & \pminus 0 & \pminus 0 & \pminus 0 & \pminus 0 & \pminus 0 & \pminus 0 & \pminus 0 & \pminus 0 & \pminus 0 & \pminus 0 & \pminus
   0 & \pminus 0 & \pminus 0 & \pminus 0 & \pminus 0 & \pminus 0 & \pminus 0 & \pminus 0 & \pminus 0 & \pminus 0 & \pminus 0 & \pminus 0 \\
 0 & \pminus 0 & \pminus 0 & \pminus 0 & \pminus 0 & \pminus 0 & \pminus 0 & \pminus 0 & \pminus 0 & \pminus 0 & \pminus 0 & \pminus 0 & \pminus 0 & \pminus 0 & \pminus 0 & \pminus 0 & \pminus 0 & \pminus 0 & \pminus 0 & \pminus
   0 & \pminus 0 & \pminus 0 & \pminus 0 & \pminus 0 & \pminus 0 & \pminus 0 & \pminus 0 & \pminus 0 & \pminus 0 & \pminus 0 & \pminus 0 \\
 0 & \pminus 0 & \pminus 0 & \pminus 0 & \pminus 0 & \pminus 0 & \pminus 0 & \pminus 0 & \pminus 0 & \pminus 0 & \pminus 0 & \pminus 0 & \pminus 0 & \pminus 0 & \pminus 0 & \pminus 0 & \pminus 0 & \pminus 0 & \pminus 0 & \pminus
   0 & \pminus 0 & \pminus 0 & \pminus 0 & \pminus 0 & \pminus 0 & \pminus 0 & \pminus 0 & \pminus 0 & \pminus 0 & \pminus 0 & \pminus 0 \\
\end{array}
\right)
}\,,
\end{align}

\begin{align}
\MM_{12} = \scalemath{0.53}{
\left(
\begin{array}{ccccccccccccccccccccccccccccccc}
 0 & \pminus 0 & \pminus 0 & \pminus 0 & \pminus 0 & \pminus 0 & \pminus 0 & \pminus 0 & \pminus 0 & \pminus 0 & \pminus 0 & \pminus 0 & \pminus 0 & \pminus 0 & \pminus 0 & \pminus 0 & \pminus 0 & \pminus 0 & \pminus 0 & \pminus
   0 & \pminus 0 & \pminus 0 & \pminus 0 & \pminus 0 & \pminus 0 & \pminus 0 & \pminus 0 & \pminus 0 & \pminus 0 & \pminus 0 & \pminus 0 \\
 0 & \pminus 0 & \pminus 0 & \pminus 0 & \pminus 0 & \pminus 0 & \pminus 0 & \pminus 0 & \pminus 0 & \pminus 0 & \pminus 0 & \pminus 0 & \pminus 0 & \pminus 0 & \pminus 0 & \pminus 0 & \pminus 0 & \pminus 0 & \pminus 0 & \pminus
   0 & \pminus 0 & \pminus 0 & \pminus 0 & \pminus 0 & \pminus 0 & \pminus 0 & \pminus 0 & \pminus 0 & \pminus 0 & \pminus 0 & \pminus 0 \\
 0 & \pminus 0 & \pminus 0 & \pminus 0 & \pminus 0 & \pminus 0 & \pminus 0 & \pminus 0 & \pminus 0 & \pminus 0 & \pminus 0 & \pminus 0 & \pminus 0 & \pminus 0 & \pminus 0 & \pminus 0 & \pminus 0 & \pminus 0 & \pminus 0 & \pminus
   0 & \pminus 0 & \pminus 0 & \pminus 0 & \pminus 0 & \pminus 0 & \pminus 0 & \pminus 0 & \pminus 0 & \pminus 0 & \pminus 0 & \pminus 0 \\
 0 & \pminus 0 & \pminus 0 & \pminus 0 & \pminus 0 & \pminus 0 & \pminus 0 & \pminus 0 & \pminus 0 & \pminus 0 & \pminus 0 & \pminus 0 & \pminus 0 & \pminus 0 & \pminus 0 & \pminus 0 & \pminus 0 & \pminus 0 & \pminus 0 & \pminus
   0 & \pminus 0 & \pminus 0 & \pminus 0 & \pminus 0 & \pminus 0 & \pminus 0 & \pminus 0 & \pminus 0 & \pminus 0 & \pminus 0 & \pminus 0 \\
 0 & \pminus 0 & \pminus 0 & \pminus 0 & \pminus 0 & \pminus 0 & \pminus 0 & \pminus 0 & \pminus 0 & \pminus 0 & \pminus 0 & \pminus 0 & \pminus 0 & \pminus 0 & \pminus 0 & \pminus 0 & \pminus 0 & \pminus 0 & \pminus 0 & \pminus
   0 & \pminus 0 & \pminus 0 & \pminus 0 & \pminus 0 & \pminus 0 & \pminus 0 & \pminus 0 & \pminus 0 & \pminus 0 & \pminus 0 & \pminus 0 \\
\end{array}
\right)
}\,,
\end{align}

\begin{align}
\MM_{13} = \scalemath{0.53}{
\left(
\begin{array}{ccccccccccccccccccccccccccccccc}
 0 & \pminus 0 & \pminus 0 & \pminus 0 & \pminus 0 & \pminus 0 & \pminus 0 & \pminus 0 & \pminus 0 & \pminus 0 & \pminus 0 & \pminus 0 & \pminus 0 & \pminus 0 & \pminus 0 & \pminus 0 & \pminus 0 & \pminus 0 & \pminus 0 & \pminus
   0 & \pminus 0 & \pminus 0 & \pminus 0 & \pminus 0 & \pminus 0 & \pminus 0 & \pminus 0 & \pminus 0 & \pminus 0 & \pminus 0 & \pminus 0 \\
 0 & \pminus 0 & \pminus 0 & \pminus 0 & \pminus 0 & \pminus 0 & \pminus 0 & \pminus 0 & \pminus 0 & \pminus 0 & \pminus 0 & \pminus 0 & \pminus 0 & \pminus 0 & \pminus 0 & \pminus 0 & \pminus 0 & \pminus 0 & \pminus 0 & \pminus
   0 & \pminus 0 & \pminus 0 & \pminus 0 & \pminus 0 & \pminus 0 & \pminus 0 & \pminus 0 & \pminus 0 & \pminus 0 & \pminus 0 & \pminus 0 \\
 0 & \pminus 0 & \pminus 0 & \pminus 0 & \pminus 0 & \pminus 0 & \pminus 0 & \pminus 0 & \pminus 0 & \pminus 0 & \pminus 0 & \pminus 0 & \pminus 0 & \pminus 0 & \pminus 0 & \pminus 0 & \pminus 0 & \pminus 0 & \pminus 0 & \pminus
   0 & \pminus 0 & \pminus 0 & \pminus 0 & \pminus 0 & \pminus 0 & \pminus 0 & \pminus 0 & \pminus 0 & \pminus 0 & \pminus 0 & \pminus 0 \\
 -1 & \pminus 0 & \pminus 0 & -2 & \pminus 0 & \pminus 0 & \pminus 0 & \pminus 0 & \pminus 0 & \pminus 0 & \pminus 0 & \pminus 0 & \pminus 0 & \pminus 0 & \pminus 0 & \pminus 0 & \pminus 0 & \pminus 0 & \pminus 0
   & \pminus 0 & \pminus 0 & \pminus 0 & \pminus 0 & \pminus 0 & \pminus 0 & \pminus 0 & \pminus 0 & \pminus 0 & \pminus 0 & \pminus 0 & \pminus 0 \\
 0 & \pminus 0 & \pminus 0 & \pminus 0 & \pminus 0 & \pminus 0 & \pminus 0 & \pminus 0 & \pminus 0 & \pminus 0 & \pminus 0 & \pminus 0 & \pminus 0 & \pminus 0 & \pminus 0 & \pminus 0 & \pminus 0 & \pminus 0 & \pminus 0 & \pminus
   0 & \pminus 0 & \pminus 0 & \pminus 0 & \pminus 0 & \pminus 0 & \pminus 0 & \pminus 0 & \pminus 0 & \pminus 0 & \pminus 0 & \pminus 0 \\
\end{array}
\right)
}\,,
\end{align}

\begin{align}
\MM_{14} = \scalemath{0.53}{
\left(
\begin{array}{ccccccccccccccccccccccccccccccc}
 0 & \pminus 0 & \pminus 0 & \pminus 0 & \pminus 0 & \pminus 0 & \pminus 0 & \pminus 0 & \pminus 0 & \pminus 0 & \pminus 0 & \pminus 0 & \pminus 0 & \pminus 0 & \pminus 0 & \pminus 0 & \pminus 0 & \pminus 0 & \pminus 0 & \pminus
   0 & \pminus 0 & \pminus 0 & \pminus 0 & \pminus 0 & \pminus 0 & \pminus 0 & \pminus 0 & \pminus 0 & \pminus 0 & \pminus 0 & \pminus 0 \\
 -1 & -2 & \pminus 0 & \pminus 0 & \pminus 0 & \pminus 0 & \pminus 0 & \pminus 0 & \pminus 0 & \pminus 0 & \pminus 0 & \pminus 0 & \pminus 0 & \pminus 0 & \pminus 0 & \pminus 0 & \pminus 0 & \pminus 0 & \pminus 0
   & \pminus 0 & \pminus 0 & \pminus 0 & \pminus 0 & \pminus 0 & \pminus 0 & \pminus 0 & \pminus 0 & \pminus 0 & \pminus 0 & \pminus 0 & \pminus 0 \\
 0 & \pminus 0 & \pminus 0 & \pminus 0 & \pminus 0 & \pminus 0 & \pminus 0 & \pminus 0 & \pminus 0 & \pminus 0 & \pminus 0 & \pminus 0 & \pminus 0 & \pminus 0 & \pminus 0 & \pminus 0 & \pminus 0 & \pminus 0 & \pminus 0 & \pminus
   0 & \pminus 0 & \pminus 0 & \pminus 0 & \pminus 0 & \pminus 0 & \pminus 0 & \pminus 0 & \pminus 0 & \pminus 0 & \pminus 0 & \pminus 0 \\
 0 & \pminus 0 & \pminus 0 & \pminus 0 & \pminus 0 & \pminus 0 & \pminus 0 & \pminus 0 & \pminus 0 & \pminus 0 & \pminus 0 & \pminus 0 & \pminus 0 & \pminus 0 & \pminus 0 & \pminus 0 & \pminus 0 & \pminus 0 & \pminus 0 & \pminus
   0 & \pminus 0 & \pminus 0 & \pminus 0 & \pminus 0 & \pminus 0 & \pminus 0 & \pminus 0 & \pminus 0 & \pminus 0 & \pminus 0 & \pminus 0 \\
 0 & \pminus 0 & \pminus 0 & \pminus 0 & -2 & -4 & \pminus 0 & \pminus 0 & \pminus 0 & \pminus 0 & \pminus 0 & \pminus 0 & \pminus 0 & \pminus 0 & \pminus 0 & \pminus 0 & \pminus 0 & \pminus 0 & \pminus 0
   & \pminus 0 & \pminus 0 & \pminus 0 & \pminus 0 & \pminus 0 & \pminus 0 & \pminus 0 & \pminus 0 & \pminus 0 & \pminus 0 & \pminus 0 & \pminus 0 \\
\end{array}
\right)
}\,,
\end{align}

\begin{align}
\MM_{15} = \scalemath{0.53}{
\left(
\begin{array}{ccccccccccccccccccccccccccccccc}
 0 & \pminus 0 & \pminus 0 & \pminus 0 & \pminus 0 & \pminus 0 & \pminus 0 & \pminus 0 & \pminus 0 & \pminus 0 & \pminus 0 & \pminus 0 & \pminus 0 & \pminus 0 & \pminus 0 & \pminus 0 & \pminus 0 & \pminus 0 & \pminus 0 & \pminus
   0 & \pminus 0 & \pminus 0 & \pminus 0 & \pminus 0 & \pminus 0 & \pminus 0 & \pminus 0 & \pminus 0 & \pminus 0 & \pminus 0 & \pminus 0 \\
 0 & \pminus 0 & \pminus 0 & \pminus 0 & \pminus 0 & \pminus 0 & \pminus 0 & \pminus 0 & \pminus 0 & \pminus 0 & \pminus 0 & \pminus 0 & \pminus 0 & \pminus 0 & \pminus 0 & \pminus 0 & \pminus 0 & \pminus 0 & \pminus 0 & \pminus
   0 & \pminus 0 & \pminus 0 & \pminus 0 & \pminus 0 & \pminus 0 & \pminus 0 & \pminus 0 & \pminus 0 & \pminus 0 & \pminus 0 & \pminus 0 \\
 0 & \pminus 0 & \pminus 0 & \pminus 0 & \pminus 0 & \pminus 0 & \pminus 0 & \pminus 0 & \pminus 0 & \pminus 0 & \pminus 0 & \pminus 0 & \pminus 0 & \pminus 0 & \pminus 0 & \pminus 0 & \pminus 0 & \pminus 0 & \pminus 0 & \pminus
   0 & \pminus 0 & \pminus 0 & \pminus 0 & \pminus 0 & \pminus 0 & \pminus 0 & \pminus 0 & \pminus 0 & \pminus 0 & \pminus 0 & \pminus 0 \\
 0 & \pminus 0 & \pminus 0 & \pminus 0 & \pminus 0 & \pminus 0 & \pminus 0 & \pminus 0 & \pminus 0 & \pminus 0 & \pminus 0 & \pminus 0 & \pminus 0 & \pminus 0 & \pminus 0 & \pminus 0 & \pminus 0 & \pminus 0 & \pminus 0 & \pminus
   0 & \pminus 0 & \pminus 0 & \pminus 0 & \pminus 0 & \pminus 0 & \pminus 0 & \pminus 0 & \pminus 0 & \pminus 0 & \pminus 0 & \pminus 0 \\
 0 & \pminus 0 & \pminus 0 & \pminus 0 & \pminus 0 & \pminus 0 & \pminus 0 & \pminus 0 & \pminus 0 & \pminus 0 & \pminus 0 & \pminus 0 & \pminus 0 & \pminus 0 & \pminus 0 & \pminus 0 & \pminus 0 & \pminus 0 & \pminus 0 & \pminus
   0 & \pminus 0 & \pminus 0 & \pminus 0 & \pminus 0 & \pminus 0 & \pminus 0 & \pminus 0 & \pminus 0 & \pminus 0 & \pminus 0 & \pminus 0 \\
\end{array}%
\right)
}\,.
\end{align}

\bibliographystyle{JHEP}
\bibliography{references}

\end{document}